\documentclass[11pt]{article}

\usepackage[utf8]{inputenc}
\usepackage[T1]{fontenc}
\usepackage[english]{babel}
\usepackage[a4paper,nohead]{geometry}
\usepackage{amsmath}
\usepackage[mathscr]{euscript}
\usepackage{amsfonts}
\usepackage{amssymb}
\usepackage{amsthm}
\usepackage{mathtools}
\usepackage{enumitem}
\usepackage{xspace}
\usepackage{cite}

\usepackage[bookmarks=true,
           bookmarksopen=true,hidelinks]{hyperref}

\newtheorem{theorem}{Theorem}[section] 
\newtheorem{definition}[theorem]{Definition}
\newtheorem{lemma}[theorem]{Lemma}

\newtheorem{proposition}[theorem]{Proposition}

\theoremstyle{remark}
\newtheorem{remark}[theorem]{Remark}
\numberwithin{equation}{section}
\numberwithin{footnote}{section}

\newcommand {\R} {\mathbb R}

\newcommand{\Eqref}[1]{Eq.~\eqref{#1}}
\newcommand{\Eqsref}[1]{Eqs.~\eqref{#1}}
\newcommand{\Sectionref}[1]{Section~\ref{#1}}  
\newcommand{\Defref}[1]{Definition~\ref{#1}}
\newcommand{\Lemref}[1]{Lemma~\ref{#1}}
\newcommand{\Propref}[1]{Proposition~\ref{#1}}
\newcommand{\Theoremref}[1]{Theorem~\ref{#1}}

\newcommand{\keyword}[1]{\emph{#1}}

\newcommand{\unksymb}{v}
\newcommand{\vnksymb}{u}
\newcommand{\nunksymb}{\nu}
\newcommand{\unkgen}{\unksymb}
\newcommand{\vnkgen}{\vnksymb}
\newcommand{\nunkgen}{\nunksymb}
\newcommand{\unk}{{\unksymb_1}}
\newcommand{\unkt}{{\unksymb_2}}
\newcommand{\vnk}{{\vnksymb_1}}
\newcommand{\vnkt}{{\vnksymb_2}}
\newcommand{\nunk}{{\nunksymb_1}}
\newcommand{\nunkt}{{\nunksymb_2}}
\newcommand{\unkd}{w}

\newcommand{\sourcegen}{F}
\newcommand{\source}{\sourcegen_1}
\newcommand{\sourcet}{\sourcegen_2}
\newcommand{\switchgen}{\sigma}
\newcommand{\switch}{\sigma_1}
\newcommand{\switcht}{\sigma_2}
\newcommand{\switchNLgen}{\boldsymbol\sigma}

\newcommand{\matchtime}{\tau}
\newcommand{\finaltime}{T}

\newcommand{\FiniteMatchingMapSymbol}{\psi}
\newcommand{\AsymptoticMatchingMapSymbol}{\Psi}
\newcommand{\FiniteMatchingMapAbstr}[1]{\FiniteMatchingMapSymbol_{#1}\xspace}

\newcommand{\AsymptoticMatchingMapAbstr}[1]{\AsymptoticMatchingMapSymbol_{#1}\xspace}
\newcommand{\AsymptoticMatchingMapRegularityAbstr}[2]{\AsymptoticMatchingMapSymbol_{#1,#2}\xspace}
\newcommand{\AsymptoticMatchingMapIndexRegularityAbstr}[3]{\AsymptoticMatchingMapSymbol^{(#1)}_{#2,#3}\xspace}
\newcommand{\AsymptoticMatchingMapIndexAbstr}[2]{\AsymptoticMatchingMapSymbol^{(#1)}_{#2}\xspace}
\newcommand{\PreMatchingMapAbstr}{\AsymptoticMatchingMapSymbol^{(\mathrm{pre})}\xspace}

\newcommand{\energysymbol}{E}
\newcommand{\littleenergysymbol}{e}
\newcommand{\littleenergywk}[3]{\littleenergysymbol_{#3}[#1]\xspace} 
\newcommand{\littleenergy}[2]{\littleenergysymbol_{#2}[#1]\xspace}
\newcommand{\littleenergyswitch}[3]{\littleenergysymbol_{#2,#3}[#1]\xspace}
\newcommand{\energy}[3]{\energysymbol_{#2,#3}[#1]\xspace}
\newcommand{\energyswitch}[4]{\energysymbol_{#3,#2,#4}[#1]\xspace}

\newcommand{\GenCD}[2]{{#2}_{[#1]}}
\newcommand{\CDfinaltime}[1]{{#1}_*}
\newcommand{\LOT}[1]{{#1}_{(0)}}

\newcommand{\Lop}[1]{\mathcal L[#1]}

\newcommand{\BG}[1]{\mathring{#1}}

\title{Contracting asymptotics of the
 linearized lapse-scalar field sub-system of the Einstein-scalar field equations}
\author{
Ellery Ames\footnote{Department of Mathematics, Humboldt State University, 1 Harpst St. Arcata, CA, USA. Email: ellery.ames@humboldt.edu.}\and 
Florian Beyer\footnote{Department of Mathematics and Statistics, University of Otago, P.O. Box 56, Dunedin 9054, New Zealand. Email: fbeyer@maths.otago.ac.nz.}\and
James Isenberg\footnote{Department of Mathematics, University of Oregon, Eugene, OR 97403, USA. Email: isenberg@uoregon.edu.}
}
\date{\today}

\begin{document}

\maketitle

\begin{abstract}
We prove an asymptotic stability result for a linear coupled hyper\-bolic--elliptic system on a large class of singular background spacetimes in CMC gauge on the $n$-torus. At each spatial point these background spacetimes are perturbations of Kasner-like solutions of the Einstein-scalar field equations which are not required to be close to the homogeneous and isotropic case. We establish the existence of a homeomorphism between Cauchy data for this system and a set of functions naturally associated with the asymptotics in the contracting direction, which we refer to as asymptotic data. This yields a complete characterization of the degrees of freedom of all solutions of this system in terms of their asymptotics. Spatial derivative terms can in general not be fully neglected which yields a clarification of the notion of \emph{asymptotic velocity term dominance} (AVTD).
\end{abstract}

\section{Introduction}
\label{sec:intro}
Do cosmological solutions of Einstein's equations exhibit any generic coherent behavior in the neighborhood of the Big Bang singularity? Mathematically, a cosmological solution is a globally hyperbolic Lorentzian manifold with closed Cauchy hypersurfaces. 
Certain symmetry-defined classes of vacuum cosmological spacetimes have infinite dimensional families of solutions which exhibit asymptotically velocity term dominated (AVTD) type behavior near the singularity \cite{eardley1972,isenberg1990}. Such solutions are modeled in the limit towards the Big Bang singularity (i.e., in the contracting time direction) by solutions to an \emph{asymptotic model system} -- the so-called \emph{VTD system} -- in which spatial derivative terms are dropped from the equations. The Gowdy vacuum solutions in areal gauge \cite{kichenassamy1998,rendall2000,beyer2010b} and wave gauge \cite{ames2017}, the polarized and half polarized $T^2$-symmetric vacuum solutions in areal gauge \cite{ames2013a,isenberg1999}, and the polarized and half polarized U(1)-symmetric vacuum solutions in wave-type gauges \cite{choquet-bruhat2004,choquet-bruhat2006,isenberg2002} all have infinite dimensional families of solutions which exhibit AVTD type behavior near the singularity. 

Heuristic and numerical studies suggest that the general set of cosmological solutions do not exhibit AVTD behavior in any gauge near the singularity \cite{berger1998,berger2000,berger1998a,berger2001,weaver1998,berger1998b}. Rather, it is conjectured that general solutions have oscillatory behavior, with an infinite sequence of AVTD-like epochs. While this so-called ``BKL conjecture'' (named after the authors of \cite{belinskii1982,belinskii1970}) is  believed to be very difficult to check rigorously, it is also known to fail in cases where weak null singularities form \cite{luk2017}.

In contrast to the general cosmological setting, there are reasons to suspect that generic solutions of the Einstein system coupled to the \emph{scalar field} equations may exhibit AVTD behavior near the singularity. 
Several authors \cite{barrow1978,belinski1973,eardley1972} have noted that if the matter is modeled by a scalar field (or a fluid with a stiff equation of state) the singular dynamics is monotonic rather than oscillatory in some cases, greatly simplifying the mathematical analysis. This simplification has since been exploited, first, in work by Andersson and Rendall \cite{andersson2001}, who prove, using Fuchsian methods (see below), the existence of an infinite dimensional family of solutions to the Einstein--scalar field system with VTD asymptotics.
These solutions are not limited to being close to the homogeneous and isotropic (FLRW) solution. While the family of solutions obtained is general in the sense of function counting, the question of whether the VTD asymptotic data yield solutions corresponding to an open set of Cauchy data is not addressed by the Fuchsian methods (particularly, but not exclusively, if restricted to the analytic category). 
The question of openness and stability is addressed in remarkable recent work by Rodnianski and Speck \cite{rodnianski2014,rodnianski2018}. The authors show that AVTD behavior is nonlinearly stable in the sense that there is an open set around FLRW in which AVTD behavior holds.

In their work, Rodnianski and Speck use the degrees of freedom of the regular Cauchy problem of the Einstein-scalar field equations to establish a notion of stability. We refer to these as \emph{Cauchy data} or \emph{Cauchy degrees of freedom}. They show that certain quantities, which characterize some of the asymptotics of the solutions at the singularity, depend continuously (in a sense which the authors make precise) on these Cauchy degrees of freedom -- at least if the Cauchy data are close to FLRW data; some more details are given below. This result is a major breakthrough and it is the first time that singular solutions of Einstein's equations without symmetry could be controlled in this detail.  It turns out however that their result does not provide sufficient control of the asymptotics to address the reverse problem: \emph{Do the Cauchy degrees of freedom of the solutions also depend continuously (in some sense) on the asymptotics?} Is there is a description of the asymptotics of general solutions in terms of \emph{asymptotic data} which are in one-to-one correspondence with the Cauchy data? If yes, do the VTD equations, that is, the asymptotic model equations where spatial derivatives are dropped, imply the same asymptotic degrees of freedom? Notice that spatial derivatives turn out to be negligible for the result by Rodnianski and Speck, but may very well be significant to address this problem here because it requires a much tighter control of the asymptotics. 

Investigating these questions for the full Einstein--scalar field system with its large number of degrees of freedom appears out of reach in light of the technical arguments required for the results of \cite{rodnianski2014,rodnianski2018}. The main simplification which we introduce in this paper is therefore to \emph{freeze some of the degrees of freedom}. By this we mean that some of the equations are not imposed and the corresponding unknowns are treated as fixed (but arbitrary) background fields. As we explain in detail in \Sectionref{sec:lsf}, we have decided to freeze the degrees of freedom associated with the spatial metric $\gamma^{ab}$ and the trace-free part of the second fundamental form ${\chi_a}^b$ in the Einstein-scalar system in CMC gauge with zero shift. These quantities are consequently treated as arbitrary, but fixed, fields which we write as $\BG\gamma^{ab}$ and ${\BG\chi_a}^{\hspace{1ex}b}$. The remaining degrees of freedom, which our paper consequently focusses on, are those associated with the scalar field $\phi$, its conjugate momentum $\pi = t \partial_t \phi/\alpha$, and the lapse $\alpha$, which are governed by the coupled hyperbolic-elliptic system
\begin{gather}
    \label{eq:scalarfieldnonlinear}
  -t\partial_t\pi
  +(1-\alpha)\pi
  +\alpha t^2\BG\gamma^{ab}D_a D_b\phi  
  +t^2\BG\gamma^{ab}D_a \alpha D_b\phi=0,\\
  \label{eq:lapsenonlinear}
  t^2\BG\gamma^{ab}D_a D_b\alpha
  -\left(t^2 {\BG\chi_a}^{\hspace{1ex}b}{\BG\chi_b}^{\hspace{1ex}a}
  +\frac {1}{n}
  +\pi^2\right)\alpha
  +1=0,\\
  \label{eq:pinonlinear}
  -t\partial_t\phi+\alpha\pi=0,
\end{gather}
for $t>0$.
Notice that $D_a$ is the covariant derivative associated with $\BG\gamma$.
We refer to \Eqsref{eq:scalarfieldnonlinear} -- \eqref{eq:pinonlinear}  as the \emph{lapse-scalar field system}. We expect this system to have two Cauchy degrees of freedom, one associated with each of $\phi$ and $\pi$. As $\alpha$ is a solution of the elliptic equation \eqref{eq:lapsenonlinear} there should not be any freedom associated with it. 
It turns out that one can identify two \emph{asymptotic} degrees of freedom -- referred to as \emph{asymptotic data} -- which fully characterize the asymptotics at $t=0$. Moreover, by suitably taking spatial derivatives into account we establish the existence of a homeomorphism between this asymptotic data and Cauchy data in a natural topology. 
Crucially, the asymptotics of the fields $\phi$ and $\pi$ rely on taking into account the coupling between the hyperbolic part \Eqsref{eq:scalarfieldnonlinear}, \eqref{eq:pinonlinear} and the elliptic part \Eqref{eq:lapsenonlinear} of the lapse-scalar field equations. 
This feature is not seen, for example in the study of wave equations on cosmological background spacetimes in \cite{alho2019}, due to the lack of a coupled elliptic equation in that setting. 

We expect that subsystem \Eqsref{eq:scalarfieldnonlinear} --
\eqref{eq:pinonlinear} captures the asymptotics of the scalar field
and the lapse in the full system if $\gamma^{ab}$ and
${\chi_a}^{\hspace{1ex}b}$ have asymptotics that are consistent with
the given background fields $\BG\gamma^{ab}$ and
${\BG\chi_a}^{\hspace{1ex}b}$. 
If the background fields $\BG\gamma^{ab}$ and ${\BG\chi_a}^{\hspace{1ex}b}$ are chosen to be close to the FRLW solution, then these asymptotics should be consistent with those established by Rodnianski and Speck.
For example in \cite{rodnianski2018} it is proved that for each solution of the linearized Einstein-scalar field system which is sufficiently close to the FLRW solution, there is a function $\Psi_{Bang}(x)$ such that\footnote{See \cite{rodnianski2018} for more information regarding the norms and the constants $C,c>0$. Similar statements hold for the nonlinear problem \cite{rodnianski2014}.}
\begin{gather*}
  \|t\partial_t\varphi(t,\cdot)-\Psi_{Bang}\|\le C t^{2/3-c\eta},\quad
  \|\partial_i\varphi(t,\cdot)-\log t\partial_i\Psi_{Bang}\|\le C, 
\end{gather*}
in terms of some smallness parameter $\eta>0$ where $\varphi$ is the
linearized scalar field. The asymptotic quantity $\Psi_{Bang}$ depends
continuously on the Cauchy data. The reverse problem which largely
motivates our work here is: Given an arbitrary smooth function $\Psi_{Bang}$,
is there always a uniquely determined solution $\varphi$ which
approaches $\Psi_{Bang}$ in the sense of these estimates? It is clear
that this cannot be the case. As mentioned above, the scalar field
$\varphi$ comes with {two degrees of
  freedom}, 
and one must therefore go beyond that \emph{single} asymptotic datum
$\Psi_{Bang}$. In fact, it turns out that $\Psi_{Bang}$ is completely
determined by the asymptotics of $\BG\gamma^{ab}$ and
${\BG\chi_a}^{\hspace{1ex}b}$ and not by the scalar field equation at
all. The quantity $\Psi_{Bang}$ is therefore considered as a
\emph{frozen field} here. This implies that the actual degrees of
freedom associated with the scalar field are not covered by the above
estimates at all. In our analysis below, which allows us to identify
the actual degrees of freedom, it turns out to be useful to subtract
this ``frozen asymptotic datum'' $\Psi_{Bang}$ (which we label as $A$
below) off from the scalar field quantity.

The particular goal for our work is to find and characterize a
homeomorphism $\Psi$ between Cauchy data and asymptotic data for the
lapse-scalar field system under conditions as general as possible, in
particular not restricting to near-FLRW backgrounds. In doing this we
wish to derive as much detail regarding the asymptotics as
possible. Since detail is therefore the paramount concern here, we
restrict to the linearization of the lapse-scalar field system;
see \Sectionref{sec:lsf}. We conjecture that
the estimates we establish here for the linearized system will suffice
to derive essentially the same statements for the nonlinear system (in
some small data setting). We will obtain the corresponding estimates
for the nonlinear system in future work exploiting recent results
regarding a large class of Fuchsian-type equations in
\cite{oliynyk2016,beyer2019a}.

Our main result, which is given in rigorous form in \Theoremref{thm:maintheorem}, can be stated in rough terms as follows. For given $\BG\gamma^{ab}$ and ${\BG\chi_a}^{\hspace{1ex}b}$, which at each spatial point decay to the time-dependent gravitational data for a Kasner-scalar space-time (see \Sectionref{sec:Kasnerscalarfieldspacetimes}), 
we consider the linearization of the system \Eqsref{eq:scalarfieldnonlinear} -- \eqref{eq:pinonlinear} on the $n$-torus as discussed in \Sectionref{sec:lsf}. We prove that there is a homeomorphism 
with respect to a natural topology between Cauchy data (prescribed at a regular time $\finaltime>0$) and asymptotic data (associated with the limit at the singular time $t=0$). 
The existence of such a homeomorphism does not only yield that the full degrees of freedom of the solution set are equivalently characterized by the Cauchy data or  asymptotic data. 
It also follows that Cauchy data depend continuously on the asymptotic data and vice versa which precludes chaotic behavior. Most importantly, however, both this map and its inverse are open maps which implies that any \emph{perturbation} of Cauchy data (or asymptotic data), i.e., any change within an open set, guarantees that the corresponding asymptotic data (or Cauchy data) also change within some open set. 
We interpret this as a form of stability related to (but not the same as) asymptotic stability. We remark that the framework of \textit{asymptotic matching problems}, which we develop in this paper and which we use for the proof, is not only useful to establish the \emph{existence} of the homeomorphism. It also yields a practical way to \emph{calculate} it numerically by approximating the \emph{asymptotic} matching problem by \emph{finite} ones with (in principle) arbitrary accuracy.

The \emph{Fuchsian method} has been used in a number of cases
\cite{kichenassamy1998,rendall2000,beyer2010b,
  ames2013a,ames2013b,kichenassamy2007k,claudel1998a} to identify
\emph{asymptotic} degrees of freedom -- \emph{asymptotic data} --
(often determined heuristically by formal expansions at the
singularity) and then for rigorously validating this identification
(by proving the existence of a solution of the equation which realizes
each choice of asymptotic data in some well-defined sense). It
typically yields \emph{continuous} maps from asymptotic data to Cauchy
data. The main drawback is that the results based on the Fuchsian
method make no statements about
potential openness of the maps.
This issue has been addressed for the first time in comprehensive work on linear wave equations by Ringstr\"{o}m \cite{ringstrom2017}. 
Among other things, Ringstr\"{o}m proves the existence (and also the non-existence) of homeomorphisms between Cauchy data and asymptotic data for a large class (defined by several technical conditions which we do not describe here) of linear wave equations with spatially homogeneous coefficients. 
Before proceeding, we make a couple of further remarks relating our main theorem to the results described above.

\begin{remark}
  The method we use in this paper extends some of the general theory
  of \cite{ringstrom2017} in important ways.  First, the system of
  equations we consider here has spatially non-homogeneous
  coefficients; indeed, the background metric is allowed to approach
  (with suitable decay estimates) a different solution of the
  Kasner--scalar field system at each spatial point. Second,
  Ringstr\"{o}m studies the case in which the asymptotics are
  determined by solutions of model asymptotic equations in which the
  spatial derivative terms are negligible.
However, to fully characterize solutions of the linearized lapse-scalar field system by their asymptotics and obtain a homeomorphism between Cauchy data and asymptotic data, certain spatial derivative terms must however be accounted for and included in the asymptotic model equation. Finally our results are not restricted to wave equations (as opposed to \cite{ringstrom2017}) as we take the elliptic lapse equation into account as well.
\end{remark}

\begin{remark}
\label{rmk:ImportanceOfSpatialDerivatives}
\Theoremref{thm:maintheorem} provides a characterization of the asymptotics of solutions to the linearized lapse-scalar field system that contains the full set of degrees of freedom for solutions of these equations. A surprising result is that while spatial derivative terms do not contribute to the very leading-order terms (consistent with AVTD behavior), the full set of degrees of freedom can only be described in terms of the asymptotics by incorporating spatial derivative terms of sufficiently significant order. 
This shows that, at least for the linearized lapse-scalar field system, the VTD-equations (where these spatial derivative terms are dropped) do not suffice to describe the full asymptotic behavior of generic solutions.
\end{remark}

\begin{remark}
Our results are not limited to backgrounds which are FLRW or near FLRW, but may include highly inhomogeneous anisotropic backgrounds. As a consequence we must deal with additional terms which are not present in the corresponding estimates in \cite{rodnianski2018} where the exact FLRW background is assumed. In order to deal with these additional terms we find it useful to (I) carefully organize the estimates as discussed in \Sectionref{sec:LSFProofs}, and, (II) work consistently with the conjugate momentum $\pi=t\partial_t\phi/\alpha$ (defined by \Eqref{eq:pinonlinear}) as opposed to $t\partial_t\phi$. As a result, we are, in particular, able to remove all time derivatives of $\alpha$ from the estimates, which would otherwise spoil some of the arguments. The removal of these terms is key to the derivation of useful estimates for both evolutions \emph{towards the singular time} $t = 0$ (the decreasing time direction) as well as \emph{away from the singular time} (the increasing time direction). 
As opposed to the works \cite{rodnianski2014,rodnianski2018}, where
only estimates in the decreasing time direction required, both time
directions are necessary for constructing the homeomorphism determined
above.
\end{remark}

The outline of the paper is as follows. In \Sectionref{sec:background} we provide some necessary preliminaries. In particular we derive the lapse-scalar field sub-system from the ADM formulation of the Einstein-scalar field equations in CMC gauge with zero shift in \Sectionref{sec:lsf}. \Sectionref{sec:Kasnerscalarfieldspacetimes} is devoted to  Kasner-scalar field spacetimes which play an essential role for our analysis. In \Sectionref{sec:mainresults} we present and discuss our main result. Before we discuss the proof in several sub-steps in \Sectionref{sec:LSFProofs}, we introduce a new methodology in \Sectionref{sec:modelproblemmatching}, i.e., a systematic way to \emph{asymptotically match} solutions of (in principle) any system of equations of interest with certain asymptotic \emph{model equations}, essentially following the paradigm used to define the AVTD property in \cite{isenberg1990}. This approach is the key to our proof of our main result in \Sectionref{sec:LSFProofs}. We believe that it is a robust approach which has the potential to apply to more general classes of problems. In particular it allows us to reach a satisfactory level of detail in describing the asymptotics of the solutions of our system of equations.

\section{Preliminaries}
\label{sec:background}

\subsection{The lapse-scalar field system}
\label{sec:lsf}

In this section we briefly discuss the origin of the lapse-scalar field system \Eqsref{eq:scalarfieldnonlinear} -- \eqref{eq:pinonlinear}. The starting point is the standard ADM formulation of the Einstein-(minimally coupled) scalar field system (with zero potential) under the CMC-zero shift gauge condition 
\begin{equation}
    \label{eq:gaugecond}
    K=-1/t,\quad \beta^a=0.
\end{equation}
Here, and throughout the paper Latin indices $a, b, \ldots$ take values $1,2, \ldots, n$ (and therefore represent purely spatial tensors), while Greek indices $\mu, \nu, \ldots$ take values $0, 1, \ldots, n$ (and therefore represent spacetime tensors). With the above gauge choice, the Einstein-scalar field system can be written as in \cite{rodnianski2018}:
\begin{enumerate}
\item Constraint equations:
\begin{align}
  \label{eq:nonlineareq1}
  \mathrm{Scal}[\gamma]-{\chi_a}^{b}{\chi_b}^{a}
  +\frac {n-1}{nt^2}
  &=\frac 1{\alpha^2} \dot\phi^2+\gamma^{a b} D_a\phi D_b\phi,\\
  D_b {\chi_a}^{b}&=-\frac 1{\alpha} \dot\phi D_a\phi,\quad
  {\chi_a}^{a}=0.
\end{align}  
\item Evolution equations for $\gamma^{{a}b}$ and ${\chi_{a}}^{b}$:
\begin{align}
  \label{eq:nonlineargammaevol}
  \dot\gamma^{{a}b}&=2\alpha {\chi_{c}}^{a}\gamma^{c b}
                   -\frac{2\alpha}{n t} \gamma^{{a}b},\\
\label{eq:nonlinearkappaevol}
  \dot {\chi_{a}}^{b}
&=-\gamma^{b c}D_{a} D_c\alpha
  -\frac \alpha t {\chi_{a}}^{b}
  +\frac {\alpha-1}{n t^2} {\delta_{a}}^{b}
  +\alpha {\mathrm{Ric}[\gamma]_{a}}^{b}
  -\alpha \gamma^{b c}D_{a}\phi D_c\phi.
\end{align}
\item  Scalar field equation:
\begin{equation}
\label{eq:scalarfieldsecondorder}
  -\frac 1\alpha \frac{d}{dt}\left(\frac 1\alpha \dot
    \phi\right)
  -\frac 1{\alpha t} \dot\phi
  +\gamma^{{a}b}D_{a} D_b\phi
  +\frac 1\alpha\gamma^{{a}b}D_{a} \alpha D_b\phi=0.
\end{equation}
\item Lapse equation:
\begin{equation}
  \label{eq:nonlineareqlastnontrival}
  \gamma^{{a}b}D_{a} D_b\alpha=\frac{\alpha-1}{t^2} 
  +\alpha \mathrm{Scal}[\gamma] -\alpha \gamma^{{a}b}D_{a}\phi D_b \phi.  
\end{equation}
\end{enumerate}
Here, $\gamma^{{a}b}$ is the $3$-metric, ${\chi_{{a}}}^{b}$ the tracefree part of the second fundamental form, $\alpha$ is the lapse and $\phi$ is the scalar field. A dot refers to the time derivative $\partial_t$, and $D_{a}$ is the Levi-Civita connection compatible with $\gamma$. In order to extract the lapse-scalar field system \Eqsref{eq:scalarfieldnonlinear} -- \eqref{eq:pinonlinear} now we first eliminate $\mathrm{Scal}[\gamma]$ from \Eqref{eq:nonlineareqlastnontrival} using \Eqref{eq:nonlineareq1}. This equation together with 
 \Eqref{eq:scalarfieldsecondorder} then takes the form
  \begin{gather}
    \label{eq:scalarfieldnonlinearorig}
  -t\partial_t(t\partial_t\phi)
  +(1-\alpha)t\partial_t\phi 
  +\frac {t\partial_t\alpha\, t\partial_t\phi}{\alpha} 
  +\alpha^2t^2\gamma^{{a}b}D_{a} D_b\phi  
  +\alpha t^2\gamma^{{a}b}D_{a} \alpha D_b\phi=0,\\
  \label{eq:lapsenonlinearorig}
  t^2\gamma^{{a}b}D_{a} D_b\alpha
  -\alpha t^2{\chi_{a}}^{b}{\chi_b}^{{a}}
  -\frac {\alpha-n}{n}
  -\frac 1{\alpha} (t\partial_t\phi)^2
  =0.
\end{gather}
\Eqsref{eq:scalarfieldnonlinear} -- \eqref{eq:pinonlinear} are
obtained from \eqref{eq:scalarfieldnonlinearorig} --
\eqref{eq:lapsenonlinearorig} by defining
$\pi=t\partial_t\phi/\alpha$, replacing $t\partial_t\phi$ by
$\pi\alpha$ and by considering the metric and the trace-free part of
the second fundamental form as arbitrary, but fixed, fields
$\BG\gamma^{ab}$ and ${\BG\chi_a}^{\hspace{1ex}b}$.  Below we shall
demand that these fields have certain asymptotics related to the form
introduced in \Sectionref{sec:Kasnerscalarfieldspacetimes} in the
limit $t\searrow 0$.  If we had written the equations in terms of
$t\partial_t\phi$ instead of $\pi$, then the time-derivative of
$\alpha$ would need to be handled, which becomes an issue if the
background space-time is not presumed to be spatially
homogeneous. Recall that in working with
\Eqsref{eq:scalarfieldnonlinear} -- \eqref{eq:pinonlinear}, we do not
impose any of the remaining equations above in all of what follows.

In addition to $\BG\gamma^{ab}$ and ${\BG\chi_a}^{\hspace{1ex}b}$, fix now arbitrary smooth fields $\BG\phi(t,x)$, $\BG\alpha(t,x)$, $\BG\pi(t,x)$
such that
\begin{gather}
    \label{eq:scalarfieldnonlinearbg}
  -t\partial_t \BG\pi
  +(1-\BG\alpha) \BG\pi
  +\BG\alpha t^2 \BG\gamma^{{a}b}D_{a} D_b \BG\phi  
  +t^2 \BG\gamma^{{a}b}D_{a} \BG\alpha D_b \BG\phi=-f^{(1)},\\
  \label{eq:lapsenonlinearbg}
  t^2 \BG\gamma^{{a}b}D_{a} D_b \BG\alpha
  -t^2 {\BG\chi_a}^{\hspace{1ex}b}{\BG\chi_b}^{\hspace{1ex}a} \BG\alpha
  -\frac {\BG\alpha}{n}
  -{\BG\alpha} \BG\pi^2
  +1=-f^{(2)},\\
  \label{eq:pinonlinearbg}
  -t\partial_t \BG\phi+\BG\alpha \BG\pi=-f^{(3)},
\end{gather}
for some smooth functions $f^{(1)}(t,x)$, $f^{(2)}(t,x)$ and $f^{(3)}(t,x)$. If these latter functions are zero than we interpret $(\BG\phi, \BG\alpha, \BG\pi)$ as a \emph{background solution} of the lapse-scalar field system.
In general we shall impose specific conditions on the $t \searrow 0$ asymptotics of these residuals $f^{(1)}, f^{(2)}, f^{(3)}$ in \Theoremref{thm:maintheorem} below. Given this, the fields
\begin{equation}
  \label{eq:defdiffquant}
  u=\pi-\BG\pi,\quad \nu=\alpha-\BG\alpha,\quad \varphi=\phi-\BG\phi,
\end{equation}
satisfy the system
\begin{gather}
    \label{eq:scalarfieldnonlineardiff}
  \begin{split}
  -t\partial_tu
  -\BG\pi\nu
  &+(1-\BG\alpha) u  
  +\BG\alpha t^2 \BG\gamma^{{a}b}D_{a} D_b\varphi\\  
  &+\nu t^2 \BG\gamma^{{a}b}D_{a} D_b \BG\phi
  +t^2 \BG\gamma^{{a}b}D_{a} \BG\alpha D_b\varphi
  +t^2 \BG\gamma^{{a}b}D_{a} \nu D_b \BG\phi\\
  &+u\nu+\nu t^2 \BG\gamma^{{a}b}D_{a} D_b\varphi
  +t^2 \BG\gamma^{ab}D_a \nu D_b\varphi
  =f^{(1)},
\end{split}\\
  \label{eq:lapsenonlineardiff}
  t^2 \BG\gamma^{ab}D_a D_b\nu
  -\left(t^2 {\BG\chi_a}^{\hspace{1ex}b}{\BG\chi_b}^{\hspace{1ex}a}+\frac {1}{n}+\BG\pi^2\right) \nu
  -2\BG\pi\BG\alpha u
  -{\nu} u^2 -2\BG\pi{\nu} u-\BG\alpha u^2
  =f^{(2)},\\
  \label{eq:pinonlineardiff}
   -t\partial_t\varphi+\BG\alpha u+\BG\pi\nu+\nu u=f^{(3)}.
\end{gather}
The \emph{linearized lapse-scalar field equations} 
are then obtained by deleting all nonlinear terms from \Eqref{eq:scalarfieldnonlineardiff} -- \eqref{eq:pinonlineardiff}.
The given background $(\BG\phi, \BG\alpha, \BG\pi)$,
which is an approximate solution of the the lapse-scalar field
equations \Eqsref{eq:scalarfieldnonlinearbg}-\eqref{eq:pinonlinearbg},
motivates two problems of interest: The dynamics of its linear
and its non-linear perturbations $(u,\nu,\varphi)$. As mentioned
above, this paper focuses exclusively on the dynamics of its linear
perturbations, and not on its nonlinear perturbations.

\subsection{Kasner-scalar field spacetimes}
\label{sec:Kasnerscalarfieldspacetimes}

\newcommand{\KSF}[1]{\overline{#1}}

The background solutions of interest to this paper are related to a simple family of solutions: the \emph{Kasner-scalar field spacetimes} (referred to as (generalized) Kasner spacetimes in \cite{rodnianski2018}). In this section we briefly summarize their most important properties.  

The Kasner-scalar field spacetimes are spatially homogeneous (but in general very anisotropic) solutions $(\KSF\gamma^{ab}, {\KSF\chi_a}^{b},\KSF\alpha,\KSF\phi)$ of the Einstein-scalar field system with spatial manifold $M=\mathbb T^n$. 
 In zero shift CMC gauge (cf. \Eqref{eq:gaugecond}) we consider here the spatial metric takes the form
\begin{equation}
    \label{eq:defKasnerringmetric}
    \KSF\gamma^{ab}=\mathrm{diag}\left(t^{-2q_1},\ldots, t^{-2q_n}\right)
  \end{equation}
written in terms of 
standard Cartesian coordinates on $M$
where the \emph{Kasner exponents} $q_1$, \ldots, $q_n$ are real numbers  subject to
\begin{equation}
  \label{eq:KSBG5}
  \sum_{i=1}^n q_i=1,\quad \sum_{i=1}^n q_i^2=1-A^2,
\end{equation}
for any\footnote{Without loss of generality we assume that $A$ is non-negative throughout this whole paper. If $A$ is negative we can replace $\KSF\phi$ by $-\KSF\phi$ since the Einstein-scalar field system with zero potential is invariant under the transformation $\phi\mapsto-\phi$.}
\begin{equation}
  \label{eq:Arange}
  A\in [0,A_+],\quad A_+:=\sqrt{1-\frac 1n}.
\end{equation}
The scalar field is given by
\begin{equation}
\KSF\phi=A\log t+B
\end{equation}
for any real constant $B$. Furthermore we have
\begin{equation}
  \KSF\alpha=1,
\end{equation}
and
\begin{equation}
    \label{eq:KSBG4}
    {\KSF\chi_a}^{b}    
=\frac 1t\mathrm{diag}\left(\frac 1n-q_1,\ldots, \frac 1n-q_n\right). 
  \end{equation}
Notice that this implies the useful formula
\begin{equation}
  \label{eq:KSBG6}
  t^2 {\KSF\chi_a}^{b}{\KSF\chi_b}^{a}=1-\frac 1n-A^2.
\end{equation}
Two well-known special cases are (1) the isotropic FLRW-case given by $A=A_+$ and $q_1=\ldots=q_n=1/n$, and (2) the \emph{vacuum} Kasner solutions characterized by $A=0$\footnote{Note that in the vacuum case $\phi=B=const$ has no dynamics and no gravitational interaction.}.

For the discussion below, it is useful to establish certain bounds on the set of admissible Kasner exponents for any given $A\in [0,A_+]$. 
Setting $p_i=q_i-1/n$, \Eqref{eq:KSBG5} transforms into
\begin{equation}
    \label{eq:KSBG5p}
    \sum_{i=1}^n p_i=0,\quad \sum_{i=1}^n p_i^2=1-A^2-1/n=A_+^2-A^2.
  \end{equation}
All vectors $(p_1,\ldots,p_n)\in\R^n$ consistent with this are on the intersection of the plane through the {origin} perpendicular to the vector
$n_1=(1,\ldots,1)/\sqrt n$ 
and the $(n-1)$-sphere around the origin with radius $\sqrt{A_+^2-A^2}$.
Combining the square of the first identity in \Eqref{eq:KSBG5p} 
\begin{equation*}
p_1^2=\sum_{i,j=2}^n p_ip_j
=-\frac 12\sum_{i,j=2}^n (p_i-p_j)^2+(n-1)\sum_{i=2}^np_i^2
\end{equation*}
with the second identity eventually yields
\begin{equation*}
p_1^2=\frac{n-1}n(A_+^2-A^2) -\frac 1{2n}\sum_{i,j=2}^n (p_i-p_j)^2.
\end{equation*}
The largest possible value 
\[|p_1|=\sqrt{\frac{n-1}n}\sqrt{A_+^2-A^2}=A_+\sqrt{A_+^2-A^2}\]
is therefore obtained when 
\[p_i=-p_1/(n-1)\]
for all $i=2,\ldots,n$.
We conclude
that for any given $A\in [0,A_+]$
the largest and the smallest possible Kasner exponents are
\begin{equation}
  \label{eq:boundsonKasner}
q_{upper}=1-A_+^2 + A_+ \sqrt{A_+^2-A^2},\quad
q_{lower}=1-A_+^2 - A_+ \sqrt{A_+^2-A^2}.
\end{equation}
Given any collection $(q_1,\ldots,q_n)$ of Kasner exponents, we write 
\begin{equation}
\label{eq:defqmax}
q_{max}=\max\{q_1,\ldots,q_n\},\quad q_{min}=\min\{q_1,\ldots,q_n\}.
\end{equation}
Given $n$ and $A$ satisfying \Eqref{eq:Arange}, which Kasner-scalar field solution, cf.\ \Eqref{eq:KSBG5}, has the smallest value of $q_{max}$? The above considerations imply that this is achieved when one of the Kasner exponents takes the value of $q_{lower}$ and all other Kasner exponents take the value 
\begin{equation}
  \label{eq:defqmax0}
  q_{max,0}=\frac 1n+\frac{A_+}{n-1}\sqrt{A_+^2-A^2}
  =\left(1-A_+^2\right)\left(1+\sqrt{1-A^2/A_+^2}\right).
\end{equation}

\section{Main result}
\label{sec:mainresults}
\renewcommand{\unksymb}{\varphi}

Before stating our main theorem, \Theoremref{thm:maintheorem} below, we introduce further notation and conventions. As explained above we focus on the linearization of the lapse-scalar field system \Eqref{eq:scalarfieldnonlineardiff} -- \eqref{eq:pinonlineardiff}.
In the analysis of this system, it is convenient to introduce a
\emph{switching parameter} $\switchgen$, which multiplies certain
terms. In a first study of this section and
\Theoremref{thm:maintheorem}, it is sufficient to think of the case
$\switchgen = 1$. The parameter $\switchgen$ is used to ``switch''
between the full linearized lapse-scalar field system (corresponding
to $\switchgen = 1$) and an \emph{asymptotic model system}
(corresponding to $\switchgen= 0$) which is used in the proof of
\Theoremref{thm:maintheorem} to derive the asymptotic formula of the
solutions given below. As we argue below, allowing any value for the
switching parameter may be of interest beyond the proof (see our
discussion of \emph{asymptotic matching problems}
in \Sectionref{sec:modelproblemmatching}) which is why we include
$\switchgen$ in all of what follows.  With this, the system of
interest for the variables $u, \nu$ and $\varphi$ takes the form
\begin{gather}
    \label{eq:scalarfieldlinearMatch}
  \begin{split}
  -t\partial_tu
  -A\nu&-\switchgen (\BG\pi-A)\nu
  +\switchgen(1-\BG\alpha) u  
  +\switchgen\BG\alpha t^2\BG\gamma^{ab}D_a D_b\varphi\\  
  &+\switchgen\nu t^2\BG\gamma^{ab}D_a D_b \BG\phi
  +\switchgen t^2\BG\gamma^{ab}D_a \BG\alpha D_b\varphi
  +\switchgen t^2\BG\gamma^{ab}D_a \nu D_b \BG\phi
  = f^{(1)},
\end{split}\\
  \label{eq:lapselinearMatch}
  \begin{split}
  \switchgen t^2\BG\gamma^{ab}D_a D_b\nu
  -\nu
  -2A u
  &-\switchgen\left(\left(t^2{\BG\chi_a}^{\hspace{1ex}b}{\BG\chi_b}^{\hspace{1ex}a}+\frac {1}{n}+A^2-1\right)+(\BG\pi^2-A^2)\right) \nu\\
  &-2\switchgen(\BG\pi \BG\alpha-A) u
  = f^{(2)},
\end{split}\\
  \label{eq:pilinearMatch}
  -t\partial_t\varphi+u+A\nu
  +\switchgen(\BG\alpha-1) u+\switchgen(\BG\pi-A)\nu= f^{(3)}.
\end{gather}
Note that $\switchgen$ multiplies the spatial derivative terms as well
as other terms that turn out to be of ``higher order'' in an
appropriate sense, which we make precise below. 

Before we state our main result, we need a few definitions. First we
equip $M=\mathbb T^n$ with the time-independent flat {reference
  metric} $\delta_{ab}$. In terms of Cartesian coordinates on $M$ we
assume that $\delta_{ab}$ takes the form
\begin{equation}
  \label{eq:defdelta}
  \delta_{ab}=\mathrm{diag}\left(1,\ldots,1\right).
\end{equation}
In all of what follows, all spatial index operations are performed with the metric $\delta_{ab}$. 
Let $\partial_a$ be the (time-independent) covariant derivative associated with $\delta_{ab}$.
For any smooth time-dependent tensor
fields\footnote{\label{ftnote:tensornotation} Arbitrary smooth
  (time-dependent or time-independent) tensor fields $S$ of rank
  $(0,r)$ on $M$ are written in either the index-free form as, $S$, or,
  if we want to emphasize the rank as, $S_{[r]}$, or, using abstract
  indices as, $S_{a_1\ldots a_r}$.} $S$, $\tilde S$ and $\hat S$ on $M$ of the same arbitrary rank $(0,r)$ we write
\begin{equation}
  \label{eq:deltanorm}
  (S,\tilde S)_{\delta}=S_{a\ldots b} \tilde S^{a\ldots b},\quad
  \left|S\right|^2_\delta=\left(S,S\right)_{\delta},
\end{equation}
and
\[S=\hat S+O(f)\]
for some smooth function $f(t,x)$
provided that for each non-negative integers $k$ and $\ell$ there is a constant $C>0$ such that\footnote{This notion of the $O$-symbol can be relaxed. In fact, some of the quantities below do not require control over any time-derivatives and/or only over a finite number of spatial derivatives. Nevertheless, we use this strict version of the $O$-symbol in order to simplify the statement of the main theorem.} 
\[\left|(t\partial_t)^\ell\partial^k (S(t,x)-\hat S(t,x))\right|_{\delta}\le C f(t,x),\]
for all $x\in M$ and all $t\in (0,\finaltime]$ where $\finaltime>0$ is
some \emph{final time}.  Here, $\partial^k (S-\hat S)$ is the short
hand notation for the tensor field
$\partial_{a_1}\cdots\partial_{a_k}(S_{b_1\ldots b_r}-{\hat
  S}_{b_1\ldots b_r})$ of rank $(0,k+r)$. This notion allows us now to
introduce the class of background fields relevant for
\Theoremref{thm:maintheorem}.
\begin{definition}[Asymptotically point-wise Kasner--scalar field background]
\label{def:APKSF}
Let $\BG\gamma^{ab}$ be a smooth time-dependent Riemannian metric, ${\BG\chi_a}^{\hspace{1ex}b}$ a smooth time-dependent symmetric $(1,1)$-tensor field, and  ${\BG\alpha}, {\BG\pi}, \BG\phi$ be smooth time-dependent scalar fields on $M$. We say that the collection of fields $(\BG\gamma^{ab}, {\BG\chi_a}^{\hspace{1ex}b}, {\BG\alpha}, {\BG\pi}, \BG\phi)$ on $M=\mathbb T^n$ is an \textbf{asymptotically point-wise Kasner--scalar field background with decay} $\mathbf{\beta}$ for some smooth time-independent positive function $\beta$ on $M$ provided:
\begin{enumerate}
\item There are smooth functions $q_1(x), \ldots, q_n(x)$ and $A(x)$
  on $M$, 
such that \Eqsref{eq:KSBG5} and \eqref{eq:Arange} hold at each $x\in M$,
and smooth time-dependent fields $h^{ab}$, $\tilde h_{ab}$ on $M$ \footnote{\label{footnote:inversemetric} Recall that spatial index manipulations are performed with the flat reference metric $\delta_{ab}$. In our notation, $\KSF\gamma^{-1}_{ab}$ (the inverse of $\KSF\gamma^{ab}$) does therefore in general differ from $\KSF\gamma_{ab}$.} 
such that
\begin{equation}
     \label{eq:LSFasldkj}
     \BG\gamma^{ab}(t,x)=\KSF\gamma^{ab}(t,x)+h^{ab},\quad
     \BG\gamma^{-1}_{ab}(t,x)=\KSF\gamma^{-1}_{ab}(t,x)+\tilde h_{ab}.
   \end{equation}
where
\begin{equation}
  \label{eq:gammaass2}
  h^{ab}\KSF\gamma^{-1}_{bc}=O(t^\beta), \quad
  \tilde h_{ab}\KSF\gamma^{bc}=O(t^\beta),    
\end{equation}
and where 
$\KSF\gamma^{ab}(t,x)$ is defined by \Eqref{eq:defKasnerringmetric} at each $x\in M$.
\item We have
\[t {\BG\chi_a}^{\hspace{1ex}b}=t {\KSF\chi_a}^{b}+O(t^\beta),\]
where ${\KSF\chi_a}^{b}$ is defined by \Eqref{eq:KSBG4} at each $x\in M$. 
\item
The functions ${\BG\alpha}$, ${\BG\pi}$ and $\BG\phi$ on $M$ satisfy
    \begin{equation}
    \label{eq:alphaSpiSassumps}
    \BG\alpha=1+ O(t^\beta),\quad \BG\pi=A+ O(t^\beta), \quad \BG\phi=(A+ O(t^\beta))\log t,
  \end{equation}
where the function $A$ is determined by \Eqref{eq:KSBG5} at each $x\in M$.
\end{enumerate}
\end{definition}
Given an asymptotically point-wise Kasner--scalar field background we
refer to the functions $q_1,\ldots,q_n$ as the \emph{Kasner exponents}
and $A$ as the \emph{scalar field strength}. We refer to $q_{max}$,
$q_{min}$, $q_{upper}$ and $q_{lower}$ by the formulas
\Eqref{eq:defqmax} and \eqref{eq:boundsonKasner} at each $x\in M$.
Notice that the exact Kasner-scalar field solutions
in \Sectionref{sec:Kasnerscalarfieldspacetimes} are special examples
of asymptotically point-wise Kasner--scalar field backgrounds.  It is
important to point out however that although the asymptotically
point-wise Kasner--scalar field background fields asymptote at each
spatial point to a solution of the Einstein--scalar field equations,
the fields themselves are not required to be solutions to these
equations. 
In addition, we note that this class captures the
essential features of the \emph{Kasner footprint maps} introduced in \cite{rodnianski2014,rodnianski2018},
and, as such, the generalization of our results to backgrounds of this
type is straightforward.

The last steps before we can state the main theorem are now to introduce a smooth function $\xi(x)$ on $M$ with range in $[0,1]$ by (recall \Eqref{eq:Arange})
\begin{equation}
  \label{eq:defxi}
  A=A_+\sqrt{1-\xi^2},
\end{equation}
 and then define at each $x\in M$,
\begin{equation}
  \label{eq:deflambdac}
  \lambda_c(x)=
  \begin{cases}
    2A^2(x)=2(1-\xi^2(x))A_+^2 & \text{if $\xi(x)\in [0,1/3]$},\\
    \frac{(1+\xi(x))^3}{4\xi(x)}A_+^2 & \text{if $\xi(x)\in [1/3,1]$}.
  \end{cases}
\end{equation}
In order to simplify some of the following formulas, we also define
\begin{align}
  \label{eq:defS1}
  S^{(1)}(t,x)=&\int_t^\finaltime \left(f^{(1)}(s,x)-A(x)f^{(2)}(s,x)\right) \left(\frac ts\right)^{2A^2(x)}s^{-1}ds\\
  \label{eq:defS2}
  S^{(2)}(t,x)=& -(2A^2(x)-1)\int_0^tS^{(1)}(s,x)s^{-1}ds\\
&-\int_0^t\left(f^{(3)}(s,x)+A(x)f^{(2)}(s,x)\right)s^{-1}ds,\notag
\end{align}
and
\begin{equation}
  \label{eq:LSF17}
  \Lop{g}=\switchgen\left({\BG\alpha} t^2\BG\gamma^{ab}D_a D_b g
+t^2\BG\gamma^{ab}D_b \BG\alpha D_a g\right),
\end{equation}
for any smooth functions $f^{(1)}$, $f^{(2)}$, $f^{(3)}$ and $g(t,x)$ for which these integrals are finite.
\begin{theorem}
  \label{thm:maintheorem}
Pick a sufficiently small constant $\finaltime$ in $(0,1]$, an integer $n\ge 3$ and
an arbitrary asymptotically point-wise Kasner--scalar field background 
\[\Gamma=(\BG\gamma^{ab}, {\BG\chi_a}^{\hspace{1ex}b}, {\BG\alpha}, {\BG\pi}, \BG\phi)\] with decay $\beta$, where $\beta$ is an arbitrary smooth positive time-independent function on $M=\mathbb T^n$ with
\begin{equation}
\label{eq:LSFtheominequFinal}
\beta>\lambda_c-\min\{2A^2,2(1-q_{max})\}
\end{equation}
where $\lambda_c$ is defined in \Eqref{eq:deflambdac}.
Suppose that 
\begin{equation}
  \label{eq:LSFLSFmodLem293482}
  \lambda_c<\min\{4(1-q_{max}),2A^2+2(1-q_{max})\}
\end{equation}
  at each point in $M$. 
Suppose the source terms $f^{(1)}(t,x)$,
  $f^{(2)}(t,x)$ and $f^{(3)}(t,x)$ in
  \Eqsref{eq:scalarfieldlinearMatch} -- \eqref{eq:pilinearMatch}   are 
  smooth functions on $(0,\finaltime]\times M$ such that
  \begin{equation*}
    \int^{\finaltime}_0 \Bigl(\|s^{-\lambda_s} f^{(1)}\|_{\delta,H^{k+4}(M)}
    + \|s^{-\lambda_s} f^{(2)}\|_{\delta,H^{k+4}(M)}
    +\|s^{-\lambda_s} f^{(3)}\|_{\delta,H^{k+5}(M)}\Bigr)s^{-1} ds<\infty,
\end{equation*}
  for all integers $k\ge 0$ and for some smooth function $\lambda_s(x)$ on $M$ with
\begin{equation}
    \label{eq:LSFLSF1cond4}
    \lambda_s>\max\{\lambda_c-2(1-q_{max}), \lambda_c-\beta, 0\}.
  \end{equation}  
   Then,
  for each constant $\switchgen\in[0,1]$, there is a homeomorphism in the $C^\infty$-topology
  \[\Psi: (C^\infty(M))^2\rightarrow (C^\infty(M))^2,\quad 
    (\unkgen_{(0)}, \unkgen_{(1)})
    \mapsto
    (\CDfinaltime{\vnkgen}, \CDfinaltime{\unkgen})    
    \]
    with the following property.
    For each integer $k\ge 0$ and smooth function $\lambda(x)$ with
\begin{equation}
  \label{eq:LSFLSFmodLem29348}
  \lambda<\min\{\lambda_s+2(1-q_{max}),4(1-q_{max}),2A^2+2(1-q_{max}), \beta+2(1-q_{max}),2A^2+\beta\},
\end{equation}
there is a constant $C>0$, such that for any $(\unkgen_{(0)}, \unkgen_{(1)})\in (C^\infty(M))^2$, the solution
$(\vnkgen,\nunkgen, \unkgen)$  of
    \Eqsref{eq:scalarfieldlinearMatch} -- \eqref{eq:pilinearMatch}    
    determined by Cauchy data $\unkgen(\finaltime,\cdot)=\CDfinaltime{\unkgen}$ and $\vnkgen(\finaltime,\cdot)=\CDfinaltime{\vnkgen}$
with  $(\CDfinaltime{\vnkgen}, \CDfinaltime{\unkgen})=\Psi (\unkgen_{(0)}, \unkgen_{(1)})$
 satisfies
    \begin{align}      
\Bigl\|t^{-\lambda}\Bigl(&\vnkgen(t,\cdot)
+2A^2t^{2A^2} \unkgen_{(1)}
    -\int_t^\finaltime \Lop{\unkgen_{(0)}}(s,\cdot) \left(\frac ts\right)^{2A^2}s^{-1}ds- S^{(1)}(t,\cdot)
\Bigr)\Bigr\|_{H^k(M)}^2\notag\\
\label{eq:LSFLSFmodCor23132}
+\Bigl\|t^{-\lambda}\Bigl(&\unkgen(t,\cdot)
-\unkgen_{(0)}-(2A^2-1)t^{2A^2} \unkgen_{(1)}\\
  &\!\!\!\!\!\!\! +(2A^2-1)\int_0^t\int_\tau^\finaltime \Lop{\unkgen_{(0)}}(s,\cdot) \left(\frac ts\right)^{2A^2}s^{-1}ds\,\tau^{-1} d\tau-S^{(2)}(t,\cdot)
\Bigr)\Bigr\|_{H^k(M)}^2
\le  C\notag
\end{align}
for all $t\in (0,\finaltime]$ and all integers $k\ge 0$. The constant $C>0$ may depend on $k$, $\lambda$, $\finaltime$, $f^{(1)}$, $f^{(2)}$, $f^{(3)}$, $\unkgen_{(0)}$, $\unkgen_{(1)}$ and $\Gamma$.
\end{theorem}

The \hyperlink{thmproof:maintheorem}{proof} of \Theoremref{thm:maintheorem} is discussed in detail in \Sectionref{sec:LSFProofs}. In the remainder of this section we make a few remarks and discuss consequences of \Theoremref{thm:maintheorem}. 
First let us note that the estimate \eqref{eq:LSFLSFmodCor23132} for $u$ and $\varphi$ can be complemented by an estimate for $\nu$ which can be obtained by means of \Eqref{eq:preestimat3N} for $\sourcegen^{(2)}=f^{(2)}$ (see \Eqsref{eq:defenergies1N} -- \eqref{eq:defEnergies2}).

On the one hand, this theorem states that any solution of
\Eqsref{eq:scalarfieldlinearMatch} -- \eqref{eq:pilinearMatch}
launched by \emph{Cauchy data}
$(\CDfinaltime{\vnkgen}, \CDfinaltime{\unkgen})$ is described
asymptotically by \Eqsref{eq:LSFLSFmodLem29348} and
\eqref{eq:LSFLSFmodCor23132} for \emph{asymptotic data}
$(\unkgen_{(0)}, \unkgen_{(1)})=\Psi^{-1}(\CDfinaltime{\vnkgen},
\CDfinaltime{\unkgen})$. On the other hand, it states that for any
choice of {asymptotic data} $(\unkgen_{(0)}, \unkgen_{(1)})$ there
exists a solution of the Cauchy problem for
\Eqsref{eq:scalarfieldlinearMatch} -- \eqref{eq:pilinearMatch} with
Cauchy data
$(\CDfinaltime{\vnkgen}, \CDfinaltime{\unkgen})=\Psi (\unkgen_{(0)},
\unkgen_{(1)})$ which realizes these asymptotic data via
\Eqsref{eq:LSFLSFmodCor23132}.  The full degrees of freedom of the
entire solution set are therefore parametrized by either Cauchy data
(``the state of the system at a finite positive time''
$t=\finaltime>0$), or equivalently, by asymptotic data (the ``state of
the system at the singular time'' $t=0$). This map between asymptotic
data and Cauchy data $\Psi$ is invertible, and, both directions are
open maps with respect to the $C^\infty$-topology. As discussed above,
this implies that the solutions are \emph{asymptotically stable} under
perturbations of either the asymptotic data or the Cauchy data.

\Eqref{eq:LSFLSFmodCor23132} yields the asymptotic behavior of arbitrary solutions up to  $O(t^\lambda)$ (in the limit $t\searrow 0$). We verify that \Eqref{eq:LSFLSFmodLem29348} allows us to choose $\lambda$ a little larger than $2A^2$ by checking that each element in the list of upper bounds in \Eqref{eq:LSFLSFmodLem29348} is larger than $2A^2$:
  \begin{itemize}
  \item $\lambda_s+2(1-q_{max})>2A^2$ because of \Eqsref{eq:LSFLSF1cond4} and \eqref{eq:deflambdac}.
    \item $4(1-q_{max})>2A^2$ because 
      \begin{align*}
        4(1-q_{max})-2A^2&\ge 4(1-q_{upper})-2A^2= 4A_+^2(1-\xi)-2A_+^2(1-\xi^2)\\
                         &= 2A_+^2(1-\xi)(2-(1+\xi))=2A_+^2(1-\xi)^2>0,
      \end{align*}
      for all $\xi\in [0,1)$ using \Eqsref{eq:defxi} and \eqref{eq:boundsonKasner}.
    \item $2A^2+2(1-q_{max})>2A^2$ because $q_{max}<1$ for all $\xi<1$.
    \item $\beta+2(1-q_{max})>2A^2$ because
      \begin{align*}
        &\beta+2(1-q_{max})-2A^2\\
        &>
        \begin{cases}
          \underbrace{\lambda_c-2A^2}_{\ge 0}+\underbrace{ 2(1-q_{max})-2A^2}_{\ge 0} & \text{if $2(1-q_{max})\ge 2A^2$},\\
          \lambda_c-2(1-q_{max})+2(1-q_{max}) -2A^2 & \text{if $2(1-q_{max})\le 2A^2$}
        \end{cases}\\
        &\ge 0,
      \end{align*}   
      as a consequence of \Eqref{eq:LSFtheominequFinal}.
    \item $2A^2+\beta >2A^2$ because $\beta>0$.
  \end{itemize}
Because of this, \Eqref{eq:LSFLSFmodCor23132} is  ``correct up to $O(t^{2A^2+\epsilon})$'' for any sufficiently small $\epsilon>0$. It
can therefore be interpreted as the asymptotic representation of the full degrees of freedom of the solution set as it contains both the first asymptotic datum, $\unkgen_{(0)}$, associated with the $t$-power $0$, and the second asymptotic datum, $\unkgen_{(1)}$, associated with the $t$-power $2A^2$. 
Hence it does make sense to call $(\unkgen_{(0)}, \unkgen_{(1)})$  \emph{asymptotic data}. Observe that these two quantities are associated with the following limits of an arbitrary solution:
  \begin{align}
    \label{eq:limit1}
    \unkgen_{(0)}(x)&=\lim_{t\searrow 0} \unkgen(t,x),\\
    -2A^2(x)\unkgen_{(1)}(x)&=\lim_{t\searrow 0} \left(\vnkgen(t,x) -\int_t^\finaltime \Lop{\unkgen_{(0)}}(s,x) \left(\frac ts\right)^{2A^2(x)}s^{-1}ds- S^{(1)}(t,x)\right)t^{-2A^2(x)}.\notag
  \end{align}
Let us stress that the limit of the first integral on the right-hand side multiplied with $t^{-2A^2(x)}$ by itself would in general not be finite (see below). 

\label{refereeanswer}
At a first glance it may be surprising that the limit of $\unkgen$ should always be finite as implied by \Eqref{eq:limit1}. If we interpret \Eqref{eq:scalarfieldlinearMatch} heuristically as an equation of the form
$t\partial_t u=O(t^\epsilon)$, general solutions should be $u=O(1)$  if $\epsilon>0$. \Eqref{eq:pilinearMatch} then implies that $\unkgen=O(\log t)$. It therefore is conceivable that general linearized scalar field solutions contribute to the leading asymptotics in the same way as the background field $\BG\phi=A\log(t)+\ldots$. 
However, this is not the case and the reason is that the coupling of \Eqref{eq:scalarfieldlinearMatch}  to \Eqref{eq:lapselinearMatch} cannot be neglected. In fact, using the same heuristic notion as before, it turns out that the equation which describes the asymptotics of $u$ correctly is ``\eqref{eq:scalarfieldlinearMatch} $-A\times$ \eqref{eq:lapselinearMatch}'', chosen to cancel the $\nu$-term. This is an equation of the form $t\partial_t u-2A^2 u=O(t^\epsilon)$  which has only strictly decaying solutions if $A$ and $\epsilon$ are positive. Our proof gives a rigorous justification of these claims. It is interesting that in a first step of our rigorous analysis we are indeed not able to establish any decay for $u$; see the ``rough estimate'' \eqref{eq:LSFbackward1}. We find that only once this ``rough estimate'' is ``improved'' (at the cost of some differentiability) the optimal decay rate for $u$ appears; see \Eqref{eq:LSFimproveddecayubackward}.

One of the most interesting conclusions which we can draw from \Eqref{eq:LSFLSFmodCor23132} is that all terms whose $t$-powers are smaller than $2A^2$ are \emph{significant for the asymptotic characterization of the full degrees of freedom}. 
Obviously it does not only contain the ``pure asymptotic data terms'', i.e., the term associated with the $t$-powers $0$ and $2A^2$. Of particular interest are the terms involving $\Lop{\unkgen_{(0)}}$ which originate in spatial derivative terms in the equations.
The other terms $S^{(1)}$ and $S^{(2)}$  are generated by the source terms $f^{(1)}$, $f^{(2)}$, $f^{(3)}$ of the equations.
In principle, there might be additional terms generated by  fields in $\Gamma$. The assumption \eqref{eq:LSFtheominequFinal} on the decay rate $\beta$ however makes sure that no contributions from fields in $\Gamma$ show show up in \Eqref{eq:LSFLSFmodCor23132} at orders compatible with \Eqref{eq:LSFLSFmodLem29348}. 
We justify now that the above terms generated by \emph{spatial derivatives are significant for the asymptotic characterization of the full degrees of freedom} in general -- a statement which demonstrates that the notion of AVTD cannot be applied naively. We see this as follows. The $t$-powers associated with those spatial derivative terms in \Eqref{eq:LSFLSFmodCor23132} are all of the form $2(1-q)$ where $q$ is one of the Kasner exponents. The most dominant term among these is therefore the one associated with the $t$-power $2(1-q_{max})$. If $2(1-q_{max})>2A^2$ at all points $x\in M$, then all spatial derivative terms would be negligible. If $2A^2\ge 2(1-q_{max})$ at some point $x\in M$, however, then some terms generated by spatial derivatives \emph{are} significant for the asymptotic characterization of the full degrees of freedom. We can shed light on this by considering the following sharp bounds (recall definition of $q_{upper}, q_{max}$ in \Eqsref{eq:boundsonKasner}-\eqref{eq:defqmax} and $q_{max, 0}$ in \Eqref{eq:defqmax0})
  \begin{equation*}
    2(1-q_{max})\ge 2(1-q_{upper})=2 A_+^2 (1-\xi),
  \end{equation*}
  and
  \begin{equation*}
    2(1-q_{max})\le 2(1-q_{max,0})=2(A_+^2-(1-A_+^2)\xi).
  \end{equation*}
  
The possible range for $2(1-q_{max})$ therefore depends on $\xi$, but so does $2A^2=2A_+^2(1-\xi^2)$. Whether spatial derivatives are hence significant in the sense above or not depends on the following bounds
  \begin{equation}
    \label{eq:boundsonKasnerIntLow}
    \frac{2(1-q_{max})}{2A^2}\ge \frac{2 A_+^2 (1-\xi)}{2A_+^2(1-\xi^2)}=\frac{1}{1+\xi}> \frac 12,
  \end{equation}
if $\xi\in[0,1)$,
and
\begin{equation*}
    \frac{2(1-q_{max})}{2A^2}
    \le \frac{A_+^2-\left(1-A_+^2\right)\xi}{A_+^2(1-\xi^2)}
    =1-\frac{1-(n-1) \xi}{(n-1) \left(1-\xi ^2\right)}\xi  .
  \end{equation*}
The first bound tells us that, given any $n\ge 3$ (and hence $A_+$ by \Eqref{eq:Arange}) and $\xi\in [0,1)$ (and hence $A$ by \Eqref{eq:defxi}), we can always choose the Kasner exponents
such that some of the terms generated by \emph{spatial derivatives are significant for the asymptotic characterization of the full degrees of freedom} in the sense above. This is the case in particular if $q_{max}$ is sufficiently close to $q_{upper}$. This bound also implies that all positive integer multiples of $2(1-q_{max})$ (except possibly for $2(1-q_{max})$ itself) are larger than $2A^2$. We thus conclude that only the most leading spatial derivative terms can ever be significant in this sense. We therefore do not need to consider any further spatial derivative terms of potentially higher order.
The other bound tells us that, given any $n\ge 3$ (and hence $A_+$ by \Eqref{eq:Arange}) and $\xi\in [0,1)$ (and hence $A$ by \Eqref{eq:defxi}), spatial derivative terms are {significant in this sense for \emph{all} possible choices of Kasner exponents}  if $\xi\in [0,1/(n-1)]$. Only if $\xi>1/(n-1)$, there exist certain choices of Kasner exponents for which spatial derivative terms are \emph{completely insignificant}; this occurs if the Kasner exponents are such that $q_{max}$ is sufficiently close to the smallest possible value given by \Eqref{eq:defqmax0}.

This discussion suggests that \Eqref{eq:LSFLSFmodCor23132} is the correct asymptotic formula for all solutions of our equations and represents the full degrees of freedom for $\xi\in [0,1)$ \emph{even if we violate \Eqref{eq:LSFLSFmodLem293482}}. 
What is this restriction \Eqref{eq:LSFLSFmodLem293482} and where does it come from? 
\begin{itemize}
\item Given arbitrary $n\ge 3$ (and hence $A_+$ by \Eqref{eq:Arange}), then \Eqref{eq:LSFLSFmodLem293482} holds for all $\xi\in [0,\Xi)$ and all admissible choices of Kasner exponents provided 
 $\Xi\approx 0.58$ (the smallest real root of $\xi ^3+19 \xi ^2-13 \xi +1$ in the interval $(1/3,1)$).
\item Given arbitrary $n\ge 3$, then \Eqref{eq:LSFLSFmodLem293482} holds for all $\xi\in [0,\Xi)$ and all admissible choices of Kasner exponents provided $q_{max}$ is sufficiently close to \Eqref{eq:defqmax0} and
 $\Xi\approx 0.82$ (the smallest real root of $\xi ^3+19 \xi ^2-13 \xi +1$ in the interval $(1/3,1)$).
\item \Eqref{eq:LSFLSFmodLem293482} holds for all $\xi\in [0,1)$, all sufficiently large $n\ge 3$ and all admissible choices of Kasner exponents provided $q_{max}$ is sufficiently close to \Eqref{eq:defqmax0}.
\end{itemize}
We claim that \Eqref{eq:LSFLSFmodLem293482} is of technical origin. Recall that the main idea in the above discussion is that all significant terms are of $O(t^\lambda)$ with $\lambda$ smaller or equal than $2A^2$. As we discuss in detail in  \Sectionref{sec:LSFProofs}, our estimates  however demand the asymptotic formula \Eqref{eq:LSFLSFmodCor23132} to be correct up to order $O(t^\lambda)$ with $\lambda$ a little larger than $\lambda_c$ (which can be significantly larger than $2A^2$; see \Eqref{eq:deflambdac}). Replacing 
 $2A^2$ by $\lambda_c$ in \Eqref{eq:boundsonKasnerIntLow} and considering  $\xi>1/3$ (which implies that $\lambda_c>2A^2$), we get the much less favorable estimate
\begin{equation*}
    \frac{2(1-q_{max})}{\lambda_c}\ge \frac{8 \xi (1-\xi)}{(1+\xi)^3}.
  \end{equation*}
This is the origin for \Eqref{eq:LSFLSFmodLem293482}.
We believe that the technical problem leading to this can be overcome  as follows (even though we have not looked at the details yet). As we see below, the quantity $\lambda_c$ shows up for the first time in one of the ``rough estimates'' in \Propref{prop:LSF2} below. The same arguments which we then employ to ``improve'' a different one of these rough estimates in \Propref{prop:LSF4} should lead to a corresponding ``improvement'' of the particular estimates which give rise to  $\lambda_c$ as well. In this way it should be possible to ``push $\lambda_c$ arbitrarily close to $2A^2$'' iteratively on \emph{any} fixed sub-interval for $\xi$ in $(0,1]$. The larger this sub-interval is (i.e., the closer we wish $\xi$ to be to the vacuum value $1$),  the more differentiability we  expect to lose in this process. 

We close this section with remarks anticipating some of the discussion given in \Sectionref{sec:modelproblemmatching}. Recall that we have introduced the switching parameter $\switchgen$ in \Eqsref{eq:scalarfieldlinearMatch} -- \eqref{eq:pilinearMatch} and we have argued that one can think of the case $\switchgen=1$ at first reading. It turns out that this parameter plays an important role for our proof. In fact, the asymptotic formula provided by \Eqref{eq:LSFLSFmodCor23132} turns out to be   essentially the general solution of the equations obtained by  setting $\switchgen=0$ in \Eqsref{eq:scalarfieldlinearMatch} -- \eqref{eq:pilinearMatch}. In a sense which we make precise in the following, the homeomorphism constructed in \Theoremref{thm:maintheorem} can be interpreted as a (invertible) map between  solutions of the equations with $\switchgen=0$ and solutions of the equations with an arbitrary value of $\switchgen\in [0,1]$ which ``match'' as in  \Eqref{eq:LSFLSFmodCor23132} (or vice versa). Given this homeomorphism for two different values $\switch$ and $\switcht$ in $[0,1]$ (but everything else is the same) and composing the inverse of the latter homeomorphism with the first one clearly yields another homeomorphism. This ``matches'' the solutions of the $\switch$-version of the equation with solutions of the $\switcht$-version. The idea of \emph{asymptotic matching problems} is discussed  in the next section. 
We note that this formalism does not only allow us to prove \emph{existence} of the homeomorphism, but also provides a useful approximation scheme to calculate it. This is achieved by approximating solutions of \emph{asymptotic} matching problems by solutions of \emph{finite} matching problems, as explained in \Sectionref{sec:modelproblemmatching}.

\section{Matching problems}
\label{sec:modelproblemmatching}

\renewcommand{\unksymb}{v}

\subsection{Asymptotic and finite matching problems}
\label{sec:generalmatchingproblems}
In this section we present a framework for characterizing the asymptotics of solutions of, in principle, any given (system of) evolution PDE. This framework assumes that the sought asymptotics are themselves solutions of a, in general, simpler ``effective'' version of the original equation -- a \emph{(asymptotic) model equation}.  In general relativity, the classical problem prototype of this perspective is that of finding AVTD solutions as defined rigorously in \cite{isenberg1990}. In order to introduce our framework consider a PDE of the  schematic form 
  \begin{equation}
  \label{eq:abstracteqgenN}
  A[\unkgen](t,x)+\switchgen L[\unkgen](t,x)=0,
\end{equation}
where  $\unkgen(t,x)$ is the unknown defined on some time interval
$t\in(0,\finaltime]$, $\finaltime>0$, and for all $x\in M$. In general, $\unkgen$ may be a time-dependent section in some vector bundle $E$ over $M$.  
At this stage, we only assume that $A$ and $L$ in \Eqref{eq:abstracteqgenN} are linear differential operators with smooth coefficients. The quantity $\switchgen$ is a real number restricted to a bounded closed interval $I$ of $\R$ (which in general needs be chosen such that the character of the equation is not changed).

The main idea of the splitting manifested by the two terms in \Eqref{eq:abstracteqgenN} is that the operator $A$ contains all those terms  which are  \emph{significant} asymptotically (like the ``velocity terms'' in the AVTD setting above), while $L$ contains all terms which are \emph{negligible} (for example spatial derivatives in the context of AVTD). 
Of particular interest are the cases $\switchgen=1$ -- the \keyword{full equation} -- and $\switchgen=0$ -- the \keyword{model equation} (or \keyword{effective equation}). Notice that going from the full equation to the model equation means that one neglects all terms given by the operator $L$. Since the terms in $L$ are effectively ``switched off'' when we set $\switchgen=0$  we call $\switchgen$ the \emph{switching parameter} as in \Sectionref{sec:mainresults}. For different choices of the parameter $\switchgen$ we often speak of \emph{different versions of the equation}.   The hope is that the model equation describes the asymptotics of the full equation sufficiently accurately in the limit $t\searrow 0$. Below we present a method to verify this.

\begin{definition}[(Asymptotic) matching problem]
  \label{def:AMatchingProblem}
  Pick $\switch,\switcht\in I =[0,1]$. Given any solution $\unk$ (of some prescribed regularity) of the version of \Eqref{eq:abstracteqgenN} given by $\switchgen=\switch$, find a solution $\unkt$ (of some prescribed regularity) of the version of \Eqref{eq:abstracteqgenN} given by $\switchgen=\switcht$ such that 
  \begin{equation}
    \label{eq:matchingcond}
    \|\unk(t,\cdot)-\unkt(t,\cdot)\|\rightarrow 0,\quad\text{as $t\searrow 0$},
  \end{equation}
  for some (possibly time-dependent) norm $\|\cdot\|$. If this can be done for each $\unk$ in some set $\Omega$ and if the corresponding $\unkt$ is always uniquely determined we say the asymptotic matching problem is \keyword{well-defined}. A well-defined asymptotic matching problem gives rise to a map, which we call the \keyword{(asymptotic) matching map}, $\AsymptoticMatchingMapAbstr{\switch\rightarrow\switcht} : \unk\mapsto\unkt $ with domain $\Omega$.
 \end{definition} 

In the sense of the map in  \Defref{def:AMatchingProblem}, we call the $\switchgen=\switch$-version of \Eqref{eq:abstracteqgenN} the \emph{source equation} and the $\switchgen=\switcht$-version  the \emph{target equation}.
We have intentionally not yet assumed any particular PDE type for \Eqref{eq:abstracteqgenN} at this point, apart from the linearity of the differential operators appearing in the equation. It is clear however that the question of whether the asymptotic matching problem is well-defined or not depends strongly on the PDE type as well as the choice of norm and regularity requirements in the definition.

The maps $\AsymptoticMatchingMapAbstr{1\rightarrow 0}$ and $\AsymptoticMatchingMapAbstr{0\rightarrow 1}$ are of particular interest. If the model equation ($\switchgen=0$) is sufficiently simple and if its solution can be parametrized by ``data'' we shall call these ``asymptotic data''.  The former is therefore essentially the map from Cauchy data (for the full equation) to asymptotic data. 
Provided the equation has a well-posed initial value problem, the latter is the map from asymptotic data to Cauchy data and is strongly related to the \keyword{Fuchsian method} mentioned above. We remark that, indeed, it turns out to be useful to  write \Eqref{eq:abstracteqgenN} and \Defref{def:AMatchingProblem} in terms of  \emph{general} switching parameters $\switchgen$, $\switch$ and $\switcht$ (as opposed to restricting to special values $1$ or $0$). In doing this one is able to essentially treat all specific cases in a uniform manner.

Suppose that the asymptotic matching map $\AsymptoticMatchingMapAbstr{\switch\rightarrow\switcht}$ in \Defref{def:AMatchingProblem} exists for some $\switch,\switcht\in I$. A natural question is the following: Is this map invertible, and perhaps even a homeomorphism in some appropriate sense? If this is the case, then the set of all solutions of the $\switchgen=\switch$-version of the equation is in one-to-one correspondence with the set of all solutions of the $\switchgen=\switcht$-version of the equation and the correspondence is described in terms of the asymptotics. This yields a strong characterization of the asymptotics.

How does \Theoremref{thm:maintheorem} fit into this framework? In order to set up an asymptotic matching problem, and
therefore to choose $A$ and $L$ in
\Eqref{eq:abstracteqgenN}, one can start off by means of heuristic considerations. First we hope to be able to choose the background fields $\BG\phi(t,x)$, $\BG\alpha(t,x)$, $\BG\pi(t,x)$ in \Eqsref{eq:scalarfieldnonlinearbg} -- \eqref{eq:pinonlinearbg} to carry the main singular behavior of solutions of the lapse-scalar field system in the sense that  the functions $f^{(1)}(t,x)$, $f^{(2)}(t,x)$ and $f^{(3)}(t,x)$ determined by \Eqsref{eq:scalarfieldnonlinearbg} -- \eqref{eq:pinonlinearbg} are \emph{slightly less singular} in the limit $t\searrow 0$. Given this,  the fields $u$, $\nu$ and $\varphi$ defined by \Eqref{eq:defdiffquant} should be less singular as well and therefore be potentially small close to $t=0$. It is therefore generally expected that
solutions of \Eqsref{eq:scalarfieldnonlineardiff} -- \eqref{eq:pinonlineardiff} are predominantly driven by \emph{linear terms}, and, that nonlinearities contribute only at higher order in the limit $t\searrow 0$. 

In fact, this question of how significant nonlinear terms are in comparison to linear terms can be phrased as the following modified version of an asymptotic matching problem, adapted to allow for nonlinearities, with switching parameter $\switchNLgen$:
\begin{gather*}
  \begin{split}
  -t\partial_tu
  -\BG\pi\nu
  &+(1-\BG\alpha) u  
  +\BG\alpha t^2\BG\gamma^{ab}D_a D_b\varphi\\  
  &+\nu t^2\BG\gamma^{ab}D_a D_b \BG\phi
  +t^2\BG\gamma^{ab}D_a \BG\alpha D_b\varphi
  +t^2\BG\gamma^{ab}D_a \nu D_b \BG\phi\\
  &+\switchNLgen\left( u\nu+\nu t^2\BG\gamma^{ab}D_a D_b\varphi
  +t^2\BG\gamma^{ab}D_a \nu D_b\varphi\right)
  = f^{(1)},
\end{split}\\
  t^2\BG\gamma^{ab}D_a D_b\nu
  -\left(t^2{\BG\chi_a}^{\hspace{1ex}b}{\BG\chi_b}^{\hspace{1ex}a}+\frac {1}{n}+\BG\pi^2\right) \nu
  -2\BG\pi\BG\alpha u
  -\switchNLgen\left({\nu} u^2 +2\BG\pi{\nu} u+\BG\alpha u^2\right)
  = f^{(2)},\\
  -t\partial_t\varphi+\BG\alpha u+\BG\pi\nu+\switchNLgen\nu u=f^{(3)}.
\end{gather*}
If this modified asymptotic matching problem were well-defined in the sense of \Defref{def:AMatchingProblem}, this would give a clear and precise justification for considering ``the dynamics to be driven by the linear system $\switchNLgen=0$''. As mentioned above, we do not analyze this question in the present paper. Instead, the major task for this paper here is  to characterize the asymptotics of solutions of this linear system $\switchNLgen=0$ as precisely as possible and under conditions as general as possible. 

\subsection{A strategy to analyze matching problems}
\label{sec:generalmatchingproblemsstrategy}
How can the asymptotic matching map in \Defref{def:AMatchingProblem} be constructed rigorously (if it exists)? In this paper, we provide a pathway which applies to equations \Eqref{eq:abstracteqgenN} whose Cauchy problem is well-posed in the sense that for each smooth Cauchy data imposed at each initial time $\matchtime\in (0,\finaltime]$ there is a unique smooth solution defined on $(0,\finaltime]\times M$. 
 A prominent class of evolution equations ruled out by this are  \emph{parabolic} problems. In fact we restrict our attention here to evolution equations which are \emph{essentially hyperbolic} (in particular hyperbolic-elliptic problems of the kind of interest here).
In this paper, we denote Cauchy data for solutions $\unkgen$ of \Eqref{eq:abstracteqgenN} as $\GenCD{\matchtime}{\unkgen}$ if they are imposed at initial time $\matchtime\in (0,\finaltime]$. In the special case $\matchtime=\finaltime$, we often write $\CDfinaltime{\unkgen}$ instead of $\GenCD{\finaltime}{\unkgen}$. Notice that if \Eqref{eq:abstracteqgenN} is first-order in time, as we predominantly assume below, then $\GenCD{\matchtime}{\unkgen}=\unkgen(\matchtime,\cdot)$ and $\CDfinaltime{\unkgen}=\unkgen(\finaltime,\cdot)$.
It is clear that if the Cauchy problem is well-posed in this sense, the asymptotic matching map, $\AsymptoticMatchingMapAbstr{\switch\rightarrow\switcht}$ defined in \Defref{def:AMatchingProblem}, (if it exists) may be understood as a map between Cauchy data imposed at $t=\finaltime$; i.e.,  $\AsymptoticMatchingMapAbstr{\switch\rightarrow\switcht}: \CDfinaltime{\unk}\mapsto\CDfinaltime{\unkt}$.

With the Cauchy problem at our disposal, the idea is now to approximate an asymptotic matching map by a sequence of \emph{finite matching maps}. 

\begin{definition}[Finite matching problem]
  \label{def:FMatchingProblem}
  Suppose that the Cauchy problem of \Eqref{eq:abstracteqgenN} is well-posed in the sense above for any value of $\switchgen\in I$. Pick $\switch,\switcht\in I$. For any $\matchtime\in (0,\finaltime]$ and any smooth solution $\unk$ of the version of \Eqref{eq:abstracteqgenN} given by $\switchgen=\switch$ defined on $(0, \finaltime]$, the finite matching problem with matching time $\tau$ is to find the uniquely determined smooth solution $\unkt$ of the version of \Eqref{eq:abstracteqgenN} given by $\switchgen=\switcht$ such that 
  \begin{equation}
    \label{eq:finitematchingcond}
    \GenCD{\matchtime}{\unk}=\GenCD{\matchtime}{\unkt}.
  \end{equation}
  The map of the schematic form 
\begin{equation}
  \label{eq:defmatchingmapfinite}
  \FiniteMatchingMapAbstr{\switch\rightarrow\switcht}: (\matchtime,\CDfinaltime{\unk})\mapsto\CDfinaltime{\unkt}
\end{equation}
is called the \keyword{finite matching map}.
  \end{definition}
While the \emph{finite} matching map can always be found under the assumptions of \Defref{def:FMatchingProblem}, the hope is  that the \emph{asymptotic} matching map $\AsymptoticMatchingMapAbstr{\switch\rightarrow\switcht}$ can be constructed as the limit
\begin{equation}
  \label{eq:finitelimit}
  \AsymptoticMatchingMapAbstr{\switch\rightarrow\switcht}=\lim_{\matchtime\searrow 0} \FiniteMatchingMapAbstr{\switch\rightarrow\switcht}(\matchtime,\cdot).
\end{equation}
We point out that a very similar idea was originally put forward in \cite{beyer2010b,ames2013a,ames2013b} to solve singular initial value problems underlying the Fuchsian method.

To solve asymptotic matching problems as described above, it turns out to be necessary to control the solutions to a hierarchy of Cauchy problems, which we describe in the remainder of this section.
Let us generalize \Eqref{eq:abstracteqgenN} slightly by adding a so far arbitrary smooth source term $\sourcegen(t,x)$ to the right-hand side 
\begin{equation}
  \label{eq:abstracteqgenNN}
  A[\unkgen](t,x)+\switchgen L[\unkgen](t,x)=\sourcegen(t,x).
\end{equation}
As in \Defref{def:AMatchingProblem} and \Defref{def:FMatchingProblem}, we pick $\switch,\switcht\in I$
and consider any two smooth solutions $\unk$ and $\unkt$ of
\begin{gather}
  \label{eq:abstracteq}
  A[\unk]+\switch L[\unk]=\source,\quad \unk(\finaltime,\cdot)=\CDfinaltime{\unk},\\
  \label{eq:abstracteqt}
  A[\unkt]+\switcht L[\unkt]=\sourcet,\quad \unkt(\finaltime,\cdot)=\CDfinaltime{\unkt},
\end{gather}
where the two smooth source term functions $\source(t,x)$ and $\sourcet(t,x)$ are at this stage allowed to be different (in fact, we shall see below that it is crucial to allow them to be different in order to ``modify'' an asymptotic matching problem suitably). 
Note that prior to solving the matching problem (\Defref{def:AMatchingProblem} or \Defref{def:FMatchingProblem}) the solution $\unkt$ and data $\CDfinaltime{\unkt}$ are not known.
Given the solution $\unk$ of the Cauchy problem \Eqref{eq:abstracteq},
the quantity
\begin{equation}
  \label{eq:defw}
  \unkd=\unk-\unkt
\end{equation}
describes the difference of $\unk$ and the unknown solution $\unkt$ of \Eqref{eq:abstracteqt} whose Cauchy data $\CDfinaltime{\unkt}$ we aim to find.
This
is therefore a solution of
\begin{equation}
  \label{eq:abstracteqw2}
  A[\unkd]+\switcht L[\unkd]=(\switcht-\switch) L[\unk]+\source-\sourcet=:F_3,\quad
  \unkd(\matchtime,\cdot)=0.
\end{equation}
A solution $\unk$ to the Cauchy problem \Eqref{eq:abstracteq} together with a solution to the Cauchy problem \Eqref{eq:abstracteqw2} in principle allow one to construct the finite matching map, and thus determine the data $\CDfinaltime{\unkt}$.
In order to derive continuity estimates (which may eventually allow us to take the limit \Eqref{eq:finitelimit})
 we must consider two such functions $\unkd$ and $\widetilde\unkd$, each defined as in \Eqref{eq:defw} 
with two (possibly different) sets of smooth source terms $\source$, $\sourcet$, $\widetilde\source$ and $\widetilde\sourcet$ and two (possibly different) matching times\footnote{Without loss of generality we  always assume  $\matchtime\le\widetilde\matchtime$.} $\matchtime,\widetilde\matchtime\in (0,\finaltime]$. This leads to the Cauchy problems
\begin{gather}
  \label{eq:abstracteqw3}
  \begin{split}
  A[\unkd-\widetilde\unkd]+\switcht L[\unkd-\widetilde\unkd]=(\switcht-\switch) L[\unk-\widetilde\unk]+(\source-\widetilde\source)-(\sourcet-\widetilde\sourcet)=:F_4,\\ (\unkd-\widetilde\unkd)(\widetilde\matchtime,\cdot)=\unkd(\widetilde\matchtime,\cdot).
\end{split}\\
\label{eq:abstracteqw4}
  A[\unk-\widetilde\unk]+\switch L[\unk-\widetilde\unk]=\source-\widetilde\source=:F_5,
  \quad (\unk-\widetilde\unk)(\finaltime,\cdot)=\CDfinaltime{\unk}-\CDfinaltime{\widetilde\unk}.
\end{gather}

We note that the Cauchy problems listed above \Eqsref{eq:abstracteq}, \eqref{eq:abstracteqt}, \eqref{eq:abstracteqw2}, \eqref{eq:abstracteqw3} and \eqref{eq:abstracteqw4} involve equations which are essentially of the same form as \Eqref{eq:abstracteqgenNN}, but with Cauchy data imposed at various times. Each equation has the same $A$ and $L$, with either $\switchgen=\switch$ or $\switchgen=\switcht$, and having different source terms $\source$, $\sourcet$, $F_3$, $F_4$ or $F_5$.
In all of what follows it is therefore key to find estimates that are uniform in  $\switchgen\in I$ and in the initial time and where the particular choice of $\sourcegen$ and of Cauchy data appears in some particular way. We require these estimates to hold whether we evolve to the future or to the past within the time-interval $(0,\finaltime]$. In our particular application here we find two different  types of estimates, one where the initial condition is imposed at the ``end'' of the time interval $t=\finaltime$ (as for \Eqsref{eq:abstracteq}, \eqref{eq:abstracteqt} and \eqref{eq:abstracteqw4}), and second,   where the initial condition is imposed at the ``beginning'' of the time interval, i.e., at the matching time $t=\matchtime$ (as for \Eqsref{eq:abstracteqw2} and \eqref{eq:abstracteqw3}).

\subsection{Heuristics of an example problem}
\label{sec:EPDheuristics}

In order to develop some intuition for the issues we encounter in proving \Theoremref{thm:maintheorem} let us consider  
 the following example which resembles some of the features of the scalar field part of \Eqsref{eq:scalarfieldnonlinear} -- \eqref{eq:pinonlinear}. A $1+1$-\emph{Euler-Poisson-Darboux equation} takes the form
\begin{equation}
  \label{eq:EPD}
  t\partial_t(t\partial_t \unkgen)-a\, t\partial_t \unkgen-t^{2(1-p)}\partial_x^2\unkgen=g
\end{equation}
We note that $g(t,x)$ in \Eqref{eq:EPD} could be any smooth source term function. We shall assume here that $a$ and $p$ are constants with
$a\in (0,\infty)$ and $p\in (-\infty,1)$.
In consistency with the AVTD phenomenology, we expect that solutions of \Eqref{eq:EPD} should be dominated by the first two terms (possibly including the source term $g$) in the limit $t\searrow 0$, while the spatial derivative term should be negligible.
The choice
\begin{equation}
  \label{eq:EPDmatch}
A[\unkgen]=t\partial_t(t\partial_t \unkgen)-a\, t\partial_t \unkgen-g,\quad L[\unkgen]=-t^{2(1-p)}\partial_x^2\unkgen
\end{equation}
puts \Eqref{eq:EPD} into the form \Eqref{eq:abstracteqgenN}.
In order to preserve the wave-character of the equation and to avoid arbitrary large wave speeds, we restrict $\switchgen$ to  $I:=[0,1]$. 

For {any} value of $\switchgen\in I$ the formal expansions of solutions are of the form\footnote{The special case $a=2(1-p)$ is ignored in all of what follows.}
\begin{equation}
  \label{eq:EPDgenformalexp}
  \unkgen(t,x)=\unkgen_{(0)}(x)+t^a \unkgen_{(1)}(x)+\switchgen\frac{\partial_x^2 \unkgen_{(0)}(x)}{2(1-p)(2(1-p)-a)} t^{2(1-p)}+O(t^{2(1-p)+a}) +O(t^{4(1-p)}),
\end{equation}
as long as $g$ decays sufficiently fast\footnote{For the heuristic discussion in this section we shall not formulate specific decay conditions for $g$.}. The quantities $\unkgen_{(0)}(x)$ and $\unkgen_{(1)}(x)$ are smooth functions. 
Let us now pick any $\switch,\switcht\in I$ and consider the asymptotic matching problem as in \Defref{def:AMatchingProblem}. Let us first pick any constant
\begin{equation}
  \label{eq:EPDlambda1O}
  \lambda<\min\{a,2(1-p)\}
\end{equation}
and  set
  $\|\cdot\|=\|t^{-\lambda}\left(\cdot\right)\|_{L^2(M)}$
for the norm in \Eqref{eq:matchingcond}.
Intuitively, if the corresponding asymptotic matching problem is to be well-defined in this norm, the difference between $\unk$ and $\unkt$ must decay faster than $t^\lambda$ as $t \searrow 0$.
 Then, for any given solution $\unk$ of the $\switch$-version of \Eqref{eq:EPDmatch}, we can find a solution $\unkt$ of the $\switcht$-version such that \Eqref{eq:matchingcond} holds. This function $\unkt$ has to have the same coefficient $\unkgen_{(0)}$ in \Eqref{eq:EPDgenformalexp} as $\unk$. However, this does not fix $\unkt$ uniquely  because \Eqref{eq:matchingcond} the choice \Eqref{eq:EPDlambda1O} does not impose any restrictions on the coefficient  $\unkgen_{(1)}$ in \Eqref{eq:EPDgenformalexp}. Indeed it can be proved rigorously that this asymptotic matching problem is therefore not well-defined. Roughly speaking, the problem is that $\lambda$ is too small to cover both asymptotic degrees of freedom represented by the first two terms in the formal expansion \Eqref{eq:EPDgenformalexp}.

Let us therefore attempt to pick $\lambda$ larger than $a$. Let us first assume that $a<2(1-p)$ and pick
\begin{equation}
  \label{eq:EPDlambda2O}
  \lambda\in (a,2(1-p)).
\end{equation}
In this case, the issue above is resolved:
Given any solution $\unk$ of the $\switch$-version of \Eqref{eq:EPDmatch}, we can find a \emph{unique} solution $\unkt$ of the $\switcht$-version such that \Eqref{eq:matchingcond} holds by matching both $\unkgen_{(0)}$ and $\unkgen_{(1)}$ in \Eqref{eq:EPDgenformalexp} (one can prove that this can be done in this case). Indeed if $a<2(1-p)$, the asymptotic matching problem for the choice \eqref{eq:EPDlambda2O} is well-defined for any pair of switching parameters $\switch,\switcht\in I$. 
In addition, the asymptotic matching map turns out to be bijective, i.e., $(\AsymptoticMatchingMapAbstr{\switch\rightarrow\switcht})^{-1}=\AsymptoticMatchingMapAbstr{\switcht\rightarrow\switch}$, and a homeomorphism in the $C^\infty$-topology. 

\Eqref{eq:EPDgenformalexp} however also suggests that  spatial derivatives may cause trouble if $a>2(1-p)$. In this case, even if we choose $\lambda>a$, the third term in \Eqref{eq:EPDgenformalexp} is not ``negligible'' in comparison to the second term. Moreover, this third term depends explicitly on the choice of $\switchgen$. It is therefore impossible to \emph{fully} match two versions of \Eqref{eq:EPDmatch} given by two different switch parameters. 

As we show below, this is precisely the situation we encounter in the proof of our main theorem.
Is there a way to resolve this? Start again from any  solution $\unk$ of the $\switch$-version of \Eqref{eq:EPDmatch} and its formal expansion \Eqref{eq:EPDgenformalexp}. Suppose that we can determine the leading coefficient $\unkgen_{(0)}$ somehow. Then define 
\begin{equation}
  \label{eq:defmodEPD}
  \unk_{mod}=(\switch-K)\frac{\partial_x^2 \unkgen_{(0)}}{2(1-p)(2(1-p)-a)} t^{2(1-p)}
\end{equation}
for (so far) any $K\in\R$ and calculate
  \begin{align}
    t\partial_t&(t\partial_t \unk_{mod})-a\, t\partial_t \unk_{mod}-\switch t^{2(1-p)}\partial_x^2 \unk_{mod}\notag\\
\label{eq:defgmodEPD}
&=(\switch-K)\partial_x^2 \unkgen_{(0)}(x) t^{2(1-p)}
-\underbrace{\switch t^{2(1-p)}\partial_x^2 \unk_{mod}}_{=:\switch(\switch-K)g^{(mod)}_1=O(t^{4(1-p)})}.
  \end{align}
Setting now
\begin{equation}
  \label{eq:defmodEPD2}
\widetilde\unk=\unk-\unk_{mod}
\end{equation}
implies that 
\begin{equation}
  \label{eq:EPDmod1}
    t\partial_t(t\partial_t \widetilde\unk)-a\, t\partial_t \widetilde\unk-\switch t^{2(1-p)}\partial_x^2 \widetilde\unk\\
=g+\switch (\switch-K) g^{(mod)}_1 -(\switch-K)\partial_x^2 \unkgen_{(0)}(x) t^{2(1-p)}.
\end{equation}
We claim that any solution of this equation with leading term $\unkgen_{(0)}$, in particular, $\widetilde\unk$ given by \Eqref{eq:defmodEPD2}, can be fully matched to some solution of
\begin{equation}
  \label{eq:EPDmod2}
    t\partial_t(t\partial_t \widetilde\unkt)-a\, t\partial_t \widetilde\unkt-\switcht t^{2(1-p)}\partial_x^2 \widetilde\unkt\\
=g +\switcht (\switcht-K) g^{(mod)}_2 -(\switcht-K)\partial_x^2 \unkgen_{(0)}(x) t^{2(1-p)}
\end{equation}
for any sufficiently fast decaying $g^{(mod)}_2(t,x)$.
This is so because any such solution $\widetilde\unk$ has the formal expansion 
\begin{equation*}
  \widetilde\unk(t,x)=\unkgen_{(0)}(x)+t^a \unkgen_{(1)}(x)+K\frac{\partial_x^2 \unkgen_{(0)}(x)}{2(1-p)(2(1-p)-a)} t^{2(1-p)}+O(t^{2(1-p)+a}) +O(t^{4(1-p)});
\end{equation*}
i.e., the spatial derivative term does not depend on the choice of switch parameter anymore.
Indeed, for any such $\widetilde\unk$ we can therefore find $\widetilde\unkt$ as a solution of \Eqref{eq:EPDmod2} such that
\[\|t^{-\lambda}(\widetilde\unk(t,\cdot)-\widetilde\unkt(t,\cdot))\|_{L^2(M)}\rightarrow 0\]
for any $\lambda\in (a,\min\{2(1-p)+a,4(1-p)\})$ as required by  \Defref{def:AMatchingProblem}.

The choice of $K$ in this argument is free. So how do we pick $K$, and, also $g^{(mod)}_1$ and $g^{(mod)}_2$? This depends on the actual problem of interest. If the \emph{source equation} of the matching problem is the \emph{full equation} $\switchgen=\switch=1$, then we would typically choose $K=\switch$ so that both additional terms in \Eqref{eq:EPDmod1} disappear (irrespective of the choice of $g^{(mod)}_1$)  and we recover the original full source equation.
The additional terms in the \emph{target equation} \eqref{eq:EPDmod2} do in general however not disappear. In the case of interest $\switchgen=\switcht=0$, i.e., when the {target equation} is the \emph{model equation}, the term involving $g^{(mod)}_2$ is identically zero.
On the other hand, if the \emph{source equation} is the \emph{model equation} $\switchgen=\switch=0$, then we would typically pick $K=\switcht$. The function $g^{(mod)}_1$ drops out, and the \emph{target equation} becomes the original equation. For all of these particular cases the terms involving $g^{(mod)}_1$ and $g^{(mod)}_2$ vanish. As a consequence, in \Sectionref{sec:LSFProofs} below where we only concern ourselves with these cases, these terms are ignored from the start.
In the cases in which one of $\switch, \switcht$ is $0$ and the other is $1$, the result of these choices for $K$ is to modify the source term of the \emph{model equation}.
So, intuitively because \emph{spatial derivatives are not fully negligible} if $a>2(1-p)$, it is necessary to modify the equations of interest in the way above in order to obtain a well-defined matching problem. This idea shall be applied below in the proof of our main theorem.

\section{Proof of the main result for the linearized lapse-scalar field system}
\label{sec:LSFProofs}

\renewcommand{\unksymb}{\varphi}

\subsection{Main steps of the proof}
The  purpose of this section is to \hyperlink{thmproof:maintheorem}{prove} \Theoremref{thm:maintheorem}. Our proof exploits the ideas in \Sectionref{sec:generalmatchingproblems} and follows the strategy in \Sectionref{sec:generalmatchingproblemsstrategy} incorporating the insights from \Sectionref{sec:EPDheuristics}.
 \Eqsref{eq:scalarfieldlinearMatch} -- \eqref{eq:pilinearMatch} are of the form \Eqref{eq:abstracteqgenNN} with
\begin{align}
  \label{eq:LSFLSFasldkasdk1}
A[(\vnkgen,\nunkgen,\unkgen)]&=-\Bigl(t\partial_t\vnkgen
  +A\nunkgen, \nunkgen
  +2A \vnkgen, t\partial_t\unkgen-\vnkgen-A\nunkgen\Bigr),\\
\label{eq:LSFLSFasldkasdkL}
L[(\vnkgen,\nunkgen,\unkgen)]&=\Bigl(
 -(\BG\pi-A)\nunkgen
  +(1-\BG\alpha) \vnkgen+\BG\alpha t^2\BG\gamma^{ab}D_a D_b\unkgen
  +\nunkgen t^2\BG\gamma^{ab}D_a D_b \BG\phi
  \\
  &+t^2\BG\gamma^{ab}D_a \BG\alpha D_b\unkgen+t^2\BG\gamma^{ab}D_a \nunkgen D_b \BG\phi,\notag\\
  &\,\,\,t^2\BG\gamma^{ab}D_a D_b\nunkgen-\bigl(\bigl(t^2{\BG\chi_a}^{\hspace{1ex}b}{\BG\chi_b}^{\hspace{1ex}a}+\frac {1}{n}+A^2-1\bigr)+(\BG\pi^2-A^2)\bigr) \nunkgen  
  -2(\BG\pi \BG\alpha-A) \vnkgen, \notag\\
&\,\,\,(\BG\alpha-1) u+(\BG\pi-A)\nu
\Bigr), \notag\\
\label{eq:LSFLSFasldkasdklast}
                 \sourcegen&=\left(f^{(1)}, f^{(2)}, f^{(3)}\right).
\end{align}
The plan of attack is to analyze the hierarchy of Cauchy problems \Eqsref{eq:abstracteq}, \eqref{eq:abstracteqw4}, \eqref{eq:abstracteqw2} and \eqref{eq:abstracteqw3} as outlined in \Sectionref{sec:generalmatchingproblemsstrategy}.

We first notice that all these Cauchy problems are well-posed. This can be proved using the information provided in \cite{rodnianski2014,rodnianski2018} and with standard arguments incorporating \Propref{prop:LSF2}. 
Given this, 
we proceed to obtain ``rough (a-priori) estimates'' for solutions of the Cauchy problem which are independent of the value of $\switchgen$ and the particular choice of $\sourcegen$. 
Before stating these rough estimates, which are summarized in \Propref{prop:LSF2} below, we introduce the following norms and \emph{energies}.
First recall the notation \Eqref{eq:deltanorm} for any smooth (time-dependent or time-independent)  $(0,r)$-tensor fields $S$ and $\tilde S$ on $M$. Recall also that all index operations are performed with the reference metric $\delta_{ab}$ in \Eqref{eq:defdelta}, that $\partial_a$ is the covariant derivative associated with $\delta_{ab}$, and, that $\partial^k S$ represents the $(0,r+k)$ tensor field $\partial_{a_1}\cdots\partial_{a_k} S_{b_1\ldots b_r}$.
Based on this we define
\begin{equation}
  \label{eq:tensorL2norm1}
  \|S\|^2_\delta=\int_M |S|_\delta^2 dx,\quad
  \|S\|^2_{\delta,H^k(M)}=\sum_{l=0}^k\|\partial^l S\|^2_{\delta},
\end{equation}
where we integrate with respect to the volume element defined by $\delta_{ab}$. Given any asymptotically point-wise Kasner--scalar field background 
  $\Gamma=(\BG\gamma^{ab}, {\BG\chi_a}^{\hspace{1ex}b}, {\BG\alpha}, {\BG\pi}, \BG\phi)$ and the corresponding metric $\KSF\gamma^{ab}$ (see \Defref{def:APKSF}, and \Eqref{eq:defKasnerringmetric}), we also write
\begin{equation}
  \label{eq:tensorL2norm2}  
  \|t^{-\lambda} \partial S\|^2_{\delta,t^2\KSF\gamma}=\int_M t^2\KSF\gamma^{cd}\left(t^{-\lambda}\partial_c S, t^{-\lambda}\partial_dS\right)_\delta dx,
\end{equation}
for all $t\in (0,\finaltime]$. Moreover, we define the following \emph{energies} as
\begin{align}
    \label{eq:defenergies1N}
    \littleenergyswitch{S,\tilde S}\switchgen{\lambda}(t)&:=\frac 12\int_M\left(\left|t^{-\lambda} S\right|^2_\delta 
+\switchgen t^2\KSF\gamma^{cd}\left(t^{-\lambda} \partial_c\tilde S,t^{-\lambda}  \partial_d\tilde S\right)_{\delta}\right) dx,\\
    \label{eq:defenergies2}
    \littleenergyswitch{S}\switchgen{\lambda}(t)&
                                                  :=\littleenergyswitch{S,S}\switchgen{\lambda}(t),\quad
    \littleenergywk{S}{k}{\lambda}(t):=\littleenergyswitch{S,0}\switchgen{\lambda}(t),
  \end{align}
for any smooth function $\lambda(x)$.
Given any smooth time-dependent {functions} $u$, $\varphi$, $\nu$ on $M$ (not necessarily solutions of \Eqsref{eq:scalarfieldlinearMatch} -- \eqref{eq:pilinearMatch}), we also set
\begin{align}
  \label{eq:defEnergies}
  \energyswitch{u, \varphi}{\switchgen}{k}{\lambda}(t)&=\sum_{l=0}^k \littleenergyswitch{\partial^l u, \partial^l\varphi}\switchgen{\lambda}(t),
\quad \energyswitch{\nu}{\switchgen}{k}{\lambda}(t)=\sum_{l=0}^k \littleenergyswitch{\partial^l\nu}\switchgen{\lambda}(t),\\
\label{eq:defEnergies2}
  \energy{u}{k}{\lambda}(t)&=\sum_{l=0}^k \littleenergy{\partial^l u}{\lambda}(t). 
\end{align}

\begin{proposition}[Rough a-priori estimates for solutions of the linearized lapse-scalar field system]
  \label{prop:LSF2}
  Consider an arbitrary asymptotically point-wise Kasner--scalar field background 
  $\Gamma=(\BG\gamma^{ab}, {\BG\chi_a}^{\hspace{1ex}b}, {\BG\alpha}, {\BG\pi}, \BG\phi)$ with positive decay $\beta$.
Let $v=(u,\nu,\varphi)$ be any smooth solution of \Eqref{eq:abstracteqgenNN} with $A$ and $L$ given by \Eqsref{eq:LSFLSFasldkasdk1} and \eqref{eq:LSFLSFasldkasdkL}, and, 
\begin{equation}
  \label{eq:abstrst}
  \sourcegen=(\sourcegen^{(1)}(t,x), \sourcegen^{(2)}(t,x), \sourcegen^{(3)}(t,x))
\end{equation}
for arbitrary smooth  $\sourcegen^{(1)}(t,x)$, $\sourcegen^{(2)}(t,x)$ and $\sourcegen^{(3)}(t,x)$ and $\switchgen\in I=[0,1]$.
Pick any sufficiently small $\epsilon>0$ and any integer $k\ge 0$. Then there is a constant $C>0$ such that for any $t_*\in (0,\finaltime]$ the following estimates hold:
\begin{enumerate}
\item For all $t\in[t_*,\finaltime]$ and for any smooth function $\lambda>\lambda_c$ (see \Eqref{eq:deflambdac}), we have\footnote{As is standard in the literature, ``constants'' $C$ represent positive quantities whose exact values are not important and may change in each calculation step. They may in general depend on other quantities which we only list if they are relevant for the current argument. By default, we always assume that $C$ \emph{is uniform in all quantities which appear explicitly in a particular estimate}. All exceptions to this rule are listed explicitly in the text. If $C$ does \emph{not} depend on any other quantity (a quantity which does not appear explicitly in the estimate), we may also make a remark in the text.} 
\begin{equation}
  \label{eq:LSFforwards1}
  \begin{split}
    &\!\!\!\!\!\!\!\!\!\energyswitch{u,\varphi}\switchgen{k}{\lambda}(t)
\le  C\Bigl(\energyswitch{u,\varphi}\switchgen{k}{\lambda} (t_*)\\
      &\!\!\!\!\!\!\!\!\!\!\!\!\!\!+\int^t_{t_*}\Bigl(\|s^{-\lambda} \sourcegen^{(1)}\|^2_{\delta,H^k(M)} 
      +\|s^{-\lambda} \sourcegen^{(2)}\|^2_{\delta,H^k(M)} 
      +\|s^{-\lambda} \sourcegen^{(3)}\|^2_{\delta,H^{k+1}(M)}\Bigr) s^{-1+2\epsilon} ds 
      \Bigr).
  \end{split}
\end{equation}
\item For all $t\in[t_*,\finaltime]$ and for any smooth function $\lambda\ge 0$ on $M$, we have
\begin{equation}
\label{eq:LSFforward2}
\begin{split}
  &\sqrt{\energy{\unkgen}{k}{\lambda}(t)}
\le \sqrt{\energy{\unkgen}{k}{\lambda}(t_*)}\\
 &\quad+C
\int_{t_*}^t\Bigl(\sqrt{\energy{\vnkgen}{k}{\lambda}(s)}
+\|s^{-\lambda} \sourcegen^{(2)}\|_{\delta,H^k(M)}
+\|s^{-\lambda} \sourcegen^{(3)}\|_{\delta,H^k(M)}      
  \Bigr) s^{-1} ds.
\end{split}
\end{equation}
\item For all $t\in (0,t_*]$ and for any smooth function $\lambda<0$ on $M$, we have
\begin{equation}
  \label{eq:LSFbackward1}
  \begin{split}
    &\!\!\!\!\!\!\!\!\!\energyswitch{u,\varphi}\switchgen{k}{\lambda}(t)
      \le  C\Bigl(\energyswitch{u,\varphi}\switchgen{k}{\lambda+\epsilon} (t_*)\\
      &\!\!\!\!\!\!\!\!\!\!\!\!\!\!\!\!\!+\int_t^{t_*}\Bigl(\|s^{-\lambda} \sourcegen^{(1)}\|^2_{\delta,H^k(M)}
      +\|s^{-\lambda} \sourcegen^{(2)}\|^2_{\delta,H^k(M)}
      +\|s^{-\lambda} \sourcegen^{(3)}\|^2_{\delta,H^{k+1}(M)}\Bigr) s^{-1-2\epsilon} ds
      \Bigr).
  \end{split}
\end{equation}
\item For all $t\in(0,t_*]$ and for any smooth function $\lambda\le 0$ on $M$, we have
\begin{equation}
\label{eq:LSFbackward2}
\begin{split}
  &\sqrt{\energy{\varphi}{k}{\lambda}(t)}
\le \sqrt{\energy{\varphi}{k}{\lambda}(t_*)}\\
 &\quad +C
\int^{t_*}_t\Bigl(\sqrt{\energy{u}{k}{\lambda}(s)}+\|s^{-\lambda} \sourcegen^{(2)}\|_{\delta,H^{k}(M)}+\|s^{-\lambda} \sourcegen^{(3)}\|_{\delta,H^{k}(M)}      
  \Bigr) s^{-1} ds.
\end{split}
\end{equation}
\end{enumerate}
The constants $C$ may depend on $\finaltime$, $k$, $\lambda$, $\epsilon$ and $\Gamma$.
Finally, for all $t\in (0,\finaltime]$ and for any smooth $\lambda$ on $M$, we have
  \begin{equation}
  \label{eq:preestimat3N}
  \energyswitch{\nu}\switchgen{k}{\lambda}(t)
  \le C \left(\energy{u}k{\lambda}(t) 
+\|t^{-\lambda} \sourcegen^{(2)}\|^2_{\delta,H^k(M)}
 \right).
\end{equation}
where the constant $C>0$ may depend on $\finaltime$, $k$ and $\Gamma$.
\end{proposition}
The proof can be found in \Sectionref{sec:LSFproof1}. As we explain in the proof it turns out that in order to identify the optimal bounds given for $\lambda$ for \Eqsref{eq:LSFforwards1} and \eqref{eq:LSFbackward1}, it is necessary to find the right combination of the hyperbolic and elliptic estimates which are derived from the equations for $u$ and $\nu$ separately in a first step.
Also notice that \Propref{prop:LSF2} does yet not provide optimal control of the asymptotics as a result of the restrictions for $\lambda$ in both \Eqsref{eq:LSFforwards1} and \eqref{eq:LSFbackward1}. In fact, \Eqref{eq:LSFbackward1} suggests that $u$ might not even be bounded at $t=0$.
 By sacrificing some differentiability, it turns out that we can
 indeed recover the optimal decay for $u$, and this step is crucial in
 the proof of \Theoremref{thm:maintheorem}. This result is summarized
 in the following proposition. 

\begin{proposition}[Improved decay estimates for solutions of the linearized lapse-scalar field system]
  \label{prop:LSF4}
  Consider an arbitrary asymptotically point-wise Kasner--scalar field background 
  $\Gamma=(\BG\gamma^{ab}, {\BG\chi_a}^{\hspace{1ex}b}, {\BG\alpha}, {\BG\pi}, \BG\phi)$ with positive decay $\beta$.
 Given any $\CDfinaltime{u},\CDfinaltime{\varphi}\in C^\infty(M)$, let $v=(u,\nu,\varphi)$ be the smooth solution of \Eqref{eq:abstracteqgenNN} with $A$ and $L$ given by \Eqsref{eq:LSFLSFasldkasdk1} and \eqref{eq:LSFLSFasldkasdkL}, and, $\sourcegen$ given by \Eqref{eq:abstrst},
satisfying the initial condition
    $u(\finaltime,\cdot)=\CDfinaltime{u}$,
 $\varphi(\finaltime,\cdot)=\CDfinaltime{\varphi}$.
Pick any sufficiently small $\epsilon>0$ and any integer $k\ge 0$. Then there is a constant $C>0$, which may depend on $\finaltime$, $k$, $\lambda$, $\epsilon$ and $\Gamma$, such that
\begin{align}
  \label{eq:LSFimproveddecayubackward}
  &\energy{u}{k}{\lambda}(t)  
\le C\Bigl(
\|u_*\|^2_{\delta,H^{k+2}(M)}+\switchgen \|\varphi_*\|^2_{\delta,H^{k+3}(M)}+\|\varphi_*\|^2_{\delta,H^{k+2}(M)}
\\      
  &+\int_t^{\finaltime}\Bigl(\|s^{-\lambda} \sourcegen^{(1)}\|^2_{\delta, H^{k+2}(M)}
    +\|s^{-\lambda} \sourcegen^{(2)}\|^2_{\delta, H^{k+2}(M)}
+\|s^{-\lambda+2\kappa-\epsilon} \sourcegen^{(3)}\|^2_{\delta,H^{k+3}(M)}\Bigr) s^{-1-2\epsilon}ds
\Bigr)\notag
\end{align}
for all $t\in (0,\finaltime]$ for any smooth function $\lambda$ on $M$ with $\lambda<\min\{2(1-q_{max}),2A^2\}$. 
We define $\kappa$ to be any smooth strictly positive function strictly smaller than $1-q_{max}$ at each $x\in M$, where $q_{max}$ is defined in \Eqref{eq:defqmax}.
\end{proposition}
The proof is discussed in \Sectionref{sec:LSFproof2}. 
Thanks to the above estimates, we can now understand the issue which is heuristically motivated  in \Sectionref{sec:EPDheuristics}.
In order to solve the Cauchy problem \Eqref{eq:abstracteqw2} we need to estimate the source term $F_3$ there. If we do this for the case here we encounter integrals of the form
\[\int_{t_*}^t \|s^{-\lambda}\BG\alpha s^2\BG\gamma^{ab}D_a D_b\unkgen\|_{H^k(M)} s^{-1} ds\]
for $t\ge t_*$ which are required to be bounded in the limit $t_*\searrow 0$ for $\lambda>\lambda_c$. We have discussed above in \Sectionref{sec:mainresults} that this is only possible in the exceptional case that $2(1-q_{max})>\lambda_c$. In fact, the asymptotic matching problem is therefore in general ill-defined and we need to consider suitable ``modifications''. In order to work out these modifications as in \Sectionref{sec:EPDheuristics} we require a map which yields the asymptotic datum $\unkgen_{(0)}$ (cf. \Theoremref{thm:maintheorem}) for any given solution. The process of finding this is referred to as \emph{pre-matching}; the motivation for this name is given in the \hyperlink{propproof:LSF3}{proof} of \Propref{prop:LSF3} discussed in \Sectionref{sec:LSFproof3}. 

\begin{proposition}[Pre-matching for the linearized lapse-scalar field system]
  \label{prop:LSF3}
  Consider an arbitrary asymptotically point-wise Kasner--scalar field background 
  $\Gamma=(\BG\gamma^{ab}, {\BG\chi_a}^{\hspace{1ex}b}, {\BG\alpha}, {\BG\pi}, \BG\phi)$ with positive decay $\beta$. Suppose the source terms $f^{(1)}(t,x)$,
  $f^{(2)}(t,x)$ and $f^{(3)}(t,x)$ in
  \Eqsref{eq:scalarfieldlinearMatch} -- \eqref{eq:pilinearMatch} are 
  smooth functions on $(0,\finaltime]\times M$ such that
  \begin{equation*}    
    \int^{\finaltime}_0 \Bigl(\|s^{-\lambda_s} f^{(1)}\|_{\delta,H^{k+2}(M)}
    + \|s^{-\lambda_s} f^{(2)}\|_{\delta,H^{k+2}(M)}
    +\|s^{-\lambda_s+2\kappa} f^{(3)}\|_{\delta,H^{k+3}(M)}\Bigr)s^{-1} ds<\infty,
\end{equation*}
  holds  
  for some integer $k\ge 0$ and for some smooth function $\lambda_s(x)$ on $M$ with $\lambda_s>0$.
  Pick  $\switchgen\in I=[0,1]$.
  Then there is a map
  \[\PreMatchingMapAbstr: (C^\infty(M))^2 \rightarrow C^\infty(M), \]
  and a constant $C>0$ such that,
given any $\CDfinaltime{\vnkgen}, \CDfinaltime{\unkgen}\in C^\infty(M)$ and the  corresponding smooth solution $(\vnkgen,\nu,\unkgen)$ of the Cauchy problem of \Eqsref{eq:scalarfieldlinearMatch} -- \eqref{eq:pilinearMatch} with Cauchy data with $\unkgen(\finaltime,\cdot)=\CDfinaltime{\unkgen}$ and $\vnkgen(\finaltime,\cdot)=\CDfinaltime{\vnkgen}$,
\begin{align}  
    \label{eq:LSFLSFprematchestimate}
&\left\|t^{-\lambda}
\left(\unkgen(t,\cdot)-\PreMatchingMapAbstr(\CDfinaltime{\vnkgen}, \CDfinaltime{\unkgen})\right)
\right\|_{\delta,H^k(M)}^2
\le
C  \Bigl(
\|u_*\|^2_{\delta,H^{k+2}(M)}+\|\varphi_*\|^2_{\delta,H^{k+3}(M)}
\\      
  &+\int_0^{\finaltime}\Bigl(\|s^{-\tilde\lambda} f^{(1)}\|^2_{\delta, H^{k+2}(M)}
    +\|s^{-\tilde\lambda} f^{(2)}\|^2_{\delta, H^{k+2}(M)}+\|s^{-\tilde\lambda+2\kappa} f^{(3)}\|^2_{\delta,H^{k+3}(M)}\Bigr) s^{-1}ds
\Bigr),\notag
\end{align}
for all $t\in (0,\finaltime]$, for any smooth $0<\tilde\lambda\le \lambda_s$ and
$\lambda<\min\{\tilde\lambda, 2(1-q_{max}),2A^2\}$. 
In fact, there is only one map with this property. Moreover, for any smooth $\lambda<\min\{2(1-q_{max}),2A^2\}$, we have
\begin{equation}
  \label{eq:LSFLSFcontinuitypre129031lemma2}
  \begin{split}
    \left\|t^{-\lambda}\left[
\left(\unkgen(t,\cdot)-\PreMatchingMapAbstr (\CDfinaltime{\vnkgen}, \CDfinaltime{\unkgen})\right)
-\left(\tilde\unkgen(t,\cdot)-\PreMatchingMapAbstr (\CDfinaltime{\tilde\vnkgen}, \CDfinaltime{\tilde\unkgen}) \right)
\right]
\right\|_{\delta,H^k(M)}\\
\le
C \left(\|\CDfinaltime{\vnkgen}-\CDfinaltime{\tilde\vnkgen}\|_{\delta,H^{k+2}(M)}
  +\|\CDfinaltime{\unkgen}-\CDfinaltime{\tilde\unkgen}\|_{\delta,H^{k+3}(M)}\right),
\end{split}
\end{equation}
for any $\CDfinaltime{\vnkgen}, \CDfinaltime{\unkgen}, \CDfinaltime{\tilde\vnkgen}, \CDfinaltime{\tilde\unkgen}\in C^\infty(M)$ and all $t\in (0,\finaltime]$. The constants $C$ above may depend on $k$, $\finaltime$, $\lambda$, $\tilde\lambda$ and $\Gamma$. The function $\kappa$ is defined in \Propref{prop:LSF4}.
\end{proposition}

Once we have obtained the pre-matching map as in the previous proposition, we are now in the position to formulate (and solve) suitably modified asymptotic matching problems for the linearized lapse-scalar field system. Finding such modifications, as motivated in \Sectionref{sec:EPDheuristics}, is however not straightforward. 
The problem is that in order to modify the equations (typically the model equation), the leading asymptotic datum must be known, and that this datum itself depends on the solution of the modified equation. In fact, the notion of the asymptotic matching problem introduced in \Sectionref{sec:modelproblemmatching} relies on \emph{fixed} choices for $A$, $L$, $F_1$ and $F_2$ and that the each Cauchy problem in the hierarchy described there is well-posed. The \emph{combined} problem now of finding a solution of the asymptotic matching problem \emph{and} of finding the modification which renders the asymptotic matching problem well-defined for this solution, however, cannot be expected to be well-defined in general. The following proposition carefully separates these two problems in a way which applies in particular to the asymptotic matching problem given by \Theoremref{thm:maintheorem}.

\newcommand{\AD}[1]{{#1}_\times}
\begin{proposition}[Modified asymptotic matching problem for the linearized lapse-{sca\-lar} field system]
  \label{prop:LSFresult1}  
  Consider an arbitrary asymptotically point-wise Kasner--scalar field background 
  $\Gamma=(\BG\gamma^{ab}, {\BG\chi_a}^{\hspace{1ex}b}, {\BG\alpha}, {\BG\pi}, \BG\phi)$ with positive decay $\beta$ satisfying \Eqref{eq:LSFtheominequFinal}
such that 
\Eqref{eq:LSFLSFmodLem293482}
  holds where $\lambda_c$ is defined in \Eqref{eq:deflambdac}.
Suppose the source terms $f^{(1)}(t,x)$,
  $f^{(2)}(t,x)$ and $f^{(3)}(t,x)$ in
  \Eqsref{eq:scalarfieldlinearMatch} -- \eqref{eq:pilinearMatch} are 
  smooth functions on $(0,\finaltime]\times M$ such that
  \begin{equation*}
    \int^{\finaltime}_0 \Bigl(\|s^{-\lambda_s} f^{(1)}\|_{\delta,H^{k+4}(M)}
    + \|s^{-\lambda_s} f^{(2)}\|_{\delta,H^{k+4}(M)}
    +\|s^{-\lambda_s+2\kappa} f^{(3)}\|_{\delta,H^{k+5}(M)}\Bigr)s^{-1} ds<\infty,
\end{equation*}
  holds  
  for some integer $k\ge 0$ and for some smooth function $\lambda_s(x)$ on $M$ satisfying 
  \Eqref{eq:LSFLSF1cond4}.    
  Pick $\switch,\switcht\in I=[0,1]$,
  $K\in\R$ and any sufficiently small $\epsilon>0$.
Suppose that for some integers $k_0,k_1\ge 0$, there is a continuous map
\[\Phi: H^{k+4+k_0}(M)\times H^{k+4+k_1}(M)\rightarrow H^{k+4}(M)\times H^{k+5}(M),\quad (\AD{\vnk}, \AD{\unk})\mapsto (\CDfinaltime{\vnk}, \CDfinaltime{\unk})\]
and a map\footnote{Just to avoid any confusions, we stress that this map $\chi$ is completely unrelated to the trace-free part of the second fundamental form ${\chi_a}^b$.}
\[\chi: (C^\infty(M))^2\rightarrow C^\infty(M),\quad (\AD{\vnk}, \AD{\unk})\mapsto\LOT{\unk}\]
with the property that there is a constant $C>0$, which may depend on $k$, $\finaltime$, $\lambda$  and $\Gamma$, such that for every $(\AD{\vnk}, \AD{\unk})\in  (C^\infty(M))^2$,
\begin{align}  
&\left\|t^{-\lambda}
\left(\unk(t,\cdot)-\LOT{\unk}\right)
\right\|_{\delta,H^{k+2}(M)}^2
\le
C  \Bigl(
\|\AD{\vnk}\|^2_{\delta,H^{k+4+k_0}(M)}+ \|\AD{\unk}\|^2_{\delta,H^{k+4+k_1}(M)}
\notag\\      
\label{eq:LSFLSFLOTestimate}
  &\qquad+\int_0^{\finaltime}\Bigl(\|s^{-\lambda_s} f^{(1)}\|^2_{\delta, H^{k+4}(M)}
    +\|s^{-\lambda_s} f^{(2)}\|^2_{\delta, H^{k+4}(M)}\\
&\qquad\qquad+\|s^{-\lambda_s+2\kappa} f^{(3)}\|^2_{\delta,H^{k+5}(M)}\Bigr) s^{-1}ds
\Bigr),\notag\\
  &\energy{\vnk}{k+2}{\lambda}(t)  
\le C\Bigl(
\|\AD{\vnk}\|^2_{\delta,H^{k+4+k_0}(M)}+ \|\AD{\unk}\|^2_{\delta,H^{k+4+k_1}(M)}
\notag\\      
\label{eq:LSFLSFLOTestimate2}
  &\qquad+\int_0^{\finaltime}\Bigl(\|s^{-\lambda_s} f^{(1)}\|^2_{\delta, H^{k+4}(M)}
    +\|s^{-\lambda_s} f^{(2)}\|^2_{\delta, H^{k+4}(M)}\\
&\qquad\qquad+\|s^{-\lambda_s+2\kappa} f^{(3)}\|^2_{\delta,H^{k+5}(M)}\Bigr) s^{-1}ds
\Bigr),\notag
\end{align}
for all $t\in (0,\finaltime]$, and for any smooth function $\lambda<\min\{\lambda_s,2(1-q_{max}),2A^2\}$ on $M$. Here $v_1=(\vnk,\nunk,\unk)$ is the solution of \Eqref{eq:abstracteqgenNN} with $A$ and $L$ given by \Eqsref{eq:LSFLSFasldkasdk1} and \eqref{eq:LSFLSFasldkasdkL} and
\begin{equation}
    \label{eq:LSFLSFmodLem12}
    \begin{split}
    \sourcegen^{(1)}&= f^{(1)} +(\switch-K)\left({\BG\alpha} t^2\BG\gamma^{ab}D_a D_b\LOT{\unk}
+t^2\BG\gamma^{ab}D_b \BG\alpha D_a\LOT{\unk}\right),\\
\sourcegen^{(2)}&= f^{(2)},\quad \sourcegen^{(3)}= f^{(3)},\quad \switchgen=\switch,
\end{split}
\end{equation}
satisfying the Cauchy condition 
  \begin{equation}
\unk(\finaltime,\cdot)=\CDfinaltime{\unk},\quad \vnk(\finaltime,\cdot)=\CDfinaltime{\vnk}.
\end{equation}
Then there there is a map
\[\AsymptoticMatchingMapAbstr{\switch\rightarrow \switcht}: (C^\infty(M))^2 \rightarrow (C^\infty(M))^2\]
with the following properties:
\begin{enumerate}
\item Pick any $(\AD{\vnk}, \AD{\unk})\in (C^\infty(M))^2$ as above and consider
 the corresponding solution $v_1=(\vnk,\nunk,\unk)$ as above. The solution 
$v_2=(\vnkt,\nunkt,\unkt)$ of \Eqref{eq:abstracteqgenNN} with $A$ and $L$ given by \Eqsref{eq:LSFLSFasldkasdk1} and \eqref{eq:LSFLSFasldkasdkL} and
\begin{equation}
    \label{eq:LSFLSFmodLem22}
    \begin{split}
    \sourcegen^{(1)}&= f^{(1)} +(\switcht-K)\left({\BG\alpha} t^2\BG\gamma^{ab}D_a D_b\LOT{\unk}
+t^2\BG\gamma^{ab}D_b \BG\alpha D_a\LOT{\unk}\right),\\
\sourcegen^{(2)}&= f^{(2)},\quad \sourcegen^{(3)}= f^{(3)},\quad \switchgen=\switcht,
\end{split}
\end{equation}
determined by Cauchy data 
\begin{equation}
  \label{eq:LSFLSFmodLem095}
(\CDfinaltime{\vnkt}, \CDfinaltime{\unkt})
=\AsymptoticMatchingMapAbstr{\switch\rightarrow \switcht} (\AD{\vnk}, \AD{\unk})
\end{equation}
with $\unkt(\finaltime,\cdot)=\CDfinaltime{\unkt}$ and $\vnkt(\finaltime,\cdot)=\CDfinaltime{\vnkt}$
satisfies
\begin{align}
  \label{eq:LSFLSFmodLem23132}
&\left\|t^{-\lambda}(\vnk(t,\cdot)-\vnkt(t,\cdot))\right\|_{H^k(M)}^2
+\left\|t^{-\lambda}(\unk(t,\cdot)-\unkt(t,\cdot))\right\|_{H^k(M)}^2
  \\
&\le  C
\Bigl(  \|\AD{\vnk}\|^2_{\delta,H^{k+4+k_0}(M)}+ \|\AD{\unk}\|^2_{\delta,H^{k+4+k_1}(M)}
\notag\\
&+\int^\finaltime_{0}\Bigl(
    \left\|s^{-\lambda_s} f^{(1)}\right\|^2_{H^{k+4}(M)} 
    \left\|s^{-\lambda_s} f^{(2)}\right\|^2_{H^{k+4}(M)} 
    \left\|s^{-\lambda_s+2\kappa} f^{(3)}\right\|^2_{H^{k+5}(M)} 
\Bigr) s^{-1} ds
\Bigr) \notag
\end{align}
for all $t\in (0,\finaltime]$
and any smooth function $\lambda$ on $M$ satisfying
\Eqref{eq:LSFLSFmodLem29348}.
The constant $C$ here may depend on $k$, $\finaltime$, $\lambda$ and $\Gamma$.
\item The map $\AsymptoticMatchingMapAbstr{\switch\rightarrow \switcht}$
is continuous in the sense that
\begin{equation}
\label{eq:LSFLSFmodLem23098}
\begin{split}
&\|\AsymptoticMatchingMapAbstr{\switch\rightarrow \switcht}( \AD{\vnk}, \AD{\unk})
-\AsymptoticMatchingMapAbstr{\switch\rightarrow \switcht}(\AD{\tilde\vnk}, \AD{\tilde\unk})\|_{H^k(M)}
\\
&\le C \Bigl(
\|\AD{\vnk}-\AD{\tilde\vnk}\|_{H^{k+4+k_0}(M)}  +\|\AD{\unk}- \AD{\tilde\unk}\|_{H^{k+4+k_1}(M)}
\Bigr)
\end{split}
\end{equation}
for any smooth $\AD{\vnk}$, $\AD{\unk}$, $\AD{\tilde\vnk}$, $\AD{\tilde\unk}$. The constant $C$ here may depend on $k$, $\finaltime$, $\lambda$, $\Gamma$ and $\Phi$.
\item Given any smooth $(\AD{\vnk}, \AD{\unk})\in (C^\infty(M))^2$ and the corresponding solution $v_1=(\vnk,\nunk,\unk)$ as above, suppose there is a smooth solution $(\hat\vnkt,\hat\nunkt,\hat\unkt)$ of \Eqref{eq:abstracteqgenNN} with $A$ and $L$ given by \Eqsref{eq:LSFLSFasldkasdk1} and \eqref{eq:LSFLSFasldkasdkL} and \eqref{eq:LSFLSFmodLem22} such that
 \begin{equation}
   \label{eq:LSFLSFmodLem1239}
\left\|t^{-\lambda}\left(t\partial_t\unk(t,\cdot)-t\partial_t\hat\unkt(t,\cdot)\right) \right\|_{L^2(M)} 
+
\left\|t^{-\lambda}\left(\unk(t,\cdot)-\hat\unkt(t,\cdot)\right) \right\|_{H^1(M)} 
\le C
 \end{equation}
 for all $t\in (0,\finaltime]$ and \emph{some} $\lambda$ consistent with \Eqref{eq:LSFLSFmodLem29348} where the constant $C$ here may depend on $k$, $\finaltime$, $\lambda$, $\Gamma$, $(\vnk,\nunk,\unk)$ and $(\hat\vnkt,\hat\nunkt,\hat\unkt)$.
Then $(\hat\vnkt,\hat\nunkt,\hat\unkt)$ agrees identically with the solution $v_2=(\vnkt,\nunkt,\unkt)$ of \Eqref{eq:abstracteqgenNN} with $A$ and $L$ given by \Eqsref{eq:LSFLSFasldkasdk1}, \eqref{eq:LSFLSFasldkasdkL} and \eqref{eq:LSFLSFmodLem22}
determined by the Cauchy data \Eqref{eq:LSFLSFmodLem095}.
\item Pick any two smooth $(\AD{\vnk}, \AD{\unk})$ and $(\AD{\tilde\vnk}, \AD{\tilde\unk})$. Supposing that
  $\AsymptoticMatchingMapAbstr{\switch\rightarrow \switcht}(\AD{\vnk}, \AD{\unk})
=\AsymptoticMatchingMapAbstr{\switch\rightarrow \switcht}(\AD{\tilde\vnk}, \AD{\tilde\unk})$,
it follows that $\LOT{\unk}=\LOT{\tilde\unk}$. 
\end{enumerate}
The function $\kappa$ is defined in \Propref{prop:LSF4}.
\end{proposition}

The \hyperlink{propproof:LSFresult1}{proof} is discussed in \Sectionref{sec:LSFproof3}.
In consistency with the discussion above, it is not clear in general whether maps $\Phi$ and $\chi$ assumed in this proposition can always be found given $\switch$, $\switcht$ and $K$. In the \hyperlink{thmproof:maintheorem}{proof} of \Theoremref{thm:maintheorem} we show how one can pick these quantities in the most important cases of interest. Before we do this however let us remark that \Propref{prop:LSFresult1}   is restricted to asymptotic matching problems with modifications based on time-\emph{independent} ``leading order terms'' $\LOT{\unk}(x)$. These can be found thanks to the map $\PreMatchingMapAbstr$ in \Propref{prop:LSF3}. While this is sufficient for our purposes, more general problems may require time-dependent choices which take into account terms of higher orders of the solution at $t=0$. For this, \Propref{prop:LSF3} (by itself) would not be sufficient anymore.

\begin{proof}[Proof of \Theoremref{thm:maintheorem}.]
\hypertarget{thmproof:maintheorem}{}
For the following proof we assume $\switchgen\not=0$. The case $\switchgen=0$ is trivially covered by taking $\Psi=\mathrm{id}$. 

We start with the following considerations. Given 
any smooth $(\unkgen_{(0)}, \unkgen_{(1)})\in (C^\infty(M))^2$, 
\begin{equation}
  \label{eq:LSFLSFmodelsol}
  \begin{split}
    \vnk(t,x)=\,&-2A^2(x)t^{2A^2(x)} \unkgen_{(1)}(x)\\
    &+\int_t^\finaltime \Lop{\unkgen_{(0)}}(s,x) (t/s)^{2A^2(x)}s^{-1}ds+ S^{(1)}(t,x)\\
  \unk(t,x)=\,&\unkgen_{(0)}(x)+(2A^2(x)-1)t^{2A^2(x)} \unkgen_{(1)}(x)\\
  &\!\!\!\!\!\!\! -(2A^2(x)-1)\int_0^t\int_\tau^\finaltime \Lop{\unkgen_{(0)}}(s,x) (\tau/s)^{2A^2(x)}s^{-1}ds\,\tau^{-1} d\tau+S^{(2)}(t,x)\\
  \nunk(t,x)=\,&-2A(x)\vnk(t,x)-f^{(2)}(t,x)
\end{split}
\end{equation}
represents the general solution of \Eqref{eq:abstracteqgenNN} with \Eqsref{eq:LSFLSFasldkasdk1}, \eqref{eq:LSFLSFasldkasdkL} and 
    \eqref{eq:LSFLSFmodLem12}, $\switch=0$, $\switcht=\switchgen$ and $K=\switcht$ for $\unk_{(0)}=\unkgen_{(0)}$. The source equation is therefore the modified model equation. The target equation would agree with the original full equation if $\switchgen$ (which is a free parameter in \Theoremref{thm:maintheorem}) were chosen to be one.
Recall the definitions of $\Lop{\cdot}$,  $S^{(1)}$ and $S^{(2)}$ in \Eqsref{eq:LSF17}, \eqref{eq:defS1} and \eqref{eq:defS2}. 
Recall that all integrals are well-defined and finite.
 The corresponding
  Cauchy data $(\CDfinaltime{\vnk}, \CDfinaltime{\unk})$ are found by evaluating \Eqref{eq:LSFLSFmodelsol} at $t=\finaltime$, and, the corresponding map
$\Phi:(\unkgen_{(0)}, \unkgen_{(1)})\mapsto (\CDfinaltime{\vnk}, \CDfinaltime{\unk})$ is therefore well-defined. Using integration by parts, we can write
\begin{align}
  \mathfrak{E}[\unkgen_{(0)}]&:=-\int_0^\finaltime \Lop{\unkgen_{(0)}} s^{-1}ds\\
&=-\frac {2A^2}{2A^2-1}\unkgen_{(0)}
+\frac {2A^2}{2A^2-1}(\CDfinaltime{\unk}-S^{(2)}(\finaltime,\cdot))+\CDfinaltime{\vnk}-S^{(1)}(\finaltime,\cdot)\notag\\
\label{EPD:LSFCD2AD2NN}
\unkgen_{(1)}&=-\frac{1}{2A^2 \finaltime^{2A^2} }\left(\CDfinaltime{\vnk}-S^{(1)}(\finaltime,\cdot)\right).
\end{align}
Observe carefully that the time integral here \emph{acts only on the time-dependent
  coefficients of $\Lop{\cdot}$ and not on $\unkgen_{(0)}$ (which does not depend on
  time by assumption)}. In fact, $\mathfrak{E}$ is therefore a standard linear differential operator acting on $\unkgen_{(0)}$ by spatial derivatives only.
 Following Appendix~II of \cite{choquet-bruhat2008c}, it is easy to show that $\mathfrak{E}$ is elliptic. Since it is therefore a continuous map $H^{k+2}(M)\rightarrow H^k(M)$ for any $k>n/2-2$ (see Corollary~2.2 there), it follows that 
\begin{equation}
  \label{eq:contPhithm}
  \Phi: H^{k+2}(M)\times H^k(M)\rightarrow H^k(M)\times H^k(M),\quad (\unkgen_{(0)}, \unkgen_{(1)})\mapsto (\CDfinaltime{\vnk}, \CDfinaltime{\unk})
\end{equation}
is a continuous map.

Given this, the next step of this proof is now to apply \Propref{prop:LSFresult1} with $\switch=0$, $\switcht=\switchgen$ and $K=\switcht$ for this choice of $\Phi$ and for the map $\chi$ being the projection on the first component. In verifying the hypothesis of this proposition we first notice that $(\unkgen_{(0)}, \unkgen_{(1)})$ in \Theoremref{thm:maintheorem} plays the role of $(\AD{\vnk}, \AD{\unk})$ in \Propref{prop:LSFresult1}, $\unkgen_{(0)}$ in \Theoremref{thm:maintheorem} the role of $\LOT{\unk}$ in \Propref{prop:LSFresult1}, and, $(\CDfinaltime{\vnkgen}, \CDfinaltime{\unkgen})$ in \Theoremref{thm:maintheorem} the role of  $(\CDfinaltime{\vnkt}, \CDfinaltime{\unkt})$ in \Propref{prop:LSFresult1}. In the following we express everything in terms of the quantities from \Theoremref{thm:maintheorem}.
Regarding the hypothesis for $\Phi$ in \Propref{prop:LSFresult1}, we replace $k$ in \Eqref{eq:contPhithm} by $k+5$ and set $k_0=3$ and $k_1=1$. Now in order to check \Eqsref{eq:LSFLSFLOTestimate}, we exploit \Propref{prop:LSF3} applied to the $\switch$-system where $f^{(1)}$, $f^{(2)}$, $f^{(3)}$ are replaced by the components of the source term in \eqref{eq:LSFLSFmodLem12}. We conclude that $\unkgen_{(0)}=\PreMatchingMapAbstr(\CDfinaltime{\vnkgen}, \CDfinaltime{\unkgen})$, where $\PreMatchingMapAbstr$ is defined in \Propref{prop:LSF3}, and that for any\footnote{Notice that, as it is done in this proof, taking into account that $\kappa$ is positive in the assumption on the source terms leads to a slight improvement of  \Theoremref{thm:maintheorem}. } $\lambda<\min\{\lambda_s, 2(1-q_{max}),2A^2\}$
\begin{align*}  
&\left\|t^{-\lambda}
\left(\unk(t,\cdot)-\PreMatchingMapAbstr(\CDfinaltime{\vnkgen}, \CDfinaltime{\unkgen})\right)
\right\|_{\delta,H^{k+2}(M)}^2
\le
C  \Bigl(
\|\unkgen_{(0)}\|^2_{\delta,H^{k+4+k_0}(M)}+ \|\unkgen_{(1)}\|^2_{\delta,H^{k+4+k_1}(M)}
\\      
  &+\int_0^{\finaltime}\Bigl(\|s^{-\lambda_s} f^{(1)}\|^2_{\delta, H^{k+4}(M)}
    +\|s^{-\lambda_s} f^{(2)}\|^2_{\delta, H^{k+4}(M)}+\|s^{-\lambda_s+2\kappa} f^{(3)}\|^2_{\delta,H^{k+5}(M)}\Bigr) s^{-1}ds\\      
  &+\int_0^{\finaltime}\Bigl(\|s^{-2\kappa} \Lop{\unkgen_{(0)}}\|^2_{\delta, H^{k+4}(M)}
    \Bigr) s^{-1}ds
\Bigr),
\end{align*}
for all $t\in (0,\finaltime]$, where the last term can also be estimated by $\|\unkgen_{(0)}\|^2_{\delta,H^{k+4+k_0}(M)}$. This verifies \Eqref{eq:LSFLSFLOTestimate}.
Using \Eqref{eq:LSFimproveddecayubackward} from \Propref{prop:LSF4},
 the same arguments verify \Eqref{eq:LSFLSFLOTestimate2}.
\Propref{prop:LSFresult1} therefore yields the asymptotic matching map $(\CDfinaltime{\vnkt}, \CDfinaltime{\unkt})
=\AsymptoticMatchingMapAbstr{0\rightarrow \switchgen} (\unkgen_{(0)}, \unkgen_{(1)})$ which is $C^\infty$-continuous.
The solution 
$v_2=(\vnkt,\nunkt,\unkt)$ 
of \Eqref{eq:abstracteqgenNN} with \Eqsref{eq:LSFLSFasldkasdk1}, \eqref{eq:LSFLSFasldkasdkL} and 
    \eqref{eq:LSFLSFmodLem22}
determined by Cauchy data $\unkt(\finaltime,\cdot)=\CDfinaltime{\unkt}$ and $\vnkt(\finaltime,\cdot)=\CDfinaltime{\vnkt}$ satisfies the estimate 
\Eqref{eq:LSFLSFmodLem23132} for all $t\in (0,\finaltime]$ and for any $\lambda$ in the range \eqref{eq:LSFLSFmodLem29348}. 
Replacing the right-hand side in \Eqref{eq:LSFLSFmodLem23132} by a generic constant $C>0$, \Eqref{eq:LSFLSFmodCor23132} follows from \Eqref{eq:LSFLSFmodelsol}. 
Since \Eqref{eq:abstracteqgenNN} with \Eqsref{eq:LSFLSFasldkasdk1}, \eqref{eq:LSFLSFasldkasdkL} and 
    \eqref{eq:LSFLSFmodLem22} agrees with \Eqsref{eq:scalarfieldlinearMatch} -- \eqref{eq:pilinearMatch} (because $\switchgen=\switcht$ and $K=\switcht$),
the map $\Psi$ asserted in \Theoremref{thm:maintheorem} is given by 
\[\Psi=\AsymptoticMatchingMapAbstr{\switch\rightarrow \switcht}.\]
This map must be  injective as \Eqref{eq:LSFLSFmodCor23132} could not be satisfied for the same solution $(\vnkgen,\nunkgen, \unkgen)$  of
    \Eqsref{eq:scalarfieldlinearMatch} -- \eqref{eq:pilinearMatch} given two different choices of $(\unkgen_{(0)},\unkgen_{(1)})$.

Let us now address the surjectivity of $\Psi$. The main idea for this part of the proof is to construct a candidate for the inverse of $\Psi$ by applying \Propref{prop:LSFresult1} now 
(reversing the role of the model equation)
with $\switcht=0$, $\switch=\switchgen$ and $K=\switch$, $\Phi=\mathrm{id}$ and $\chi=\PreMatchingMapAbstr$ given by \Propref{prop:LSF3} for $\switchgen=\switch$. Observe here that since $K=\switch$, the particular choice of $\Phi$ is irrelevant and  we therefore pick the identity for simplicity. This allows us to replace all $(\AD{\vnk}, \AD{\unk})$ in \Propref{prop:LSFresult1} by $(\CDfinaltime{\vnk}, \CDfinaltime{\unk})$.
With the choice of $\chi$ here we write
 \begin{equation}
  \label{eq:LSFLSFazxcskld18273}
\LOT{\unk}=\PreMatchingMapAbstr(\CDfinaltime{\vnk}, \CDfinaltime{\unk}).
\end{equation}
The solution $v_1 = (\vnk,\nunk,\unk)$ is uniquely determined by solving the well-posed Cauchy problem of \Eqref{eq:abstracteqgenNN} with \Eqsref{eq:LSFLSFasldkasdk1}, \eqref{eq:LSFLSFasldkasdkL} and 
    \eqref{eq:LSFLSFmodLem12} with arbitrary smooth Cauchy data $(\CDfinaltime{\vnk}, \CDfinaltime{\unk})$.
Using \Propref{prop:LSF4} and \Propref{prop:LSF3} we verify the hypothesis of \Propref{prop:LSFresult1} as before. 
The $C^\infty$-continuous map $\AsymptoticMatchingMapAbstr{\switch\rightarrow \switcht}=\AsymptoticMatchingMapAbstr{\switchgen\rightarrow 0}$ asserted by \Propref{prop:LSFresult1}  therefore yields smooth $(\CDfinaltime{\vnkt}, \CDfinaltime{\unkt})=\AsymptoticMatchingMapAbstr{\switchgen\rightarrow 0}(\CDfinaltime{\vnk}, \CDfinaltime{\unk})$ such that 
the solution 
$v_2=(\vnkt,\nunkt,\unkt)$  of \Eqref{eq:abstracteqgenNN} with \Eqsref{eq:LSFLSFasldkasdk1}, \eqref{eq:LSFLSFasldkasdkL} and 
    \eqref{eq:LSFLSFmodLem22}
determined by Cauchy data $\unkt(\finaltime,\cdot)=\CDfinaltime{\unkt}$ and $\vnkt(\finaltime,\cdot)=\CDfinaltime{\vnkt}$
satisfies the estimate
\Eqref{eq:LSFLSFmodLem23132} for all $t\in (0,\finaltime]$ and any smooth $\lambda$ in the range \eqref{eq:LSFLSFmodLem29348}.
Now, this solution $v_2$ must be of the form \Eqref{eq:LSFLSFmodelsol} for some smooth $(\unkgen_{(0)}, \unkgen_{(1)})$. \Eqref{eq:LSFLSFmodLem23132} implies that $\unkgen_{(0)}$ equals \Eqref{eq:LSFLSFazxcskld18273}, and, $\unkgen_{(1)}$ is given by \Eqref{EPD:LSFCD2AD2NN} (with $\CDfinaltime{\vnk}$ replaced by $\CDfinaltime{\vnkt}$). It is straightforward to conclude from the information given in the proof of \Propref{prop:LSF3} that $\chi=\PreMatchingMapAbstr$ is $C^\infty$-continuous. We have therefore constructed a $C^\infty$-continuous map from the set of  smooth $(\CDfinaltime{\vnk}, \CDfinaltime{\unk})$ to the set of smooth $(\unkgen_{(0)}, \unkgen_{(1)})$ with the above properties. Now pick any smooth $(\CDfinaltime{\vnk}, \CDfinaltime{\unk})$ and let $(\unkgen_{(0)}, \unkgen_{(1)})$ be determined by this map. Let $(\CDfinaltime{\tilde\vnk}, \CDfinaltime{\tilde\unk})=\Psi (\unkgen_{(0)}, \unkgen_{(1)})$. The uniqueness property of $\Psi$ asserted by \Propref{prop:LSFresult1} implies that $(\CDfinaltime{\tilde\vnk}, \CDfinaltime{\tilde\unk})=(\CDfinaltime{\vnk}, \CDfinaltime{\unk})$. We conclude that $\Psi$ is therefore surjective in addition to injective, and, in fact, that both $\Psi$ and inverse $\Psi^{-1}$ are $C^\infty$-continuous.
\end{proof}

\subsection{Proof of \Propref{prop:LSF2}: Rough a-priori estimates for solutions of the linearized lapse-scalar field system}
\label{sec:LSFproof1}

This subsection is concerned with the \hyperlink{propproof:LSF2}{proof} of \Propref{prop:LSF2}. Let us start by picking an arbitrary $\switchgen\in [0,1]$ and an arbitrary asymptotically point-wise Kasner--scalar field background 
  $\Gamma=(\BG\gamma^{ab}, {\BG\chi_a}^{\hspace{1ex}b}, {\BG\alpha}, {\BG\pi}, \BG\phi)$ with positive decay $\beta$. 
Let $A$ and $L$ be given by \Eqsref{eq:LSFLSFasldkasdk1} and \eqref{eq:LSFLSFasldkasdkL}, and, $\sourcegen$  by \eqref{eq:abstrst}. Interpreting $A$ and $L$ as operators on arbitrary smooth time-dependent $(0,k)$-tensor fields, the fields
\begin{equation}
  \label{eq:defutensor}
u_{a_1\ldots a_k}=\partial_{a_1}\cdots \partial_{a_k} u,
\quad \nu_{a_1\ldots a_k}=\partial_{a_1}\cdots \partial_{a_k} \nu,
\quad \varphi_{a_1\ldots a_k}=\partial_{a_1}\cdots \partial_{a_k} \varphi,
\end{equation}
satisfy 
\begin{equation}
  \label{eq:LSFtensoreq}
  (A+\switchgen L)\left[(u_{a_1\ldots a_k}, \nu_{a_1\ldots a_k}, \varphi_{a_1\ldots a_k})\right]=(\sourcegen^{(1)}_{a_1\ldots a_k}, \sourcegen^{(2)}_{a_1\ldots a_k}, \sourcegen^{(3)}_{a_1\ldots a_k}),
\end{equation}
where 
\begin{equation}
  \label{eq:tensorialf}
  \begin{split}
  (\sourcegen^{(1)}_{a_1\ldots a_k}, \sourcegen^{(2)}_{a_1\ldots a_k}, \sourcegen^{(3)}_{a_1\ldots a_k})\,=&\left(\partial_{a_1}\cdots \partial_{a_k}\sourcegen^{(1)}, \partial_{a_1}\cdots \partial_{a_k}\sourcegen^{(2)}, \partial_{a_1}\cdots \partial_{a_k}\sourcegen^{(3)}\right)\\
&+\left[A+\switchgen L,\partial^k\right]\left[(u, \nu, \varphi)\right],
\end{split}
\end{equation}
with $[\cdot,\cdot]$ being the commutator.

The first step of the proof of \Propref{prop:LSF2} is to estimate the smooth \emph{tensorial} solutions $(u_{a\ldots b}, \nu_{a\ldots b}, \varphi_{a\ldots b})$ of \Eqref{eq:LSFtensoreq} for \emph{any} 
smooth tensorial source term fields $\sourcegen^{(1)}_{a\ldots b}$, $\sourcegen^{(2)}_{a\ldots b}$ and $\sourcegen^{(3)}_{a\ldots b}$; that is, without imposing \Eqref{eq:tensorialf}. The result is \Lemref{lem:estimatesTensorEqs}. In a second step, contained in \Lemref{lem:estimatesTensorST}, the source terms \Eqref{eq:tensorialf} are imposed. The \hyperlink{propproof:LSF2}{proof} of \Propref{prop:LSF2} is then essentially a combination of these two lemmas.
The norms defined in \Eqsref{eq:tensorL2norm1} and \eqref{eq:tensorL2norm2}, as well as energies defined in \Eqsref{eq:defenergies1N} and \eqref{eq:defenergies2} are used throughout this section.

\begin{lemma}[Estimates for smooth solutions of the tensorial equations]
  \label{lem:estimatesTensorEqs}  
  Consider an arbitrary asymptotically point-wise Kasner--scalar field background 
  $\Gamma=(\BG\gamma^{ab}, {\BG\chi_a}^{\hspace{1ex}b}, {\BG\alpha}, {\BG\pi}, \BG\phi)$ with positive decay $\beta$.
Let $(u,\nu,\varphi)$ be any smooth $(0,r)$-tensorial solution of \Eqref{eq:LSFtensoreq} with $A$ and $L$ given by \Eqsref{eq:LSFLSFasldkasdk1} and \eqref{eq:LSFLSFasldkasdkL}
for arbitrary smooth time-dependent $(0,r)$-tensor fields $\sourcegen^{(1)}(t,x)$, $\sourcegen^{(2)}(t,x)$ and $\sourcegen^{(3)}(t,x)$ on $M$ and $\switchgen\in I=[0,1]$.
Pick any sufficiently small $\epsilon>0$. Then there is a constant $C>0$ such that for any $t_*\in (0,\finaltime]$ the following estimates hold:
\begin{enumerate}
\item For all $t\in[t_*,\finaltime]$ and for any smooth function $\lambda>\lambda_c$ (see \Eqref{eq:deflambdac}), we have
\begin{equation}
  \label{eq:LSFforwardslittle}
  \begin{split}
    \littleenergyswitch{u,\varphi}\switchgen{\lambda}(t)
    &\le  C\Bigl(\littleenergyswitch{u,\varphi}\switchgen{\lambda}(t_*)\\
      &\!\!\!\!\!\!\!\!
      +\int^t_{t_*}\Bigl(\|s^{-\lambda} \sourcegen^{(1)}\|^2_\delta +\|s^{-\lambda} \sourcegen^{(2)}\|^2_\delta +\switchgen^2\|s^{-\lambda} \partial \sourcegen^{(3)}\|^2_{\delta,t^2\KSF\gamma}\Bigr) s^{-1+2\epsilon} ds\Bigr).
  \end{split}
\end{equation}
\item For all $t\in[t_*,\finaltime]$ and for any smooth function $\lambda\ge 0$ on $M$, we have
\begin{equation}
\label{eq:LSFforwardlittle2}
\begin{split}
  \sqrt{\littleenergywk{\varphi}{k}{\lambda}(t)}
\le &\sqrt{\littleenergywk{\varphi}{k}{\lambda}(t_*)} \\
&+C
\int_{t_*}^t\Bigl(\sqrt{\littleenergy{u}{\lambda}(s)}+\|s^{-\lambda} \sourcegen^{(2)}\|_{\delta}+\|s^{-\lambda} \sourcegen^{(3)}\|_{\delta}      
  \Bigr) s^{-1} ds.
\end{split}
\end{equation}
\item For all $t\in (0,t_*]$ and for any smooth function $\lambda<0$ on $M$, we have
\begin{equation}
  \label{eq:LSFbackwardslittle}
  \begin{split}
    \littleenergyswitch{u,\varphi}\switchgen{\lambda}(t)
    &\le  C\Bigl(\littleenergyswitch{u,\varphi}\switchgen{\lambda}(t_*)\\
      &\!\!\!\!\!\!\!\!\!\!
      +\int_t^{t_*}\Bigl(\|s^{-\lambda} \sourcegen^{(1)}\|^2_\delta +\|s^{-\lambda} \sourcegen^{(2)}\|^2_\delta +\switchgen^2\|s^{-\lambda} \partial \sourcegen^{(3)}\|^2_{\delta,t^2\KSF\gamma}\Bigr) s^{-1-2\epsilon} ds\Bigr).
  \end{split}
\end{equation}
\item For all $t\in(0,t_*]$ and for any smooth function $\lambda\le 0$ on $M$, we have
\begin{equation}
\label{eq:LSFbackwardlittle2}
\begin{split}
  \sqrt{\littleenergywk{\varphi}{k}{\lambda}(t)}
\le &\sqrt{\littleenergywk{\varphi}{k}{\lambda}(t_*)} \\
&+C
\int^{t_*}_t\Bigl(\sqrt{\littleenergy{u}{\lambda}(s)}+\|s^{-\lambda} \sourcegen^{(2)}\|_{\delta}+\|s^{-\lambda} \sourcegen^{(3)}\|_{\delta}      
  \Bigr) s^{-1} ds.
\end{split}
\end{equation}
\end{enumerate}
The constants $C$ may depend on $\finaltime$, $\lambda$, $\epsilon$ and $\Gamma$.
Moreover, for all $t\in (0,\finaltime]$ and for any smooth function $\lambda$ on $M$, we have
  \begin{equation}
  \label{eq:preestimat3}
\sqrt{\littleenergyswitch{\nu}\switchgen{\lambda}(t)}
  \le C \left(\sqrt{\littleenergywk{u}k{\lambda}(t)} +\|t^{-\lambda}\sourcegen^{(2)}\|_\delta\right),
\end{equation}
where the constant $C>0$ may depend on $\finaltime$ and $\Gamma$.
\end{lemma}

The \hyperlink{lemproof:estimatesTensorEqs}{proof} of this lemma is discussed towards the end of this subsection.
We remark that large parts of this proof consist of standard wave-type energy arguments using integration by parts. However, the presence of the linearized lapse $\nu$, which satisfies an elliptic equation, leads to a few peculiarities which we point out in the proof. We also remark that in contrast to the proof of Proposition~5.2 in \cite{rodnianski2018} (which restricts to estimates when the Cauchy data is imposed at the \emph{final time} $\finaltime$), we find it useful here to write the estimates in terms of the variable $u$, instead for $\partial_t\varphi$, and to keep the estimates for $u$ and $\varphi$ separate from those for $\nu$. This is crucial in order to avoid all terms involving time-derivatives of $\nu$.
We also remark that the function $\lambda(x)$ acts as the exponent of  a  \emph{time-weight}. The time weights are crucial in distinguishing the two different kinds of Cauchy problems later on: (1), where the Cauchy condition is imposed at the final time, and, (2), where the Cauchy condition is imposed at the matching time. 

The next step is to estimate the source terms. Recall footnote~\ref{ftnote:tensornotation} for our notations for tensors

\begin{lemma}[Estimates for the tensor source terms given by \Eqref{eq:tensorialf}]
  \label{lem:estimatesTensorST}  
  Consider an arbitrary asymptotically point-wise Kasner--scalar field background 
  $\Gamma=(\BG\gamma^{ab}, {\BG\chi_a}^{\hspace{1ex}b}, {\BG\alpha}, {\BG\pi}, \BG\phi)$ with positive decay $\beta$.
Pick any smooth (scalar) solution $v=(u, \nu, \varphi)$ of \Eqref{eq:abstracteqgenNN} with $A$ and $L$ given by \Eqsref{eq:LSFLSFasldkasdk1} and \eqref{eq:LSFLSFasldkasdkL}
for arbitrary smooth scalar fields $\sourcegen^{(1)}(t,x)$, $\sourcegen^{(2)}(t,x)$ and $\sourcegen^{(3)}(t,x)$ on $M$ and $\switchgen\in I=[0,1]$. Let the $(0,k)$-tensor fields
$u_{[k]}$, $\varphi_{[k]}$, $\nu_{[k]}$ be defined by \Eqref{eq:defutensor} for arbitrary integers $k\ge 1$. 
Then there is a constant $C>0$, which may depend on $k$, $\finaltime$ and $\Gamma$, such that the following estimates hold for the fields defined in \Eqref{eq:tensorialf} for any smooth $\lambda(x)$:
  \begin{gather}
    \label{eq:LSFf1}
    \begin{split}
  \sum_{l=0}^k\left\|t^{-\lambda} \sourcegen^{(1)}_{[l]}\right\|^2_{\delta}
  \le  C\Bigl(\|t^{-\lambda} \sourcegen^{(1)}\|^2_{\delta, H^k(M)}
    +\|t^{-\lambda} \sourcegen^{(2)}\|^2_{\delta, H^{k-1}(M)}\\
    +\energy{u,\varphi}{k}{\lambda-2\kappa}(t)
  +\energy{u,\varphi}{k-1}{\lambda}(t) 
  \Bigr),
\end{split}\\
  \label{eq:est3121323}
  \sum_{l=0}^k\left\|t^{-\lambda} \sourcegen^{(2)}_{[l]}\right\|^2_{\delta}
  \le C\Bigl(\|t^{-\lambda} \sourcegen^{(2)}\|^2_{\delta, H^k(M)}
  +\energy{u}{k-1}{\lambda}(t) 
+\energy{u}{k}{\lambda-2\kappa}(t)
  \Bigr),\\
  \label{eq:LSFf3}
  \sum_{l=0}^k\left\|t^{-\lambda} \sourcegen^{(3)}_{[l]}\right\|^2_{\delta}
  \le C\left(\|t^{-\lambda} \sourcegen^{(2)}\|^2_{\delta, H^{k-1}(M)}+\|t^{-\lambda} \sourcegen^{(3)}\|^2_{\delta, H^{k}(M)}
  +\energy{u}{k-1}{\lambda} (t)\right),\\
  \label{eq:LSFf3del}
  \begin{split}
  \sum_{l=0}^k\left\|t^{-\lambda} \partial \sourcegen^{(3)}_{[l]}\right\|^2_{\delta,t^2\BG\gamma}
  \le C\Bigl(\|t^{-\lambda} \sourcegen^{(3)}\|^2_{\delta, H^{k+1}(M)}
+\|t^{-\lambda} \sourcegen^{(2)}\|^2_{\delta, H^{k-1}(M)}\\
  +\energy{u,\varphi}{k-1}{\lambda}(t) + \energy{u,\varphi}{k}{\lambda-\nu-2\kappa}(t)\Bigr).
\end{split}
\end{gather}
The function $\kappa$ here is defined in \Propref{prop:LSF4}.
\end{lemma}
Before we provide the \hyperlink{lemproof:estimatesTensorST}{proof} of this lemma, we first prove the main result of this subsection, \Propref{prop:LSF2}.

\begin{proof}[Proof of \Propref{prop:LSF2}]
  \hypertarget{propproof:LSF2}{}
Pick any smooth (scalar) solution $v=(u, \nu, \varphi)$ of \Eqref{eq:abstracteqgenNN} with $A$ and $L$ given by \Eqsref{eq:LSFLSFasldkasdk1} and \eqref{eq:LSFLSFasldkasdkL}
for arbitrary smooth scalar fields $\sourcegen^{(1)}(t,x)$, $\sourcegen^{(2)}(t,x)$ and $\sourcegen^{(3)}(t,x)$ on $M$ and $\switchgen\in I=[0,1]$. Let the $(0,k)$-tensor fields
$u_{[k]}$, $\varphi_{[k]}$, $\nu_{[k]}$ be defined by \Eqref{eq:defutensor} for arbitrary integers $k\ge 1$, and $\sourcegen^{(1)}_{[k]}$, $\sourcegen^{(2)}_{[k]}$, $\sourcegen^{(3)}_{[k]}$ by \Eqref{eq:tensorialf}.

Consider first the case  $\lambda>\lambda_c$ and $k\ge 1$. Then we combine the sum of \Eqref{eq:LSFforwardslittle} from $r=0$ to $k$ with \Eqsref{eq:LSFf1}, \eqref{eq:est3121323} and \eqref{eq:LSFf3del} as follows:
\begin{equation*}
  \begin{split}
    &\energyswitch{u,\varphi}\switchgen{k}{\lambda}(t)
    \le  C\Bigl(\energyswitch{u,\varphi}\switchgen{k}{\lambda} (t_*)
    +\int^t_{t_*}\energy{u,\varphi}{k}{\lambda}(s) s^{-1+2\epsilon} ds\\
      &\quad+\int^t_{t_*}\bigl(\|s^{-\lambda} \sourcegen^{(1)}\|^2_{\delta,H^{k}(M)} 
      +\|s^{-\lambda} \sourcegen^{(2)}\|^2_{\delta,H^{k}(M)} 
      +\|s^{-\lambda} \sourcegen^{(3)}\|^2_{\delta,H^{k+1}(M)}\bigr) s^{-1+2\epsilon} ds
      \Bigr)
  \end{split}
\end{equation*}
for all $t\in [t_*,\finaltime]$ where we exploit that the function $\kappa$ is strictly positive. Notice that we get the same estimate for $k=0$ without the second term on the right-hand side and \Eqref{eq:LSFforwards1} follows directly.
If $k\ge 1$, \Eqref{eq:LSFforwards1} follows from Gr\"onwall's lemma. The resulting constant $C>0$ may depend on $k$ in addition to the quantities listed in \Lemref{lem:estimatesTensorEqs}.

Suppose next that $\lambda<0$. If $k=0$, \Eqref{eq:LSFbackward1} follows directly from \Eqref{eq:LSFbackwardslittle}. For any $k\ge 1$, a similar procedure as above yields
\begin{equation*}
  \begin{split}
    &\energyswitch{u,\varphi}\switchgen{k}{\lambda}(t)
    \le  C\Bigl(\energyswitch{u,\varphi}\switchgen{k}{\lambda} (t_*)
    +\int_t^{t_*}\energy{u,\varphi}{k-1}{\lambda}(s) s^{-1-2\epsilon} ds
+\int_t^{t_*} \energy{u,\varphi}{k}{\lambda-2\kappa}(s) s^{-1-2\epsilon} ds\\
      &\qquad+\int_t^{t_*}\bigl(\|s^{-\lambda} \sourcegen^{(1)}\|^2_{\delta,H^{k}(M)} 
      +\|s^{-\lambda} \sourcegen^{(2)}\|^2_{\delta,H^{k}(M)} 
      +\|s^{-\lambda} \sourcegen^{(3)}\|^2_{\delta,H^{k+1}(M)}\bigr) s^{-1-2\epsilon} ds      
      \Bigr)
  \end{split}
\end{equation*}
for all $t\in (0,t_*]$. 
Supposing that $\epsilon>0$ is sufficiently small there is a constant $\eta>0$ (recall that $\kappa>0$) such that Gr\"onwall's lemma implies 
\begin{equation*}
  \begin{split}
    &\energyswitch{u,\varphi}\switchgen{k}{\lambda}(t)    
      \le  C\Bigl(\energyswitch{u,\varphi}\switchgen{k}{\lambda} (t_*)
      +\frac 1{2\epsilon}\left(t_*^{2\epsilon}-t^{2\epsilon}\right)
      \sup_{s\in[t,t_*]}\energy{u,\varphi}{k-1}{\lambda+2\epsilon}(s)\\
      &\qquad+\int_t^{t_*}\Bigl(\|s^{-\lambda} \sourcegen^{(1)}\|^2_{\delta,H^{k}(M)} 
      +\|s^{-\lambda} \sourcegen^{(2)}\|^2_{\delta,H^{k}(M)} 
      +\|s^{-\lambda} \sourcegen^{(3)}\|^2_{\delta,H^{k+1}(M)}\Bigr) s^{-1-2\epsilon} ds
            \Bigr),
  \end{split}
\end{equation*}
for a constant $C>0$ which may depend on $\eta$ in addition.
If $k=1$ we can estimate the second term on the right-hand side by the $k=0$-estimate above which yields \Eqref{eq:LSFbackward1}.
For any $k\ge 2$, the same estimate holds with $k$ replaced by $k-1$ and $\lambda$ by $\lambda+2\epsilon$ (provided $\epsilon$ is sufficiently small). This can be used to estimate the second term on the right-hand above. The same argument applied repeatedly yields for any $k\ge 0$
\begin{equation*}
  \begin{split}
    &\energyswitch{u,\varphi}\switchgen{k}{\lambda}(t)    
      \le  C\Bigl(\energyswitch{u,\varphi}\switchgen{k}{\lambda+2k\epsilon} (t_*)\\
      &\quad+\int_t^{t_*}\Bigl(\|s^{-\lambda} \sourcegen^{(1)}\|^2_{\delta,H^{k}(M)} 
      +\|s^{-\lambda} \sourcegen^{(2)}\|^2_{\delta,H^{k}(M)} 
      +\|s^{-\lambda} \sourcegen^{(3)}\|^2_{\delta,H^{k+1}(M)}\Bigr) s^{-1-2(2k+1)\epsilon} ds
      \Bigr).
  \end{split}
\end{equation*}
For any fixed value of $k$ and $\epsilon$ we may therefore write $\epsilon$ instead of $(2k+1)\epsilon$ which establishes \Eqref{eq:LSFbackward1}. The constants $C>0$ here may depend on $\finaltime$, $k$, $\lambda$, $\epsilon$ and $\Gamma$.

Applying \Eqsref{eq:est3121323} and \eqref{eq:LSFf3} to \Eqref{eq:LSFforwardlittle2} for any $\lambda\ge 0$ yields \Eqref{eq:LSFforward2}. Applying \Eqsref{eq:est3121323} and \eqref{eq:LSFf3} to \Eqref{eq:LSFbackwardlittle2} for any $\lambda\le 0$ yields \Eqref{eq:LSFbackward2}. Finally, \Eqref{eq:preestimat3N} is established as part of the proof of \Lemref{lem:estimatesTensorST} in \Eqref{eq:preestimat3NN}.
\end{proof}

Before we discuss the proofs of the two lemmas of this subsection, let us introduce some more notation. Let
\begin{align}
\begin{split}
  \label{eq:coeff1}
  L^{[1,1]}&:=\switchgen(1-\BG\alpha), 
  \quad L^{[1,2]}:=-A-\switchgen(\BG\pi-A)+\switchgen t^2\BG\gamma^{ab}D_a D_b \BG\phi,  \\
  L^{[1,3],b}&:=\switchgen t^2\BG\gamma^{ab}D_a \BG\alpha, 
  \quad  L^{[1,4],b}:=\switchgen t^2\BG\gamma^{ab}D_a \BG\phi,
  \end{split} \\
\begin{split}  
 \label{eq:coeff2}
L^{[2,1]}&:=-1-\switchgen\bigl(\bigl(t^2{\BG\chi_a}^{\hspace{1ex}b}{\BG\chi_b}^{\hspace{1ex}a}+\frac {1}{n}+A^2-1\bigr)+(\BG\pi^2-A^2)\bigr), \\
  L^{[2,2]}&:=-2A -2\switchgen(\BG\pi \BG\alpha-A), 
\end{split} \\
\label{eq:coefflast}
  L^{[3,1]}&:=A +\switchgen(\BG\pi-A),  \quad
  L^{[3,2]}:=1 +\switchgen(\BG\alpha-1), 
\end{align}
and further decompose the above operators as 
\begin{equation*}
  L^{[i,j]} =: L^{[i,j]}_0 + \switchgen L^{[i,j]}_1.
\end{equation*}
For some of our discussion it is also useful to express the covariant derivative $D_a$ via the notation $\partial_a$; we also use the tensor field notation (recall the index conventions introduced in \Sectionref{sec:mainresults} and footnote~\ref{footnote:inversemetric})
\begin{equation}
  \label{eq:Ca}
\begin{split}
  {C^a}_{cb}&=\frac 12\BG\gamma^{ad}\left(\partial_c\BG\gamma^{-1}_{bd}
+\partial_b\BG\gamma^{-1}_{dc}
-\partial_d\BG\gamma^{-1}_{bc}\right),\\
  C^a&=\BG\gamma^{cb}{C^a}_{cb}=-\partial_c\BG\gamma^{ac}+\frac 12\BG\gamma^{ad}\BG\gamma^{-1}_{bc}\partial_d\BG\gamma^{bc}.
\end{split}
\end{equation}
Given this, it turns out to be convenient to write \Eqref{eq:LSFtensoreq} with $A$ and $L$ given by \Eqsref{eq:LSFLSFasldkasdk1} and \eqref{eq:LSFLSFasldkasdkL} in the form (we note that certain terms contain $\switchgen$ explicitly, while other terms contain $\switchgen$ as part of the definition of the particular $L^{[i,j]}$)\footnote{Observe carefully that we are using the index-free notation for tensorial quantities here. These equations therefore cover, but do not restrict to, scalar unknowns as a particular case.}
\begin{gather}
  \label{eq:scalarfieldequationgentens}
  \begin{split}
    -t\partial_t u    
+\switchgen{\BG\alpha} t^2\BG\gamma^{ab}\partial_a \partial_b\varphi
-\switchgen \left(t^2{\BG\alpha} C^a-L^{[1,3],a}_1\right)\partial_a\varphi\\
   +L^{[1,1]} u 
  +L^{[1,2]}\nu 
  +\switchgen L_1^{[1,4],b}\partial_b\nu=\sourcegen^{(1)},
\end{split}
\\
\label{eq:lapseequationgentens}
  \switchgen t^2\BG\gamma^{ab}\partial_a \partial_b\nu
  -\switchgen t^2C^a \partial_a\nu
  +L^{[2,1]} \nu
  +L^{[2,2]} u=\sourcegen^{(2)},\\
\label{eq:defueqtens}
-t\partial_t\varphi
  +L^{[3,1]} \nu
  +L^{[3,2]} u=\sourcegen^{(3)}.
  \end{gather}

\begin{proof}[Proof of \Lemref{lem:estimatesTensorEqs}]
  \hypertarget{lemproof:estimatesTensorEqs}{} 
Consider any smooth time-dependent $(0,r)$-tensorial solution $(u,\nu,\varphi)$ of \Eqsref{eq:scalarfieldequationgentens} -- \eqref{eq:defueqtens}. We start by
contracting \Eqref{eq:scalarfieldequationgentens} with $u^{a\ldots b}$,  extracting a total $\partial$-derivative term (``integration by parts'') and expressing the resulting factor $\partial_c  u^{a\ldots b}$ by the derivative of \Eqref{eq:defueqtens}. Then we multiply the result by $t^{-2\lambda}$ for a so far arbitrary smooth function $\lambda(x)$. After further standard but lengthy manipulations, we obtain
\begin{equation}
  \label{eq:evolutionEQidentity}
\begin{split}
    0=&-\frac 12t\partial_t\left(\left|t^{-\lambda} u\right|^2_\delta+\switchgen\frac{{\BG\alpha}t^2 }{L^{[3,2]}}\BG\gamma^{cd}\left(t^{-\lambda}\partial_c\varphi, t^{-\lambda}\partial_d\varphi\right)_\delta\right)\\
&+\left(L^{[1,1]}-\lambda\right)\left|t^{-\lambda} u\right|_\delta^2+L^{[1,2]}\left(t^{-\lambda} u, t^{-\lambda}\nu\right)_\delta\\
&+\switchgen\Bigl(-t^2\partial_c{\BG\alpha}\BG\gamma^{cd} 
-{\BG\alpha}t^2\left(\partial_c \BG\gamma^{cd}\right) 
+ \frac{{\BG\alpha}t^2}{L^{[3,2]}}\partial_c L^{[3,2]}\BG\gamma^{cd}\\
&\qquad\,- {\BG\alpha} t^2 C^d+L^{[1,3],d}_1+2 {\BG\alpha}t^2 \log t\partial_c\lambda \BG\gamma^{cd}\Bigr)\left(t^{-\lambda} u, t^{-\lambda}\partial_d\varphi\right)_\delta \\
&
+\switchgen\left(t\partial_t \left(\frac{{\BG\alpha}t^2 }{2L^{[3,2]}}\right) \BG\gamma^{cd}
+\frac{{\BG\alpha}t^2 }{2L^{[3,2]}}\left(t\partial_t\BG\gamma^{cd}\right)
-\lambda\frac{{\BG\alpha}t^2 }{L^{[3,2]}}\BG\gamma^{cd}
\right)\left(t^{-\lambda}\partial_c\varphi, t^{-\lambda}\partial_d\varphi\right)_\delta\\
&+\switchgen\frac{{\BG\alpha}t^2}{L^{[3,2]}}\partial_c  L^{[3,1]}\BG\gamma^{cd}\left(t^{-\lambda}\nu, t^{-\lambda}\partial_d\varphi\right)_\delta
+\switchgen\frac{{\BG\alpha}t^2}{L^{[3,2]}}L^{[3,1]}\BG\gamma^{cd}\left(t^{-\lambda}\partial_c\nu, t^{-\lambda}\partial_d\varphi\right)_\delta
\\
&+\switchgen L_1^{[1,4],c}\left(t^{-\lambda} u, t^{-\lambda}\partial_c\nu\right)_\delta
-\left(t^{-\lambda} u, t^{-\lambda} \sourcegen^{(1)}\right)_\delta -\switchgen\frac{{\BG\alpha}t^2}{L^{[3,2]}}\BG\gamma^{cd}\left(t^{-\lambda}\partial_c \sourcegen^{(3)}, t^{-\lambda}\partial_d\varphi\right)_\delta\\
&+\partial_c\left(\switchgen{\BG\alpha}t^2\BG\gamma^{cd}\left(t^{-\lambda} u, t^{-\lambda}\partial_d\varphi\right)_\delta\right).
\end{split}
  \end{equation}
This expression has intentionally not been integrated in space yet.
Similarly, contracting \Eqref{eq:lapseequationgentens} with $\nu^{a\ldots b}$ yields:
\begin{equation}
  \label{eq:ellipticEQidentity}
  \begin{split}
    0=&L^{[2,1]} |t^{-\lambda}\nu|^2_\delta
-\switchgen t^2\left(\partial_c \BG\gamma^{cd}+C^d-2\log t\partial_c\lambda \BG\gamma^{cd}\right) \left(t^{-\lambda}\nu, t^{-\lambda}\partial_d\nu\right)_\delta\\
&-\switchgen t^2 \BG\gamma^{cd}\left(t^{-\lambda}\partial_c\nu, t^{-\lambda}\partial_d\nu\right)_\delta
  +L^{[2,2]} \left(t^{-\lambda} u, t^{-\lambda}\nu\right)_\delta-\left(t^{-\lambda}\nu, t^{-\lambda} \sourcegen^{(2)}\right)_\delta\\
&+\partial_c\left(\switchgen t^2\BG\gamma^{cd}\left(t^{-\lambda}\nu, t^{-\lambda}\partial_d\nu\right)_\delta\right).
\end{split}
  \end{equation}

The next step is now to incorporate the asymptotics of the coefficients in \Eqsref{eq:scalarfieldequationgentens} -- \eqref{eq:defueqtens}. It follows from the hypothesis (recall \Defref{def:APKSF}) that
\begin{align}
  \label{eq:coeff1asympt}
  L^{[1,1]}&=\switchgen O(t^\beta), &
  L^{[1,2]}&=-A+\switchgen\bigl( O(t^\beta)+ O(t^{2\kappa})\bigr),\\
  L^{[1,3],b}_1&=t^2\BG\gamma^{ab}\partial_a \BG\alpha, &
  L^{[1,4],b}_1&=t^2\BG\gamma^{ab}\partial_a \BG\phi,\\
  L^{[2,1]}&=-1-\switchgen O(t^\beta),&
  L^{[2,2]}&=-2A -\switchgen O(t^\beta),\\
\label{eq:coefflastasympt}
  L^{[3,1]}&=A +\switchgen O(t^\beta),&  
  L^{[3,2]}&=1+\switchgen(\BG\alpha-1),
\end{align}
which we write in this way to be able to cancel certain terms in the following calculation.
Recall that $\kappa$ is defined in \Propref{prop:LSF4}.
A straightforward calculation involving \Eqsref{eq:gammaass2} and \eqref{eq:LSFasldkj} shows that
\begin{align*}
  -t^2\partial_c{\BG\alpha}\BG\gamma^{cd} 
-{\BG\alpha}t^2\left(\partial_c \BG\gamma^{cd}\right) 
+ \frac{{\BG\alpha}t^2}{L^{[3,2]}}\partial_c L^{[3,2]}\BG\gamma^{cd}- {\BG\alpha} t^2 C^d+L^{[1,3],d}_1+2 {\BG\alpha}t^2 \log t\partial_c\lambda \BG\gamma^{cd}\\
=\left(-\frac 12\KSF\gamma^{-1}_{bc}\partial_e\KSF\gamma^{bc}+2 \log t\partial_e\lambda + O(t^{\beta-\epsilon})\right) t^2\BG\gamma^{ed},
\end{align*}
which follows from \Eqsref{eq:Ca}, \eqref{eq:coeff1} -- \eqref{eq:coefflast} and from a calculation of the form
\begin{align*}
  \BG\gamma^{-1}_{bc}\partial_e\BG\gamma^{bc}
  &=\BG\gamma^{-1}_{bc'}\KSF\gamma^{c'd}\KSF\gamma^{-1}_{dc}\partial_e\BG\gamma^{bc}
=\BG\gamma^{-1}_{bc'}\KSF\gamma^{c'd}\KSF\gamma^{-1}_{dc}\partial_e\KSF\gamma^{bc}
+\switchgen \BG\gamma^{-1}_{bc'}\KSF\gamma^{c'd}\KSF\gamma^{-1}_{dc}\partial_eh^{bc}\\
&=\BG\gamma^{-1}_{bc'}\KSF\gamma^{c'd}\KSF\gamma^{-1}_{dc}\partial_e\KSF\gamma^{bc}
+\switchgen \BG\gamma^{-1}_{bc'}\KSF\gamma^{c'd}\partial_e\left(\KSF\gamma^{-1}_{dc}h^{bc}\right)
+\switchgen \BG\gamma^{-1}_{bc'}\partial_e\left(\KSF\gamma^{c'd}\right)\KSF\gamma^{-1}_{dc}h^{bc}
\\
&=\KSF\gamma^{-1}_{bc}\partial_e\KSF\gamma^{bc}+\switchgen O(t^{\beta-\epsilon}),
\end{align*}
where $\epsilon>0$ is any constant.
A similar calculation involving \Eqref{eq:KSBG4} yields that
\begin{align*}
  t\partial_t \left(\frac{{\BG\alpha}t^2 }{2L^{[3,2]}}\right) \BG\gamma^{cd}
+\frac{{\BG\alpha}t^2 }{2L^{[3,2]}}\left(t\partial_t\BG\gamma^{cd}\right)
-\lambda\frac{{\BG\alpha}t^2 }{L^{[3,2]}}\BG\gamma^{cd}\\
=\left(
t {\KSF\chi_a}^{c}+\left(1-\frac 1n-\lambda\right) {\delta_a}^{c} +O(t^\beta)
\right) \left({\delta_f}^{a} +O(t^\beta)\right) t^2\BG\gamma^{fd}.
\end{align*}
Incorporating these results, \Eqref{eq:evolutionEQidentity} becomes\footnote{We use the notation $\gamma^{cd}\left(\nu, \partial_d\nu\right)_\delta=\gamma^{cd}\nu^{a\ldots b} \partial_d\nu_{a\ldots b}$.}
\begin{equation}
   \label{eq:preenergyid11}
   \begin{split}
    0=&
-\frac 12t\partial_t\left(
\left|t^{-\lambda} u\right|^2_\delta
+\switchgen(1+ O(t^\beta))t^2\BG\gamma^{cd}\left(t^{-\lambda}\partial_c\varphi, t^{-\lambda}\partial_d\varphi\right)_\delta\right)\\
&-\left(\lambda+O(t^\beta)\right)\left|t^{-\lambda} u\right|_\delta^2
-\left(A+ O(t^\beta)+ O(t^{2\kappa})\right)\left(t^{-\lambda} u, t^{-\lambda}\nu\right)_\delta\\
&+\switchgen \left(-\frac 12\KSF\gamma^{-1}_{bc}\partial_e\KSF\gamma^{bc}+2 \log t\partial_e\lambda + O(t^{\beta-\epsilon})\right) t^2\BG\gamma^{ed}
\left(t^{-\lambda} u, t^{-\lambda}\partial_d\varphi\right)_\delta \\
&
+\switchgen \left(
t {\KSF\chi_f}^{c}+\left(1-\frac 1n-\lambda\right) {\delta_f}^{c} +O(t^\beta)
\right) t^2\BG\gamma^{fd}
\left(t^{-\lambda}\partial_c\varphi, t^{-\lambda}\partial_d\varphi\right)_\delta\\
&+\switchgen\left(\partial_c A+O(t^\beta)\right)t^2\BG\gamma^{cd}\left(t^{-\lambda}\nu, t^{-\lambda}\partial_d\varphi\right)_\delta\\
&+\switchgen \left(A+O(t^\beta)\right) t^2\BG\gamma^{cd}\left(t^{-\lambda}\partial_c\nu, t^{-\lambda}\partial_d\varphi\right)_\delta
+\switchgen \partial_a \BG\phi  t^2\BG\gamma^{ac}\left(t^{-\lambda} u, t^{-\lambda}\partial_c\nu\right)_\delta\\
&-\left(t^{-\lambda} u, t^{-\lambda} \sourcegen^{(1)}\right)_\delta -\switchgen (1+O(t^\beta)) t^2\BG\gamma^{cd}\left(t^{-\lambda}\partial_c \sourcegen^{(3)}, t^{-\lambda}\partial_d\varphi\right)_\delta\\
&+\partial_c\left(\switchgen{\BG\alpha}t^2\BG\gamma^{cd}\left(t^{-\lambda} u, t^{-\lambda}\partial_d\varphi\right)_\delta\right).
\end{split}
\end{equation}
For \Eqref{eq:ellipticEQidentity} we find
\begin{equation}
  \label{eq:preenergyid21}
  \begin{split}
    0=&-\left(1+O(t^\beta)\right) |t^{-\lambda}\nu|^2_\delta\\
&-\switchgen \left(\frac 12\KSF\gamma^{-1}_{bc}\partial_a\KSF\gamma^{bc}
-2\log t\partial_a\lambda+O(t^{\beta-\epsilon})\right) t^2\BG\gamma^{ad} \left(t^{-\lambda}\nu, t^{-\lambda}\partial_d\nu\right)_\delta\\
&-\switchgen t^2\BG\gamma^{cd}\left(t^{-\lambda}\partial_c\nu, t^{-\lambda}\partial_d\nu\right)_\delta
  -2(A+O(t^\beta))\left(t^{-\lambda} u, t^{-\lambda}\nu\right)_\delta\\
&-\left(t^{-\lambda}\nu, t^{-\lambda} \sourcegen^{(2)}\right)_\delta
+\partial_c\left(\switchgen t^2\BG\gamma^{cd}\left(t^{-\lambda}\nu, t^{-\lambda}\partial_d\nu\right)_\delta\right).
\end{split}
\end{equation}
Defining 
\[\tilde\littleenergysymbol
=\frac 12\int_M \left((1+O(t^\beta))\left|t^{-\lambda} \nu\right|^2_\delta 
      +\switchgen t^2\BG\gamma^{cd}\left(t^{-\lambda} \partial_c\nu,t^{-\lambda}  \partial_d\nu\right)_{\delta}\right)dx,\]
\Eqref{eq:preenergyid21} yields for all $t\in (0,\finaltime]$ and for any smooth $\lambda(x)$:
  \begin{equation*}
    \tilde\littleenergysymbol(t)\le C\left(
    \sqrt{\littleenergywk{u}k{\lambda}(t)} \sqrt{\littleenergyswitch{\nu}\switchgen{\lambda} (t)}
    +\sqrt{\littleenergyswitch{\nu}\switchgen{\lambda}(t)}\|t^{-\lambda}\sourcegen^{(2)}\|_\delta\right),
  \end{equation*}
where $C$ may depend on $\finaltime$ and $\Gamma$. 
Comparing the definition of $\tilde\littleenergysymbol$ and \Eqref{eq:defenergies2}
implies \Eqref{eq:preestimat3} for some constant $C>0$ with the same dependencies.

Setting
\begin{equation}
   U=\begin{pmatrix}
     t^{-\lambda} u,
     &t^{-\lambda}\nu, 
     &
     \sqrt{\switchgen}t^{-\lambda} t\partial_c\varphi,
     &\sqrt{\switchgen}t^{-\lambda} t\partial_c\nu
   \end{pmatrix}^T,
\end{equation}
where $(\cdot)^T$ denotes the transpose, we can cast \Eqref{eq:preenergyid11} into the form
\begin{equation}
  \label{eq:preenergyid12}
  \begin{split}
    &\frac 12t\partial_t\left(\left|t^{-\lambda} u\right|^2_\delta 
      +\switchgen \left(1+ O(t^\beta)\right) t^2\BG\gamma^{cd}\left(t^{-\lambda} \partial_c\varphi,t^{-\lambda}  \partial_d\varphi\right)_{\delta}\right)\\
    =&U^T\cdot\hat {\boldsymbol M}_1\cdot U\\
    &-\left(t^{-\lambda} u, t^{-\lambda} \sourcegen^{(1)}\right)_\delta 
    -\switchgen (1+O(t^\beta))t^2\BG\gamma^{ad}\left(t^{-\lambda}\partial_c \sourcegen^{(3)}, t^{-\lambda}\partial_d\varphi\right)_\delta
    +\partial_c\left(\ldots\right),
  \end{split}
\end{equation}
with
\begin{equation*}
  \hat {\boldsymbol M}_1=
  \begin{pmatrix}
    \hat {\boldsymbol M}_{1,11} & \hat {\boldsymbol M}_{1,12}\\
    \hat {\boldsymbol M}_{1,21} & \hat {\boldsymbol M}_{1,22}
  \end{pmatrix}
\end{equation*}
and
\begin{align*}
  \hat {\boldsymbol M}_{1,11}&=
  \begin{pmatrix}
    -\lambda & -A/2\\
    -A/2 & 0
  \end{pmatrix}+O(t^\beta) +O(t^{2\kappa}),\\
  \hat {\boldsymbol M}_{1,12}&=\hat {\boldsymbol M}_{1,21}^T=
  \begin{pmatrix}
    -\sqrt\switchgen\frac 12\left(\frac 12\KSF\gamma^{-1}_{bc}\partial_a\KSF\gamma^{bc}
-2\log t\partial_a\lambda+ O(t^{\beta-\epsilon})\right) t\BG\gamma^{ad} & \frac 12\sqrt{\switchgen}\partial_a\BG\phi t\BG\gamma^{ad}\\
    \frac 12 \sqrt{\switchgen}\left(\partial_aA +O(t^\beta)\right) t\BG\gamma^{ad} & 0^d
  \end{pmatrix},\\
  \hat {\boldsymbol M}_{1,22}&=
  \begin{pmatrix}
    \left(t {\KSF\chi_a}^{c}+\left(1-\frac 1n-\lambda\right) {\delta_a}^{c} +O(t^\beta)\right)\BG\gamma^{ad} & \frac 12 (A+O(t^\beta)) \BG\gamma^{cd}\\
    \frac 12 (A+O(t^\beta)) \BG\gamma^{cd} & 0^{cd}
  \end{pmatrix}.
\end{align*}
Above, $0, 0^d,$ and $0^{cd}$ denote the zero tensor of the corresponding rank. In the same way we find for \Eqref{eq:preenergyid21}
\begin{equation}
  \label{eq:preenergyid22}
    0=-U^T\cdot\hat {\boldsymbol M}_2\cdot U
    -\left(t^{-\lambda}\nu, t^{-\lambda} \sourcegen^{(2)}\right)_\delta
    +\partial_c\left(\switchgen t^2\BG\gamma^{cd}\left(t^{-\lambda}\nu, t^{-\lambda}\partial_d\nu\right)_\delta\right),
\end{equation}
with
\begin{equation*}
  \hat {\boldsymbol M}_2=
  \begin{pmatrix}
    \hat {\boldsymbol M}_{2,11} & \hat {\boldsymbol M}_{2,12}\\
    \hat {\boldsymbol M}_{2,21} & \hat {\boldsymbol M}_{2,22}
  \end{pmatrix}
\end{equation*}
and
\begin{align*}
  \hat {\boldsymbol M}_{2,11}&=
  \begin{pmatrix}
    0 & A\\
    A & 1
  \end{pmatrix}+O(t^\beta),\\
  \hat {\boldsymbol M}_{2,12}&=\hat {\boldsymbol M}_{2,21}^T=
  \begin{pmatrix}
    0^d & 0^d\\
    0^d & \frac 12\sqrt{\switchgen} \left(\frac 12\KSF\gamma^{-1}_{bc}\partial_a\KSF\gamma^{bc}
-2\log t\partial_a\lambda+O(t^{\beta-\epsilon})\right) t\BG\gamma^{ad}
  \end{pmatrix},\\
\hat {\boldsymbol M}_{2,22}&=
  \begin{pmatrix}
    0^{cd} & 0^{cd}\\
    0^{cd} & \BG\gamma^{cd}
  \end{pmatrix}.
\end{align*}
For the following it is useful to add the $\lambda+\mu$-multiple of \Eqref{eq:preenergyid22} to \eqref{eq:preenergyid12} using a so far unspecified smooth function $\mu(x)$ as follows:
\begin{align}
  \label{eq:preenergyid13N}
    &\frac 12t\partial_t\left(\left|t^{-\lambda} u\right|^2_\delta 
      +\switchgen (1+O(t^\beta)) t^2\BG\gamma^{cd}\left(t^{-\lambda} \partial_c\varphi,t^{-\lambda}  \partial_d\varphi\right)_{\delta}\right)\\
    =&\underbrace{U^T\cdot(\hat {\boldsymbol M}_1-(\lambda+\mu)\hat {\boldsymbol M}_2)\cdot U
    -\switchgen \partial_c (\lambda+\mu) t^2\BG\gamma^{cd}\left(t^{-\lambda}\nu, t^{-\lambda}\partial_d\nu\right)_\delta}_{=:U^T\cdot\hat {\boldsymbol M}\cdot U} \notag\\
    &-\left(t^{-\lambda} u, t^{-\lambda} \sourcegen^{(1)}\right)_\delta 
    -\switchgen (1+O(t^\beta)) t^2\BG\gamma^{cd}\left(t^{-\lambda}\partial_c \sourcegen^{(3)}, t^{-\lambda}\partial_d\varphi\right)_\delta \notag\\
    &-(\lambda+\mu)\left(t^{-\lambda}\nu, t^{-\lambda} \sourcegen^{(2)}\right)_\delta + \partial_c\left((\lambda+\mu)\switchgen t^2\BG\gamma^{cd}\left(t^{-\lambda}\nu, t^{-\lambda}\partial_d\nu\right)_\delta\right) \notag 
\end{align}
with
\begin{equation*}
  \hat {\boldsymbol M}=
  \begin{pmatrix}
    \hat {\boldsymbol M}_{11} & \hat {\boldsymbol M}_{12}\\
    \hat {\boldsymbol M}_{21} & \hat {\boldsymbol M}_{22}
  \end{pmatrix}
\end{equation*}
and
\begin{equation*}
  \hat {\boldsymbol M}_{11}=
  \begin{pmatrix}
    -\lambda & -A/2-(\lambda+\mu)A\\
    -A/2-(\lambda+\mu)A & -(\lambda+\mu)
  \end{pmatrix}+ O(t^\beta) +O(t^{2\kappa}),
\end{equation*}
\begin{equation*}
  \hat {\boldsymbol M}_{12}=\hat {\boldsymbol M}_{21}^T=\hat {\boldsymbol M}_{1,12} 
  - (\lambda + \mu) \hat {\boldsymbol M}_{2,12} +
  \begin{pmatrix}
    0^d & 0^d\\
0^d &- 1/2 \sqrt{\sigma} t \BG\gamma^{cd} \partial_c(\lambda + \mu)
\end{pmatrix}
\end{equation*}
\begin{equation*}
\hat {\boldsymbol M}_{22}=
 \begin{pmatrix}
    \left(t {\KSF\chi_a}^{c}+\left(1-\frac 1n-\lambda\right) {\delta_a}^{c} +O(t^\beta)\right)\BG\gamma^{ad} & \frac 12 (A+O(t^\beta)) \BG\gamma^{cd}\\
    \frac 12 (A+O(t^\beta)) \BG\gamma^{cd} & -(\lambda+\mu)\BG\gamma^{cd}
  \end{pmatrix}.
\end{equation*}
Let us now define
\[{\boldsymbol\Lambda}=
  \begin{pmatrix}
    {\boldsymbol\Lambda}_{1} & \boldsymbol{0}\\
   \boldsymbol{0} & {\boldsymbol\Lambda}_{2}
  \end{pmatrix},\quad 
  {\boldsymbol\Lambda}_1= \begin{pmatrix}
    1 & A \\    
    A & 1 
  \end{pmatrix},\quad
  {\boldsymbol\Lambda}_2= \begin{pmatrix}
    \BG\gamma^{cd} & 0^{cd}\\
    0^{cd} & \BG\gamma^{cd}
  \end{pmatrix}.
\]
It is important to notice that ${\boldsymbol\Lambda}$ is positive definite for all $A$ in the range \Eqref{eq:Arange}.
Writing
\begin{equation*}
  \hat {\boldsymbol M}={\boldsymbol\Lambda} ({\boldsymbol M}-\lambda\mathrm{id})\quad \Leftrightarrow\quad
  {\boldsymbol M}={\boldsymbol\Lambda}^{-1}\left(\hat {\boldsymbol M}+\lambda {\boldsymbol\Lambda}\right),
\end{equation*}
and interpreting ${\boldsymbol\Lambda}$ as a matrix representing a positive definite and symmetric bilinear form, we conclude that the symmetric bilinear form represented by the matrix $\hat {\boldsymbol M}$ is positive (or negative) definite if the endomorphism represented by the matrix ${\boldsymbol M}-\lambda\mathrm{id}$ only has positive (or negative) eigenvalues. Notice here that since $\hat {\boldsymbol M}$ is symmetric, ${\boldsymbol M}$ is self-adjoint with respect to ${\boldsymbol\Lambda}$. ${\boldsymbol M}$ is therefore diagonalizable, its eigenvalues are real and its eigenspaces are mutually orthogonal with respect to ${\boldsymbol\Lambda}$. Since all entries of the matrices ${\boldsymbol M}_{12}$ and ${\boldsymbol M}_{21}$ decay uniformly in space in the limit $t\searrow 0$ (while ${\boldsymbol M}_{11}$ and ${\boldsymbol M}_{22}$ converge to limits ${\boldsymbol M}_{11,0}$ and ${\boldsymbol M}_{22,0}$ below), we conclude that $\hat {\boldsymbol M}$ is therefore uniformly positive (or negative) definite for all  $t\in (0,\finaltime]$ and at all $x$ in the manifold $M$ if  $\finaltime>0$ is sufficiently small and if $\lambda$ is smaller than the smallest (or larger than the largest) eigenvalue of the following two matrices at each spatial point:
\begin{align}
  \label{eq:M110}
  {\boldsymbol M}_{11,0}&={\boldsymbol\Lambda}_1^{-1}(\hat {\boldsymbol M}_{11}+\lambda{\boldsymbol\Lambda}_1)
  =
  \frac 1{1-A^2}\begin{pmatrix}
    1& 
- A \\
-A & 1 
  \end{pmatrix}\begin{pmatrix}
    0& 
-\frac {1+2\mu}{2} A \\
-\frac {1+2\mu}{2} A & -\mu 
  \end{pmatrix} \\
\label{eq:M220}
  {\boldsymbol M}_{22,0}&={\boldsymbol\Lambda}_2^{-1}(\hat {\boldsymbol M}_{22}+\lambda{\boldsymbol\Lambda}_2)  
  =\begin{pmatrix}
    \mathrm{diag}\left(1-q_1,\ldots, 1-q_n\right)& \frac 12 A {\delta_a}^{c} \\
    \frac 12 A {\delta_a}^{c} & -\mu{\delta_a}^{c} 
  \end{pmatrix}, 
\end{align}
using \Eqref{eq:KSBG4}.
Notice that the eigenvalues of both matrices clearly depend on the choice of the function $\mu(x)$ and on $x$.  By rearranging the rows and columns of ${\boldsymbol M}_{22,0}$ we can bring this matrix to a block diagonal form where the only non-zero parts are  $2\times 2$-blocks of the form  
\[
  \begin{pmatrix}
    1-q_i & A/2\\
    A/2 & -\mu
  \end{pmatrix}
\]
for each $i=1,\ldots,n$ along the diagonal.
In order to facilitate the subsequent analysis, the task is now to vary $\mu$ in order 
 (1) to \emph{maximize the smallest of all eigenvalues} of \Eqsref{eq:M110} and \eqref{eq:M220}, or, (2) to \emph{minimize the largest of all eigenvalues}. With straightforward arguments exploiting the bounds \Eqref{eq:boundsonKasner} on the Kasner exponents, we find for each $x\in M$:
\begin{enumerate}[label=(\arabic{*}),leftmargin=*]
\item The smallest of all eigenvalues of \Eqsref{eq:M110} and \eqref{eq:M220} is maximized for $\mu=-1/2$. This maximal value is zero. We therefore conclude that $\hat {\boldsymbol M}$ is \emph{uniformly positive definite} for all sufficiently small $t>0$ and all $x\in M$ provided $\lambda(x)<0$ and $\mu(x)=-1/2$ for all $x\in M$.
\item Pick $A$ and $q_1$, \ldots, $q_n$ as before, and $\xi\in [0,1]$ (see \Eqref{eq:defxi}).
The largest of all eigenvalues of \Eqsref{eq:M110} and \eqref{eq:M220} 
is minimized for\footnote{Since $\xi$ can have different values at each $x\in M$, we think of $\mu_c$ as being any smooth function arbitrarily close to the values given in \Eqref{eq:muopt}.}
\begin{equation}
  \label{eq:muopt}
  \mu=\mu_c=
  \begin{cases}
    \frac 12(1-4A^2)=\frac 12(1-4(1-\eta^2)A_+^2) & \text{for $\xi\in [0,1/3]$},\\
    \frac{4 \xi ^2-A_+^2 (1-\xi ) (1+\xi)^3}{4 (1-\xi ) \xi }& \text{for $\xi\in [1/3,1]$}.
  \end{cases}
\end{equation}
This minimal value is $\lambda_c$ (see \Eqref{eq:deflambdac}).
We therefore conclude that $\hat {\boldsymbol M}$ is \emph{uniformly negative definite} for all sufficiently small $t>0$ and all $x\in M$ provided $\lambda(x)>\lambda_c(x)$ and $\mu(x)$ as above for all $x\in M$.
\end{enumerate}

Let us now assume the conditions under which $\hat {\boldsymbol M}$ is positive definite as above. Integrating \Eqref{eq:preenergyid13N} for any fixed $t\in (0,\finaltime]$ with respect to the volume element associated with $\delta_{ab}$ on $M$ yields, for any constant $\eta>0$ (recall that $\mu=-1/2$),
\begin{align*}
    t\partial_t\littleenergysymbol
    &\ge
    -2\eta\littleenergysymbol
      - \frac C{2\eta}\left(\|t^{-\lambda} \sourcegen^{(1)}\|^2_\delta 
      +\switchgen^2\|t^{-\lambda} \partial \sourcegen^{(3)}\|^2_{\delta,t^2\KSF\gamma}\right)  
    -\int_M(\lambda-1/2)\left(t^{-\lambda}\nu, t^{-\lambda} \sourcegen^{(2)}\right)_\delta dx,      
\end{align*}
where $C$ is a constant that may depend on $\finaltime$ and $\Gamma$ and
where we have set
\[\littleenergysymbol
=\frac 12\int_M\left(\left|t^{-\lambda} u\right|^2_\delta 
      +\switchgen (1+O(t^\beta)) t^2\BG\gamma^{cd}\left(t^{-\lambda} \partial_c\varphi,t^{-\lambda}  \partial_d\varphi\right)_{\delta}\right)dx.\]
For any $t_*\in (0,\finaltime]$, we divide by $t$ and integrate the result over $[t,t_*]$ for $t\in (0,t_*]$:
\begin{align*}
    t^{2\eta}\littleenergysymbol(t)
    &\le    t_*^{2\eta}\littleenergysymbol(t_*)
      +\frac C{2\eta}\int_t^{t_*}\left(
      \|s^{-\lambda} \sourcegen^{(1)}\|^2_\delta 
      +\switchgen^2\|s^{-\lambda} \partial \sourcegen^{(3)}\|^2_{\delta,t^2\KSF\gamma}\right) s^{-1+2\eta} ds\\   
    &+\int_t^{t_*}\left(\int_M(\lambda-1/2)\left(s^{-\lambda}\nu, s^{-\lambda} \sourcegen^{(2)}\right)_\delta dx\right) s^{-1+2\eta} ds.
\end{align*}
Comparing the definition of $\littleenergysymbol$ to that of $\littleenergyswitch{u,\varphi}\switchgen{\lambda}$ in \Eqref{eq:defenergies1N} allows us to conclude that
\begin{align*}
    \littleenergyswitch{u,\varphi}\switchgen{\lambda-\eta}(t)
    &\le  C\Bigl(\littleenergyswitch{u,\varphi}\switchgen{\lambda-\eta}(t_*)
      +\int_t^{t_*}\Bigl(\|s^{-\lambda+\eta} \sourcegen^{(1)}\|^2_\delta +\switchgen^2\|s^{-\lambda+\eta} \partial \sourcegen^{(3)}\|^2_{\delta,t^2\KSF\gamma}\Bigr) s^{-1} ds\Bigr)\\   
    &+\int_t^{t_*}\left(\int_M(\lambda-1/2)\left(s^{-\lambda+\eta}\nu, s^{-\lambda+\eta} \sourcegen^{(2)}\right)_\delta dx\right) s^{-1} ds,
\end{align*}
for a constant $C>0$ which may depend on $\finaltime$, $\eta$ and $\Gamma$. This inequality holds for any smooth $\lambda<0$. Since $\eta>0$ is arbitrary, it also holds for $\lambda-\eta$ replaced by $\lambda$:
\begin{equation}
  \label{eq:preenergyid13}
  \begin{split}
    \littleenergyswitch{u,\varphi}\switchgen{\lambda}(t)
    \le&  C\Bigl(\littleenergyswitch{u,\varphi}\switchgen{\lambda}(t_*)
      +\int_t^{t_*}\Bigl(\|s^{-\lambda} \sourcegen^{(1)}\|^2_\delta +\switchgen^2\|s^{-\lambda} \partial \sourcegen^{(3)}\|^2_{\delta,t^2\KSF\gamma}\Bigr) s^{-1} ds\Bigr)\\   
    &+\int_t^{t_*}\left(\int_M(\lambda-1/2)\left(s^{-\lambda}\nu, s^{-\lambda} \sourcegen^{(2)}\right)_\delta dx\right) s^{-1} ds.
  \end{split}
\end{equation}
For any constant $\epsilon>0$, 
\Eqref{eq:preestimat3} 
and the Gr\"onwall
lemma imply
\begin{equation*}
  \begin{split}
    \littleenergyswitch{u,\varphi}\switchgen{\lambda-\epsilon}(t)
    &\le  C\Bigl(\littleenergyswitch{u,\varphi}\switchgen{\lambda-\epsilon}(t_*)
      +t_*^{2\epsilon}\int_t^{t_*}\Bigl(\|s^{-\lambda} \sourcegen^{(1)}\|^2_\delta +\|s^{-\lambda} \sourcegen^{(2)}\|^2_\delta +\switchgen^2\|s^{-\lambda} \partial \sourcegen^{(3)}\|^2_{\delta,t^2\KSF\gamma}\Bigr) s^{-1} ds\Bigr).
  \end{split}
\end{equation*}
If $\epsilon>0$ is sufficiently small, this estimate must also hold when $\lambda-\epsilon$ is replaced by $\lambda$ under the same conditions for $\lambda$, which yields \Eqref{eq:LSFbackwardslittle}. 

In the case that $\hat {\boldsymbol M}$ is negative definite on the other hand (see above), the same arguments lead to the estimate
\begin{equation}
  \label{eq:preenergyid14}
  \begin{split}
    \littleenergyswitch{u,\varphi}\switchgen{\lambda}(t)
    \le&  C\Bigl(\littleenergyswitch{u,\varphi}\switchgen{\lambda}(t_*)
      +\int^t_{t_*}\Bigl(\|s^{-\lambda} \sourcegen^{(1)}\|^2_\delta +\switchgen^2\|s^{-\lambda} \partial \sourcegen^{(3)}\|^2_{\delta,t^2\KSF\gamma}\Bigr) s^{-1} ds\Bigr)\\   
    &+\int^t_{t_*}\left(\int_M(\lambda+\mu_c)\left(s^{-\lambda}\nu, s^{-\lambda} \sourcegen^{(2)}\right)_\delta dx\right) s^{-1} ds,
  \end{split}
\end{equation}
for any $\lambda>\lambda_c$ in \Eqref{eq:deflambdac} and for all $t\in[t_*,\finaltime]$ where $\mu_c$ is given by \Eqref{eq:muopt}.
The same line of arguments applied to \Eqref{eq:preenergyid14} yields \Eqref{eq:LSFforwardslittle}.

 For any sufficiently small $t\in (0,\finaltime]$ we find easily, applying similar arguments as above to \Eqref{eq:defueqtens}, that
  \begin{equation*}
  t\partial_t\sqrt{\littleenergywk{\varphi}{k}{\lambda}(t)}
\le C
\Bigl(\sqrt{\littleenergywk{u}k{\lambda}(t)}+\sqrt{\littleenergyswitch{\nu}{\switchgen}{\lambda}(t)}+\|t^{-\lambda} \sourcegen^{(3)}\|_{\delta}      
  \Bigr),
\end{equation*}
so long as $\lambda(x)\ge 0$. 
\Eqref{eq:preestimat3} therefore implies \Eqref{eq:LSFforwardlittle2}.
With the same arguments we find \Eqref{eq:LSFbackwardlittle2} for any smooth $\lambda(x)\le 0$.
\end{proof}

Having established estimates for the tensorial version of the equations, \Eqref{eq:LSFtensoreq}, we now proceed  with estimates for the tensorial source terms \Eqref{eq:tensorialf}. 

\begin{proof}[Proof of \Lemref{lem:estimatesTensorST}]
  \hypertarget{lemproof:estimatesTensorST}{}
  In order to establish this result, we apply the product estimates of the form of Prop.~3.7 in Chapter~13 of \cite{taylor2011} together with suitable estimates for the coefficients as follows.
For example, we encounter expressions like this 
\begin{align*}
  &\left(\switchgen\partial_{e\ldots f}({\BG\alpha}t^2\BG\gamma^{cd})\partial_c \varphi_{da\ldots b}\right)
\left(\switchgen\partial^{e\ldots f} ({\BG\alpha}t^2{\BG\gamma^{c}}_{d})\partial_c \varphi^{da\ldots b}\right)\\
  &=\partial_c \varphi_{da\ldots b}
\underbrace{\left(\switchgen^2\partial_{e\ldots f} ({\BG\alpha}t^2\BG\gamma^{cd})\partial^{e\ldots f} ({\BG\alpha}t^2\BG\gamma^{c'd'})\delta^{aa'}\cdots \delta^{bb'}\right)}_{=:\switchgen^2A^{cc'dd'}\delta^{aa'}\cdots \delta^{bb'}}
\partial_{c'} \varphi_{d'a'\ldots b'}.
\end{align*}
Recalling \Eqref{eq:LSFasldkj}, it suffices to establish that
$\switchgen^2A^{cc'dd'}\le C t^{2\kappa} \switchgen t^2\KSF\gamma^{cc'}\delta^{dd'}$
for each $x$ and for all sufficiently small $t>0$ for some uniform constant $C$ (which may depend on $\finaltime$ and $\Gamma$)
where the tensor $A^{cc'dd'}(t,x)$ is interpreted as a symmetric bilinear form acting on the space of $(0,2)$ tensors at $(t,x)$. Recalling that $\BG\gamma^{cd}\KSF\gamma^{-1}_{de}$ is bounded by \Eqsref{eq:gammaass2} and \eqref{eq:LSFasldkj}, that
\[t^2\BG\gamma^{cd}\delta_{de}=t^2\KSF\gamma^{cf}\left({\delta^d}_{d'}+O(t^\beta)\right)\delta_{de},\]
that ${\BG\alpha}$ satisfies \Eqref{eq:alphaSpiSassumps} and that very similar arguments imply that terms like $B^c=t^2{\BG\alpha} C^c-L^{[1,3],c}_1$ have the bound $\switchgen^2\partial_{e\ldots f}B^c\partial^{e\ldots f}B^d\le C \switchgen t^{2\kappa} t^2\KSF\gamma^{cd}$,
it follows that
\begin{equation}
    \label{eq:est092348}
    \begin{split}
  \sum_{l=0}^k&\left\|t^{-\lambda} \sourcegen^{(1)}_{[l]}\right\|^2_{\delta}\\
  \le & C\left(\|t^{-\lambda} \sourcegen^{(1)}\|^2_{\delta, H^k(M)}
    +\energy{u,\varphi}{k}{\lambda-2\kappa}(t)
  +\energy{u,\varphi}{k-1}{\lambda}(t) +\energy{\nu}{k-1}{\lambda}(t)
  \right)
\end{split}
\end{equation}
for all $t\in (0,\finaltime]$ where $\kappa$ is defined in \Propref{prop:LSF4}.
The constant $C>0$ may depend on $k$, $\finaltime$ and $\Gamma$.
The same arguments as above lead to the estimates
\begin{gather}
  \label{eq:est3121323pre}
  \begin{split}
  \sum_{l=0}^k&\left\|t^{-\lambda} \sourcegen^{(2)}_{[l]}\right\|^2_{\delta}\\
  \le & C\left(\|t^{-\lambda} \sourcegen^{(2)}\|^2_{\delta, H^k(M)}
  +\energy{u}{k-1}{\lambda}(t) +\energy{\nu}{k-1}{\lambda}(t)+ \energy{\nu}{k}{\lambda-2\kappa}(t)
  \right),
\end{split}\\
  \label{eq:est4756342pre}
  \sum_{l=0}^k\left\|t^{-\lambda} \sourcegen^{(3)}_{[l]}\right\|^2_{\delta}
  \le C\left(\|t^{-\lambda} \sourcegen^{(3)}\|^2_{\delta, H^{k}(M)}
  +\energy{u}{k-1}{\lambda-\nu} (t) +\energy{\nu}{k-1}{\lambda}(t)\right),\\
  \label{eq:est47567pre}
  \begin{split}
  \sum_{l=0}^k&\left\|t^{-\lambda} \partial \sourcegen^{(3)}_{[l]}\right\|^2_{\delta,t^2\BG\gamma}\\
  \le & C\left(\|t^{-\lambda} \sourcegen^{(3)}\|^2_{\delta, H^{k+1}(M)}
  +\energy{u}{k-1}{\lambda}(t) +\energy{\nu}{k-1}{\lambda}(t) + \energy{u}{k}{\lambda-\nu-2\kappa}(t)\right).
\end{split}
\end{gather}
This last estimate crucially depends on the fact that that $L^{[3,2]}-1=O(t^\beta)$, see \Eqsref{eq:alphaSpiSassumps} and \Eqref{eq:coefflast}.

Now let us apply
\Eqref{eq:preestimat3} to \Eqref{eq:est3121323pre} using \Eqref{eq:defEnergies}:
\begin{align*}
  \sum_{l=0}^k\left\|t^{-\lambda} \sourcegen^{(2)}_{[l]}\right\|^2_{\delta}
  \le C\Bigl(&\|t^{-\lambda} \sourcegen^{(2)}\|^2_{\delta, H^k(M)}
  +\energy{u}{k-1}{\lambda}(t) 
+\energy{u}{k}{\lambda-2\kappa}(t)\\
&+\sum_{l=0}^{k-1}\|t^{-\lambda}\sourcegen^{(2)}_{[l]}\|_\delta^2
+\sum_{l=0}^{k}\|t^{-(\lambda-2\kappa)}\sourcegen^{(2)}_{[l]}\|_\delta^2
  \Bigr).
\end{align*}
Hence, provided $t\in (0,\finaltime]$ for a sufficiently small $\finaltime$, we can redefine the constants so that \Eqref{eq:est3121323} holds (recall that $\kappa$ is strictly positive by definition). The resulting constant $C$ may depend on $T$, $k$ and $\Gamma$ as above.
Inserting \Eqref{eq:est3121323} into \Eqsref{eq:preestimat3} yields
\begin{equation}
  \label{eq:preestimat3NN}
\energyswitch{\nu}\switchgen{k}{\lambda}(t)
  \le C \left(\energy{u}k{\lambda}(t) +\|t^{-\lambda} \sourcegen^{(2)}\|^2_{\delta, H^k(M)}
 \right).
\end{equation}
This inequality can be used to rewrite \Eqref{eq:est092348} as \Eqref{eq:LSFf1}, \Eqref{eq:est4756342pre} as \Eqref{eq:LSFf3}
and \Eqref{eq:est47567pre} as \Eqref{eq:LSFf3del}.
\end{proof}

\subsection{Proof of \Propref{prop:LSF4}: Improved decay estimates}
\label{sec:LSFproof2}

Suppose that $(u,\nu,\varphi)$ is given as in the hypothesis of \Propref{prop:LSF4}. The $(0,k)$-tensor fields associated with spatial derivatives of order $k$ therefore satisfy \Eqsref{eq:scalarfieldequationgentens} -- \eqref{eq:defueqtens} with \Eqref{eq:tensorialf}  and \Eqsref{eq:coeff1} -- \eqref{eq:coefflast}. 
Using \eqref{eq:lapseequationgentens} to eliminate the $\nu$-term from \eqref{eq:scalarfieldequationgentens} yields (recall footnote~\ref{ftnote:tensornotation})
  \begin{align*}
    -t\partial_t u &+\left(L^{[1,1]}-L^{[2,2]} \frac {L^{[1,2]}}{L^{[2,1]}}\right) u\\  
=&\sourcegen^{(1)}-\frac {L^{[1,2]}}{L^{[2,1]}}\sourcegen^{(2)}
-\switchgen{\BG\alpha} t^2\BG\gamma^{ab}\partial_a \partial_b\varphi
+\switchgen (t^2{\BG\alpha} C^a-L^{[1,3],a}_1) \partial_a\varphi\\
&+\switchgen \frac {L^{[1,2]}}{L^{[2,1]}}t^2\BG\gamma^{ab}\partial_a \partial_b\nu
  -\switchgen \left(\frac {L^{[1,2]}}{L^{[2,1]}}t^2C^a+L_1^{[1,4],a}\right) \partial_a\nu.
  \end{align*}
As in the first part of the \hyperlink{lemproof:estimatesTensorEqs}{proof} of \Lemref{lem:estimatesTensorEqs} we contract this equation with the tensor field $t^{-2\lambda}u$ for any smooth function $\lambda(x)$. This yields the identity 
\begin{align*}
    \frac 12 &t\partial_t |t^{-\lambda}u|^2_{\delta} =-\Bigl(\lambda-\Bigl({L^{[1,1]}-L^{[2,2]} \frac {L^{[1,2]}}{L^{[2,1]}}}\Bigr)\Bigr) |t^{-\lambda}u|_{\delta}^2\\  
&-\left(t^{-\lambda}u,t^{-\lambda}\sourcegen^{(1)}_{[k]}\right)_{\delta}
+{\frac {L^{[1,2]}}{L^{[2,1]}}}\left(t^{-\lambda}u,t^{-\lambda}\sourcegen^{(2)}_{[k]}\right)_{\delta}\\
&+\switchgen{\BG\alpha} t^2\BG\gamma^{ab}\left(t^{-\lambda}u,t^{-\lambda}\partial_a \partial_b\varphi \right)_{\delta}
-\switchgen (t^2{\BG\alpha} C^c\BG\gamma^{-1}_{cb}-t^2\partial_b\BG\alpha)\BG\gamma^{ab} \left(t^{-\lambda}u,t^{-\lambda}\partial_a\varphi \right)_{\delta}\\
&-\switchgen \frac {L^{[1,2]}}{L^{[2,1]}}t^2\BG\gamma^{ab}\left(t^{-\lambda}u,t^{-\lambda}\partial_a \partial_b\nu \right)_{\delta}
  +\switchgen \left(\frac {L^{[1,2]}}{L^{[2,1]}}t^2C^c\BG\gamma^{-1}_{cb}+t^2\partial_b\BG\phi\right)\BG\gamma^{ab} \left(t^{-\lambda}u,t^{-\lambda}\partial_a\nu \right)_{\delta}.
  \end{align*}
From \Eqsref{eq:coeff1asympt} -- \eqref{eq:coefflastasympt} we get
\begin{equation*}
  {L^{[1,1]}-L^{[2,2]} \frac {L^{[1,2]}}{L^{[2,1]}}}=2A^2+O(t^\beta)+ O(t^{2\kappa}),\quad
  {\frac {L^{[1,2]}}{L^{[2,1]}}}=-A+O(t^\beta)+ O(t^{2\kappa}),
\end{equation*}
where $\kappa$ is defined in \Propref{prop:LSF4}.
If $\lambda<2A^2$, the preceding identity implies that
\begin{align*}
  t\partial_t \sqrt{\littleenergy{u_{[k]}}{\lambda}(t)}\ge -C\Bigl(&\|t^{-\lambda} \sourcegen_{[k]}^{(1)}\|_{\delta}
+\|t^{-\lambda} \sourcegen_{[k]}^{(2)}\|_{\delta}
+\sqrt{\littleenergy{\varphi_{[k+2]}}{\lambda-2\kappa} (t)}
+\sqrt{\littleenergy{\varphi_{[k+1]}}{\lambda-2\kappa} (t)}\\
&+\sqrt{\littleenergy{\nu_{[k+2]}}{\lambda-2\kappa} (t)}
+\sqrt{\littleenergy{\nu_{[k+1]}}{\lambda-2\kappa} (t)}
\Bigr),
\end{align*}
where $C>0$ may depend on $\finaltime$ and $\Gamma$. We replace $k$ by $l$ and sum this estimate over the integers from $0$ to $k$, and restrict $k$ to the range $k\ge 1$ (the case $k=0$ follows with similar, but simpler, arguments to those below). The first term on the right-hand side can then be estimated using the following alternative  to \Eqref{eq:est092348}, which provides sharper decay control at the cost of weaker regularity control: 
\begin{equation}
    \label{eq:est092348decayNN}
    \begin{split}
  &\sum_{l=0}^k\left\|t^{-\lambda} \sourcegen^{(1)}_{[l]}\right\|^2_{\delta}\\
  \le & C\left(\switchgen^2\|t^{-\lambda} \sourcegen^{(1)}\|^2_{\delta, H^k(M)}
    +\energy{\varphi}{k+1}{\lambda-2\kappa}(t)
  +\energy{u}{k-1}{\lambda-\beta}(t) +\energy{\nu}{k-1}{\lambda}(t)
  \right).
\end{split}
\end{equation}
The term involving $\sourcegen_{[k]}^{(2)}$ on the right-hand side of the previous energy inequality can be estimated with \Eqsref{eq:est3121323} and all energies involving $\nu$  with \Eqref{eq:preestimat3N}.
If we now tighten our assumption on $\lambda$ so that $\lambda<\min\{2\kappa,2A^2\}$ and pick any sufficiently small $\epsilon>0$, we can use \Eqref{eq:LSFbackward2} to show that
\begin{align*}
  t\partial_t &\sqrt{\energy{u}{k}{\lambda}(t)}
  \ge -C\Bigl(
                 \|t^{-\lambda} \sourcegen^{(1)}\|_{\delta, H^k(M)}
                 +\|t^{-\lambda} \sourcegen^{(2)}\|_{\delta, H^k(M)}
                 +\|t^{-\lambda+2\kappa} \sourcegen^{(2)}\|_{\delta, H^{k+2}(M)}\\
&+t^\epsilon\int^{t_*}_t\Bigl(
\|s^{-\lambda+2\kappa} \sourcegen^{(2)}\|_{\delta, H^{k+2}(M)}
+\|s^{-\lambda+2\kappa} \sourcegen^{(3)}\|_{\delta, H^{k+2}(M)}      
  \Bigr) s^{-1-\epsilon} ds\\
&+t^\epsilon\sqrt{\energy{\varphi}{k+2}{\lambda-2\kappa+\epsilon}(t_*)}
               {+\sqrt{\energy{u}{k-1}{\lambda}(t)}}
                 \underbrace{+\sqrt{\energy{u}{k}{\lambda-2\kappa}(t)}}_{\le Ct^{2\epsilon}\sup_{s\in[t,t_*]}\sqrt{\energyswitch{u,\varphi}\switchgen{k+2}{\lambda-2\kappa+2\epsilon} (s)}} \\
&
\underbrace{+\sqrt{\energy{u}{k+2}{\lambda-2\kappa} (t)}
+t^\epsilon\int^{t_*}_t\sqrt{\energy{u}{k+2}{\lambda-2\kappa+2\epsilon} (s)} s^{-1+\epsilon} ds}_{\le Ct^\epsilon\sup_{s\in[t,t_*]}\sqrt{\energyswitch{u,\varphi}\switchgen{k+2}{\lambda-2\kappa+2\epsilon} (s)}}
\Bigr)
\end{align*}
for all $t\in (0,t_*]$. The constant $C>0$ here may depend on $\finaltime$, $k$, $\lambda$ and $\Gamma$. Hence
\begin{align*}
  &\sqrt{\energy{u}{k}{\lambda}(t)} 
\le C\Bigl(\sqrt{\energy{u}{k}{\lambda}(t_*)}
+\sqrt{\energy{\varphi}{k+2}{\lambda-2\kappa+\epsilon}(t_*)}
+\sup_{s\in(0,\finaltime]}\sqrt{\energyswitch{u,\varphi}\switchgen{k+2}{\lambda-2\kappa+2\epsilon} (s)}\\
  &\quad+\int_t^{t_*}\Bigl(\|s^{-\lambda} \sourcegen^{(1)}\|_{\delta, H^k(M)}
    +\|s^{-\lambda} \sourcegen^{(2)}\|_{\delta, H^{k+2}(M)}
+\|s^{-\lambda+2\kappa-\epsilon} \sourcegen^{(3)}\|_{\delta,H^{k+2}(M)}\Bigr) s^{-1}ds
                 \\
               &\quad               +\sup_{s\in[t,t_*]}\sqrt{\energy{u}{k-1}{\lambda+\epsilon}(s)}
\Bigr).
\end{align*}
The constant $C>0$ here may depend on $\finaltime$, $k$, $\lambda$, $\epsilon$ and $\Gamma$. Notice that the last term is not present if $k=0$. If $k=1$, we can combine this estimate with the $k=0$ estimate to establish the result. 
If $k\ge 2$, we proceed inductively in a manner very similar to the \hyperlink{propproof:LSF2}{proof} of \Propref{prop:LSF2} and find
\begin{align*}
  &\sqrt{\energy{u}{k}{\lambda}(t)}  
\le C\Bigl(\sqrt{\energy{u}{k}{\lambda+\epsilon}(t_*)}
+\sqrt{\energy{\varphi}{k+2}{\lambda-2\kappa+\epsilon}(t_*)}
+\sup_{s\in(0,\finaltime]}\sqrt{\energyswitch{u,\varphi}\switchgen{k+2}{\lambda-2\kappa+\epsilon} (s)}\\
  &+\int_t^{t_*}\Bigl(\|s^{-\lambda} \sourcegen^{(1)}\|_{\delta, H^k(M)}
    +\|s^{-\lambda} \sourcegen^{(2)}\|_{\delta, H^{k+2}(M)}
+\|s^{-\lambda+2\kappa-\epsilon} \sourcegen^{(3)}\|_{\delta,H^{k+2}(M)}\Bigr) s^{-1-\epsilon}ds                 
\Bigr),
\end{align*}
where the constant $C$ has the same dependence as in the preceding equation.
In order to be able to apply \Eqref{eq:LSFbackward1} we first need to square this estimate. In general,
the H\"older inequality implies the existence of a constant $C>0$, which depends on $\finaltime$ and $2\eta_1-\eta_2$ such that for any smooth function $f(t)$ and any constant $\eta_1\in\R$
\begin{equation}
  \label{eq:Holdertimeintegrals}
  \left(\int_t^{t_*} f(s)s^{-1+\eta_1} ds\right)^2\le C \int_t^{t_*} f^2(s)s^{-1+\eta_2} ds
\end{equation}
provided $\eta_2<2\eta_1$. Here we pick $\eta_1=-\epsilon$ and $\eta_2=-3\epsilon$ 
and apply \Eqref{eq:LSFbackward1} to obtain:
\begin{align*}
  &\energy{u}{k}{\lambda}(t)  
\le C\Bigl(\energy{u}{k}{\lambda+\epsilon}(t_*)
+\energy{\varphi}{k+2}{\lambda-2\kappa+\epsilon}(t_*)
+\energyswitch{u,\varphi}\switchgen{k+2}{\lambda-2\kappa+3\epsilon} (t_*)\\      
  &+\int_t^{t_*}\Bigl(\|s^{-\lambda} \sourcegen^{(1)}\|^2_{\delta, H^{k+2}(M)}
    +\|s^{-\lambda} \sourcegen^{(2)}\|^2_{\delta, H^{k+2}(M)}
+\|s^{-\lambda+2\kappa-\epsilon} \sourcegen^{(3)}\|^2_{\delta,H^{k+3}(M)}\Bigr) s^{-1-3\epsilon}ds
\Bigr).
\end{align*}
The constant $C>0$ may depend on $\finaltime$, $k$, $\lambda$, $\epsilon$ and $\Gamma$ as above.
We can now choose $t_*=\finaltime$, replace the energies on the right-hand side by norms of Cauchy data imposed at $\finaltime$, and, replace $3\epsilon$ by $2\epsilon$.
This yields \Eqref{eq:LSFimproveddecayubackward} and completes the proof of \Propref{prop:LSF4}.

\subsection{Proofs of \Propref{prop:LSF3} and \ref{prop:LSFresult1}: The asymptotic matching problem of the linearized lapse-scalar field system}
\label{sec:LSFproof3}

\begin{proof}[Proof of \Propref{prop:LSF3}]
\hypertarget{propproof:LSF3}{}
  Considering $\switchgen\in I$, $f^{(1)}$,  $f^{(2)}$ and $f^{(3)}$ as \emph{fixed}, it is useful to rephrase the problem addressed by this proposition as the following asymptotic matching problem in the sense of \Defref{def:AMatchingProblem}:
\begin{enumerate}
\item Let $(\vnkgen,\nu,\unkgen)$ be the smooth solution of the Cauchy problem of \Eqsref{eq:scalarfieldlinearMatch} -- \eqref{eq:pilinearMatch} determined by smooth Cauchy data $(\CDfinaltime{\vnkgen},\CDfinaltime{\unkgen})$ imposed at $t=\finaltime$.
\item The functions $\vnkgen$ and $\nu$ introduced in step~1 then determine the evolution equation
\begin{equation}
  \label{eq:LSFprematchEPDeq}
  t\partial_t\unkgen^{(0)}=\switchgen^{(0)}(\vnkgen+A\nu)
\end{equation}
for a new switch parameter $\switchgen^{(0)}\in [0,1]$ and a new unknown $\unkgen^{(0)}$ (which in general is not equal to the given function $\unkgen$ from step~1). This is
derived from \Eqref{eq:pilinearMatch} using the idea that $\varphi$ should behave like a constant in time at first order. It is this first order term of $\varphi$ that is captured by \Propref{prop:LSF3}. 

\end{enumerate}
Each choice of smooth data $(\CDfinaltime{\vnkgen},\CDfinaltime{\unkgen})$ therefore determines the asymptotic matching problem for \Eqref{eq:LSFprematchEPDeq}.
Here we are clearly only interested in the case $\switch^{(0)}=1$ and $\switcht^{(0)}=0$.
We notice that \eqref{eq:LSFprematchEPDeq} is  an equation of the form \Eqref{eq:abstracteqgenNN} with \begin{equation}
  \label{eq:LSFprematchEPDAL1}
  A=t\partial_t,\quad L=0,\quad \sourcegen=\switchgen^{(0)}(\vnkgen+A\nu).
\end{equation} 
In \Eqsref{eq:abstracteq} -- \eqref{eq:abstracteqw4} we then assume
\begin{equation}
  \label{eq:LSFprematchEPDAL2}
  \source=\vnkgen+A\nu,\quad 
  \sourcet=0,\quad 
  \tilde\source=\tilde\vnkgen+A \tilde\nu,\quad 
\tilde\sourcet=0, 
\end{equation}
where $(\tilde\vnkgen,\tilde\nu,\tilde\unkgen)$ is the smooth solution of the Cauchy problem of \Eqsref{eq:scalarfieldlinearMatch} -- \eqref{eq:pilinearMatch} determined by any (possibly different) Cauchy data $(\CDfinaltime{\tilde\vnkgen},\CDfinaltime{\tilde\unkgen})$.

Let us suppose for the moment that we can indeed establish that this asymptotic matching problem is well-defined in the case $\switch^{(0)}=1$ and $\switcht^{(0)}=0$. 
The map $\PreMatchingMapAbstr$, whose existence is asserted in \Propref{prop:LSF3}, is then related to the asymptotic matching map $\AsymptoticMatchingMapAbstr{1\rightarrow 0}^{(0)}$ associated with \Eqref{eq:LSFprematchEPDeq} by
\begin{equation}
  \label{eq:LSFrelationshipmapsprematch}
  \PreMatchingMapAbstr(\CDfinaltime{\vnkgen}, \CDfinaltime{\unkgen})=\left.\AsymptoticMatchingMapAbstr{1\rightarrow 0}^{(0)}(\CDfinaltime{\unkgen^{(0)}}, \CDfinaltime{\vnkgen}, \CDfinaltime{\unkgen})\right|_{\CDfinaltime{\unkgen^{(0)}}=\CDfinaltime{\unkgen}}.
\end{equation}
Notice that here and in all of what follows we make the \emph{implicit} dependence of $\AsymptoticMatchingMapAbstr{1\rightarrow 0}^{(0)}$ on  $\CDfinaltime{\vnkgen}$ and $\CDfinaltime{\unkgen}$ via $\vnkgen$ and $\nunkgen$ in \Eqref{eq:LSFprematchEPDeq} \emph{explicit} by adding these as arguments to the map. 

Given $(\CDfinaltime{\vnkgen},\CDfinaltime{\unkgen})$ as above,
the first step is to consider the corresponding \emph{finite} matching problem with the corresponding finite matching map 
\begin{equation}
  \label{eq:LSFprematchfinitematch}
  \FiniteMatchingMapAbstr{1\rightarrow 0}^{(0)}: (0,\finaltime]\times (C^\infty(M))^3\rightarrow C^\infty(M),\, (\matchtime, \CDfinaltime{\unk^{(0)}},  \CDfinaltime{\vnkgen}, \CDfinaltime{\unkgen})\mapsto \FiniteMatchingMapAbstr{1\rightarrow 0}^{(0)} (\matchtime, \CDfinaltime{\unk^{(0)}},  \CDfinaltime{\vnkgen}, \CDfinaltime{\unkgen}).
\end{equation}
According to the general discussion above, the plan is to find estimates for the hierarchy of Cauchy problems \Eqsref{eq:abstracteq}, \eqref{eq:abstracteqw4}, \eqref{eq:abstracteqw2} and \eqref{eq:abstracteqw3} 
for \Eqsref{eq:LSFprematchEPDAL1} and \eqref{eq:LSFprematchEPDAL2}
and arbitrary $\matchtime$, $\CDfinaltime{\unk^{(0)}}$,  $\CDfinaltime{\vnkgen}$ and $\CDfinaltime{\unkgen}$ as above. 
As discussed above we first require estimates for the two smooth solutions $(\vnkgen,\nu,\unkgen)$ and $(\tilde\vnkgen,\tilde\nu,\tilde\unkgen)$ of the Cauchy problem of \Eqsref{eq:abstracteq} and \eqref{eq:abstracteqw4} for \Eqref{eq:LSFLSFasldkasdk1} -- \eqref{eq:LSFLSFasldkasdklast} with the source term 
$F=\tilde F=(f^{(1)}, f^{(2)}, f^{(3)})$ 
and the switch parameter $\switchgen\in I$.
Define
\[\unkd^{(0)}=\unk^{(0)}-\unkt^{(0)}.\] 
Picking any smooth $\lambda<\min\{2\kappa,2A^2\}$ and sufficiently small constant $\epsilon>0$, \Eqref{eq:LSFimproveddecayubackward} is the required estimate for the function $u$,
and yields the required estimate for $u-\tilde u$:
\begin{equation}
\label{eq:LSFLSFfullestimate32432}
\begin{split}
  \energy{u-\tilde\vnkgen}{k}{\lambda}(t)  
&\le C\Bigl(
\|u_*-\CDfinaltime{\tilde u}\|^2_{\delta,H^{k+2}(M)}
+ \|\varphi_*-\CDfinaltime{\tilde\varphi}\|^2_{\delta,H^{k+3}(M)}
\Bigr)
\end{split}
\end{equation}
for all $t\in (0,\finaltime]$. The constant $C$ here may depend on $\finaltime$, $k$, $\lambda$, $\epsilon$ and $\Gamma$.

Now, first we  observe  
that \Eqref{eq:LSFforward2} allows us to estimate
the Cauchy problems \Eqsref{eq:abstracteqw2} and \eqref{eq:abstracteqw3} for \Eqsref{eq:LSFprematchEPDAL1} and \eqref{eq:LSFprematchEPDAL2} for any smooth $\lambda>0$ if, respectively, (1),
$\unkgen$ is replaced by $\unkd^{(0)}$, $\vnkgen$ by $\switchgen^{(0)} \vnkgen$, and $\sourcegen^{(2)}=\switchgen f^{(2)}$ and $\sourcegen^{(3)}= 0$, and, (2), $\unkgen$ is replaced by $\unkd^{(0)}-\tilde\unkd^{(0)}$, $\vnkgen$ by $\switchgen^{(0)} (\vnkgen-\tilde\vnkgen)$ and $\sourcegen^{(2)}=\sourcegen^{(3)}=0$. For $t_*=\matchtime$ and $t_*=\tilde\matchtime$ (respectively) we find that, imposing the initial conditions in \Eqsref{eq:abstracteqw2} and \eqref{eq:abstracteqw3} (and assuming without loss of generality $\matchtime\le\tilde\matchtime$), we have
\begin{equation}
  \label{eq:LSFLSFdecayestimateprematch}
    \sqrt{\energy{\unkd^{(0)}}k{\lambda}(t)}
  \le
C 
\int_{\matchtime}^t \Bigl(\sqrt{\energy{u}{k}{\lambda}(s)}
+\|s^{-\lambda} f^{(2)}\|_{\delta,H^k(M)}\Bigr)
 s^{-1} ds
\end{equation}
and
\begin{equation*}
    \sqrt{\energy{\unkd^{(0)}-\tilde\unkd^{(0)}}k{\lambda}(t)}
  \le
C \left(
\sqrt{\energy{\unkd^{(0)}}k{\lambda}(\tilde\matchtime)}
 +\int_{\tilde\matchtime}^t
\sqrt{\energy{\vnkgen-\tilde\vnkgen}k{\lambda}(s)}
 s^{-1} ds
\right)
\end{equation*}
for all $t\in[\matchtime,\finaltime]$ and $t\in[\tilde\matchtime,\finaltime]$, respectively.
The constant $C$ here may depend on $\finaltime$, $k$, $\lambda$, and $\Gamma$.
Plugging \Eqref{eq:LSFLSFdecayestimateprematch} into the last estimate with $t=\tilde\matchtime$ 
and applying \Eqref{eq:Holdertimeintegrals} with $\eta_1=0$ and $\eta_2=-2\epsilon$ generates the following result
\begin{equation*}
  \begin{split}
    \energy{\unkd^{(0)}-\tilde \unkd^{(0)}}k{\lambda}(t)
\le
C \Bigl((\tilde\matchtime^{2\epsilon}-\matchtime^{2\epsilon})\sup_{s\in[\matchtime,\tilde\matchtime]}\energy{\vnkgen}k{\lambda+2\epsilon}(s)+ \sup_{s\in[\tilde\matchtime,\finaltime]}\energy{\vnkgen-\tilde\vnkgen}k{\lambda+2\epsilon}(s)\\
+\int_{\matchtime}^{\tilde\matchtime}
\|s^{-\lambda} f^{(2)}\|^2_{\delta,H^k(M)}
 s^{-1-2\epsilon} ds
\Bigr).
\end{split}
\end{equation*}
The constant $C$ here may depend on $\finaltime$, $k$, $\lambda$, $\epsilon$ and $\Gamma$. 
Using \Eqref{eq:LSFbackward1} and \eqref{eq:LSFLSFfullestimate32432} and assuming that $0< \lambda< \min\{\lambda_s,2\kappa,2A^2\}$ with $\lambda_s$ as given in the hypothesis, we obtain
\begin{equation*}
  \begin{split}
    &\energy{\unkd^{(0)}-\tilde \unkd^{(0)}}k{\lambda}(t)
\le
C \Bigl(
(\tilde\matchtime^{2\epsilon}-\matchtime^{2\epsilon})
\Bigl(
\|u_*\|^2_{\delta,H^{k+2}(M)}+\|\varphi_*\|^2_{\delta,H^{k+3}(M)}
\\      
  &+\int_0^{\finaltime}\Bigl(\|s^{-\lambda} f^{(1)}\|^2_{\delta, H^{k+2}(M)}
    +\|s^{-\lambda} f^{(2)}\|^2_{\delta, H^{k+2}(M)}
+\|s^{-\lambda+2\kappa-\epsilon} f^{(3)}\|^2_{\delta,H^{k+3}(M)}\Bigr) s^{-1-6\epsilon}ds
\Bigr)
\\
&+ \|u_*-\CDfinaltime{\tilde u}\|^2_{\delta,H^{k+2}(M)}
+ \|\varphi_*-\CDfinaltime{\tilde\varphi}\|^2_{\delta,H^{k+3}(M)}
+\int_{\matchtime}^{\tilde\matchtime}
\|s^{-\lambda} f^{(2)}\|^2_{\delta,H^k(M)}
 s^{-1-2\epsilon} ds
\Bigr)
\end{split}
\end{equation*}
provided $\epsilon>0$ is sufficiently small. 

If an energy needs to be replaced by a norm, the following estimate becomes handy.  For \emph{any} smooth function $u(t,x)$ and \emph{any} smooth function $\lambda(x)$ we have
\begin{align}
  &\|t^{-\lambda}u(t,\cdot)\|_{\delta,H^k(M)}^2
  =\sum_{l=0}^k\|\partial^l (t^{-\lambda}u(t,\cdot))\|^2_{\delta}\notag\\
  &\le \sum_{l=0}^k\sum_{m=0}^lC_{k,l,m}
  \left\|t^{-\lambda}\partial^m u(t,\cdot) \right\|^2_{\delta}\left\|t^\lambda\partial^{l-m}t^{-\lambda}
  \right\|^2_{\delta}
  \le C t^{-2\epsilon} \sum_{l=0}^k\|t^{-\lambda}\partial^l u(t,\cdot)\|^2_{\delta}\notag\\
  \label{eq:estimatenormbyenergy}
  &=C \energy{u}k{\lambda+\epsilon}(t),
\end{align}
for all $t\in (0,\finaltime]$ and for any $\epsilon>0$, where the constant $C>0$ may depend on $k$ and $\lambda$.

Applying this to the previous estimate and using the fact that all solutions  $\unkt^{(0)}$ of the $\switchgen^{(0)}=\switcht^{(0)}=0$-version of \Eqref{eq:LSFprematchEPDeq} are constant in time, we find
\begin{align}
    &\left\|t^{-\lambda}\left[
\left(\unk^{(0)}(t,\cdot)-\FiniteMatchingMapAbstr{1\rightarrow
    0}^{(0)}(\matchtime,\CDfinaltime{\unk^{(0)}}, \CDfinaltime{\vnkgen},
  \CDfinaltime{\unkgen})\right)
-\left(\tilde\unk^{(0)}(t,\cdot)-\FiniteMatchingMapAbstr{1\rightarrow 0}^{(0)}(\tilde\matchtime,\CDfinaltime{\tilde\unk^{(0)}}, \CDfinaltime{\tilde\vnkgen}, \CDfinaltime{\tilde\unkgen}) \right)
\right]\right\|_{\delta,H^k(M)}^2\notag\\
\label{eq:LSFestprematchepd23843}
\le&
C \Bigl(
(\tilde\matchtime^{2\epsilon}-\matchtime^{2\epsilon})
\Bigl(
\|u_*\|^2_{\delta,H^{k+2}(M)}+ \|\varphi_*\|^2_{\delta,H^{k+3}(M)}
\\      
  &+\int_0^{\finaltime}\Bigl(\|s^{-\lambda} f^{(1)}\|^2_{\delta, H^{k+2}(M)}
    +\|s^{-\lambda} f^{(2)}\|^2_{\delta, H^{k+2}(M)}
+\|s^{-\lambda+2\kappa-\epsilon} f^{(3)}\|^2_{\delta,H^{k+3}(M)}\Bigr) s^{-1-8\epsilon}ds
\Bigr)
\notag\\
&+ \|u_*-\CDfinaltime{\tilde u}\|^2_{\delta,H^{k+2}(M)}
+ \|\varphi_*-\CDfinaltime{\tilde\varphi}\|^2_{\delta,H^{k+3}(M)}\notag
+\int_{\matchtime}^{\tilde\matchtime}
\|s^{-\lambda} f^{(2)}\|^2_{\delta,H^k(M)}
 s^{-1-4\epsilon} ds
\Bigr),\notag
\end{align}
where $C$ may depend on $\finaltime$, $k$, $\lambda$, $\epsilon$ and $\Gamma$.
A uniform continuity estimate for $\FiniteMatchingMapAbstr{1\rightarrow 0}^{(0)}(\matchtime,\CDfinaltime{\unk^{(0)}}, \CDfinaltime{\vnkgen},
  \CDfinaltime{\unkgen})
-\FiniteMatchingMapAbstr{1\rightarrow 0}^{(0)}(\tilde\matchtime,\CDfinaltime{\tilde\unk^{(0)}}, \CDfinaltime{\tilde\vnkgen}, \CDfinaltime{\tilde\unkgen})$ follows directly by setting $t=\finaltime$ and rearranging.

The finite matching map \Eqref{eq:LSFprematchfinitematch} thus has a unique continuous extension to the domain $[0,\finaltime]\times H^{k}(M)\times H^{k+2}(M) \times H^{k+3}(M)$ and the co-domain $H^{k}(M)$. If this extended map is evaluated at $\matchtime=0$ we obtain the map $\AsymptoticMatchingMapIndexRegularityAbstr{0}{1\rightarrow 0}{k}: H^{k}(M)\times H^{k+2}(M) \times H^{k+3}(M)\rightarrow H^{k}(M)$, which we shall claim to be the asymptotic matching map. 
The  continuity property of this map follows from \Eqref{eq:LSFestprematchepd23843} by setting $\matchtime=\tilde\matchtime=0$.
In general this map clearly depends on $k$. However, a standard argument shows that the restriction of this map to 
the dense sub-domain  $(C^\infty(M))^3$ does \emph{not} depend on the choice of $k$.

This restriction 
is referred to as
$\AsymptoticMatchingMapIndexAbstr{0}{1\rightarrow 0}: (C^\infty(M))^3 \rightarrow C^\infty(M)$. 
In order to prove that this is the asymptotic matching map of interest we need to establish  \Eqref{eq:matchingcond} for some choice of norm. Here, however, we are interested in the related map
$\PreMatchingMapAbstr$ given by \Eqref{eq:LSFrelationshipmapsprematch}. The analogue of \Eqref{eq:matchingcond} is \Eqref{eq:LSFLSFprematchestimate} which we establish now
by reconsidering \Eqref{eq:LSFLSFdecayestimateprematch} with $\matchtime=0$ and \Eqref{eq:Holdertimeintegrals} with $\eta_1=0$ and $\eta_2=-2\epsilon$, 
and then combining this with \Eqref{eq:LSFimproveddecayubackward}; thereby obtaining
\begin{align*}
    &\energy{\unkd^{(0)}}k{\lambda}(t)
\le
C  \Bigl(
\|u_*\|^2_{\delta,H^{k+2}(M)}+\|\varphi_*\|^2_{\delta,H^{k+3}(M)}
\\      
  &+\int_0^{\finaltime}\Bigl(\|s^{-\lambda} f^{(1)}\|^2_{\delta, H^{k+2}(M)}
    +\|s^{-\lambda} f^{(2)}\|^2_{\delta, H^{k+2}(M)}
+\|s^{-\lambda+2\kappa-\epsilon} f^{(3)}\|^2_{\delta,H^{k+3}(M)}\Bigr) s^{-1-6\epsilon}ds
\Bigr),
\end{align*}
provided $0<\lambda< \min\{\lambda_s,2\kappa,2A^2\}$. Having replaced the energy on the left-hand side by a norm following \Eqref{eq:estimatenormbyenergy}, we can
we rephrase the conditions for $\lambda$ in a more useful way by exploiting the condition that $\finaltime\in (0,1]$. It is clear that this inequality holds also if $\lambda$ on the right-hand side is replaced by any smooth exponent $\tilde\lambda>0$; in order to guarantee that the right-hand side is finite we demand that $0<\tilde\lambda<\lambda_s$. The exponent $\lambda$ on the left-hand side then only needs to satisfy the upper bound $\lambda<\min\{\tilde\lambda,2\kappa,2A^2\}$. This leads to  
\Eqref{eq:LSFLSFprematchestimate}.
\Eqref{eq:LSFLSFcontinuitypre129031lemma2}  follows directly from \Eqref{eq:LSFestprematchepd23843} by setting $\matchtime=\tilde\matchtime$ and noticing that this inequality holds for any smooth $\lambda<\min\{2\kappa,2A^2\}$. 

Lastly we notice that the resulting map is uniquely determined by \Eqref{eq:LSFLSFprematchestimate}, given that all solutions of the $\switchgen^{(0)}=0$-version of \Eqref{eq:LSFprematchEPDeq} are constant.
\end{proof}

\begin{proof}[Proof of \Propref{prop:LSFresult1}]
  \hypertarget{propproof:LSFresult1}{}
We notice that \Eqref{eq:abstracteqgenNN} with $A$ and $L$ given by \Eqsref{eq:LSFLSFasldkasdk1} and \eqref{eq:LSFLSFasldkasdkL}, and, with $F$ given by \Eqref{eq:LSFLSFmodLem12} and \Eqref{eq:LSFLSFmodLem22}, respectively,
are of the form \Eqsref{eq:abstracteq} and \eqref{eq:abstracteqt} with 
\begin{align*}
  \source&=\left(f^{(1)} +(\switch-K)\left({\BG\alpha} t^2\BG\gamma^{ab}\partial_a \partial_b\LOT{\unk}
-\left(t^2{\BG\alpha} C^a-t^2\BG\gamma^{ab}\partial_b \BG\alpha\right)\partial_a\LOT{\unk}\right),  f^{(2)},  f^{(3)}\right),\\
  \sourcet&=\left(f^{(1)} +(\switcht-K) \left({\BG\alpha} t^2\BG\gamma^{ab}\partial_a \partial_b\LOT{\unk}
-\left(t^2{\BG\alpha} C^a-t^2\BG\gamma^{ab}\partial_b \BG\alpha\right)\partial_a\LOT{\unk}\right),  f^{(2)},  f^{(3)}\right).
\end{align*}
In \Eqsref{eq:abstracteq} -- \eqref{eq:abstracteqw4} we then assume that
\begin{align*}
  \tilde\source&=\left(f^{(1)} +(\switch-K)\left({\BG\alpha} t^2\BG\gamma^{ab}\partial_a \partial_b\LOT{\tilde\unk}
-\left(t^2{\BG\alpha} C^a-t^2\BG\gamma^{ab}\partial_b \BG\alpha\right)\partial_a\LOT{\tilde\unk}\right),  f^{(2)},  f^{(3)}\right),\\
  \tilde\sourcet&=\left(f^{(1)} +(\switcht-K) \left({\BG\alpha} t^2\BG\gamma^{ab}\partial_a \partial_b\LOT{\tilde\unk}
-\left(t^2{\BG\alpha} C^a-t^2\BG\gamma^{ab}\partial_b \BG\alpha\right)\partial_a\LOT{\tilde\unk}\right),  f^{(2)},  f^{(3)}\right),
\end{align*}
where $\LOT{\unk}$ and $\LOT{\tilde\unk}$ are defined as in the proposition. It is useful to write $L$ in terms of the coefficients
\Eqsref{eq:coeff1} -- \eqref{eq:coefflast}. 

Under the given hypothesis
the finite matching map (analogous to the one defined in \Defref{def:FMatchingProblem}) is taken to be of the type
\begin{equation}
  \label{eq:LSFLSFfinitematch}
  \FiniteMatchingMapAbstr{\switch\rightarrow \switcht}: (0,\finaltime]\times (C^\infty(M))^2\rightarrow (C^\infty(M))^2,\quad (\matchtime, \AD{\vnk}, \AD{\unk}) \mapsto (\CDfinaltime{\vnkt}, \CDfinaltime{\unkt}),
\end{equation}
which is certainly well-defined.
According to the general strategy,
the plan 
is now to analyze the hierarchy of Cauchy problems \Eqsref{eq:abstracteq}, \eqref{eq:abstracteqw4}, \eqref{eq:abstracteqw2} and \eqref{eq:abstracteqw3}.
We write
\[\unkd^{(1)}= \vnk-\vnkt,\quad \unkd^{(2)}= \unk-\unkt,\quad \unkd^{(3)}= \nunk-\nunkt.\]
Given \Eqsref{eq:LSFLSFLOTestimate} and \eqref{eq:LSFLSFLOTestimate2}, which can be interpreted as given estimates for solutions of the Cauchy problems \eqref{eq:abstracteq} and \eqref{eq:abstracteqw4},  let us  proceed with the analysis of the Cauchy problems \Eqsref{eq:abstracteqw2} and \eqref{eq:abstracteqw3}.
The source terms of \Eqsref{eq:abstracteqw2} and \Eqref{eq:abstracteqw3} take the form
\begin{align*}
  F_3 
  =&(\switcht-\switch)\Bigl( 
{\BG\alpha} t^2\BG\gamma^{ab}\partial_a \partial_b(\unk-\LOT{\unk})
+\left(t^2{\BG\alpha} C^a-t^2\BG\gamma^{ab}\partial_b \BG\alpha\right)\partial_a(\unk-\LOT{\unk})
\\ 
&\qquad + L^{[1,1]}_1 \vnk 
  + L^{[1,2]}_1\nunk 
  + L_1^{[1,4],b}\partial_b\nunk,\\
  &\qquad 
   t^2\BG\gamma^{ab}\partial_a \partial_b\nunk
  - t^2C^a \partial_a\nunk
  + L^{[2,1]}_1 \nunk
  + L^{[2,2]}_1 \vnk,
  \qquad 
L^{[3,1]}_1 \nunk
  + L^{[3,2]}_1 \vnk\Bigr),\\
  F_4 
  =&(\switcht-\switch)\Bigl(
{\BG\alpha} t^2\BG\gamma^{ab}\partial_a \partial_b ((\unk-\tilde\unk)+(\LOT{\unk}-\LOT{\tilde\unk}))\\
&\qquad+\left(t^2{\BG\alpha} C^a-t^2\BG\gamma^{ab}\partial_b \BG\alpha\right)\partial_a ((\unk-\tilde\unk)
-(\LOT{\unk}-\LOT{\tilde\unk}))
\\ 
&\qquad + L^{[1,1]}_1 (\vnk-\tilde\vnk) 
  + L^{[1,2]}_1(\nunk-\tilde\nunk) 
  + L_1^{[1,4],b}\partial_b(\nunk-\tilde\nunk),\\
  &\qquad
   t^2\BG\gamma^{ab}\partial_a \partial_b(\nunk-\tilde\nunk)
  - t^2C^a \partial_a(\nunk-\tilde\nunk)
  + L^{[2,1]}_1 (\nunk-\tilde\nunk)
  + L^{[2,2]}_1 (\vnk-\tilde\vnk),\\ 
  &\qquad
L^{[3,1]}_1 (\nunk-\tilde\nunk)
  + L^{[3,2]}_1 (\vnk-\tilde\vnk)\Bigr).
\end{align*}
The first aim is now to estimate these $F_3$ and $F_4$. This is straightforward. For example,
\newcommand{\SobolevdeltaNorm}[2]{\left\|#1\right\|_{\delta,H^{#2}(M)}}
\begin{align*}
  &\SobolevdeltaNorm{t^{-\lambda}{\BG\alpha} t^2\BG\gamma^{ab}\partial_a \partial_b(\unk-\LOT{\unk})}{k}^2  
\le C \Bigl(
\|\AD{\vnk}\|^2_{\delta,H^{k+4+k_0}(M)}+ \|\AD{\unk}\|^2_{\delta,H^{k+4+k_1}(M)}
\\      
  &\qquad+\int_0^{\finaltime}\Bigl(\|s^{-\lambda_s} f^{(1)}\|^2_{\delta, H^{k+4}(M)}
    +\|s^{-\lambda_s} f^{(2)}\|^2_{\delta, H^{k+4}(M)}
+\|s^{-\lambda_s+2\kappa} f^{(3)}\|^2_{\delta,H^{k+5}(M)}\Bigr) s^{-1}ds
\Bigr)
\end{align*}
using the multiplication property of Sobolev-regular functions (see for example Proposition~2.3 in Appendix~I of \cite{choquet-bruhat2008c}), 
and, using \Eqref{eq:LSFLSFLOTestimate}
for any smooth 
\begin{equation}
  \label{eq:LSFlambdcond1}
  \lambda<2\kappa+\min\{\lambda_s,2(1-q_{max}),2A^2\}, 
\end{equation}
where $\kappa$ is defined in \Propref{prop:LSF4}.
Under the same assumptions we find
\begin{equation*}
  \SobolevdeltaNorm{t^{-\lambda} t^2{\BG\alpha} C^a\partial_a(\unk-\LOT{\unk})}{k}^2
\le C
\SobolevdeltaNorm{t^{-\lambda+2\kappa-\epsilon} (\unk-\LOT{\unk})}{k+1}^2,
\end{equation*}
using \Eqref{eq:Ca}, which can therefore be estimated by the same expression as above. This expression also bounds the term $t^2\BG\gamma^{ab}\partial_b \BG\alpha\partial_a(\unk-\LOT{\unk})$ 
(note that an improved bound could be obtained by taking into account that $\partial_a\BG\alpha=O(t^\beta)$).
Exploiting the known asymptotics of the coefficients together with \Eqref{eq:estimatenormbyenergy} and \Eqref{eq:preestimat3N} with $\sourcegen^{(2)}= f^{(2)}$, we also find that
\begin{align*}
  &\SobolevdeltaNorm{t^{-\lambda}\left(L^{[1,1]}_1 \vnk 
  +L^{[1,2]}_1\nunk 
 +L_1^{[1,4],b}\partial_b\nunk\right)}{k}^2
  \\  
&\le C\Bigl(
  \energy{\vnk}{k+1}{\lambda-\mu+2\epsilon}(t)  
  +\|t^{-\lambda+\mu-2\epsilon} f^{(2)}\|^2_{\delta,H^{k+1}(M)}  
  \Bigr).
\end{align*}
This holds for any smooth function $\lambda(x)$ and $\mu(x)$ where\footnote{The function $\mu$ here is unrelated to the function $\mu$ introduced in \Eqref{eq:preenergyid13N}.} 
\begin{equation}
    \label{eq:LSFmucond2}
    \mu\in (0,\min\{\beta,2\kappa\}).
  \end{equation}
Similarly
\begin{align*}
  &\SobolevdeltaNorm{t^{-\lambda}\left(- t^2\BG\gamma^{ab}\partial_a \partial_b\nunk
  + t^2C^a \partial_a\nunk
  - L^{[2,1]}_1 \nunk
  - L^{[2,2]}_1 \vnk\right)}{k}^2
  \\
&\le C\Bigl(
  \energy{\vnk}{k+2}{\lambda-\mu}(t)   
+\|s^{-\lambda+\mu} f^{(2)}\|^2_{\delta, H^{k+2}(M)}
\Bigr),
\end{align*}
and
\begin{equation*}
  \SobolevdeltaNorm{t^{-\lambda}\left(L^{[3,1]}_1 \nunk
  +L^{[3,2]}_1 \vnk\right)}{k}^2
  \le C\Bigl(
\energy{\vnk}{k}{\lambda-\beta+\epsilon}(t)  
+\|s^{-\lambda+\beta-\epsilon} f^{(2)}\|^2_{\delta, H^{k}(M)}
\Bigr).
\end{equation*}
All the constants $C>0$ in these estimates may depend on $\finaltime$, $k$, $\lambda$, $\epsilon$ and $\Gamma$.
Putting these now together yields
\begin{align}
  \label{eq:LSFestF3sum}
    &\SobolevdeltaNorm{t^{-\lambda} F_3^{(1)}}{k}^2+\SobolevdeltaNorm{t^{-\lambda} F_3^{(2)}}{k}^2+\SobolevdeltaNorm{t^{-\lambda} F_3^{(3)}}{k+1}^2\\
    &\le C\Bigl(
    \energy{\vnk}{k+2}{\lambda-\mu+2\epsilon}(t)
 +\|\AD{\vnk}\|^2_{\delta,H^{k+4+k_0}(M)}+ \|\AD{\unk}\|^2_{\delta,H^{k+4+k_1}(M)}\notag\\
  &+\int_0^{\finaltime}\Bigl(\|s^{-\lambda_s} f^{(1)}\|^2_{\delta, H^{k+4}(M)}
    +\|s^{-\lambda_s} f^{(2)}\|^2_{\delta, H^{k+4}(M)}
+\|s^{-\lambda_s+2\kappa} f^{(3)}\|^2_{\delta,H^{k+5}(M)}\Bigr) s^{-1}ds
    \Bigr),\notag
\end{align}
and
\begin{align}
 \label{eq:LSFestF4sum}
    &\SobolevdeltaNorm{t^{-\lambda} F_4^{(1)}}{k}^2+\SobolevdeltaNorm{t^{-\lambda} F_4^{(2)}}{k}^2+\SobolevdeltaNorm{t^{-\lambda} F_4^{(3)}}{k+1}^2\\
    &\le C\Bigl(
    \energy{\vnk-\tilde\vnk}{k+2}{\lambda-\mu+2\epsilon}(t)    
      +\|\AD{\vnk}-\AD{\tilde \vnk}\|^2_{\delta,H^{k+4+k_0}(M)}+ \|\AD{\tilde \unk}-\AD{\unk}\|^2_{\delta,H^{k+4+k_1}(M)}    
    \Bigr).\notag
\end{align}
Both these estimates are valid if \Eqsref{eq:LSFlambdcond1} and \eqref{eq:LSFmucond2} hold. The constants $C>0$ in these estimates may depend on $\finaltime$, $k$, $\lambda$, $\epsilon$ and $\Gamma$.

Before we proceed, let us note the following  estimate obtained by combining \Eqref{eq:LSFforward2} with \Eqsref{eq:estimatenormbyenergy} and \eqref{eq:Holdertimeintegrals} for $\eta_1=0$ and $\eta_2=-2\epsilon$. This estimate is useful in a number of steps below. To this end, consider
\Eqref{eq:abstracteqgenNN} with $A$ and $L$ given by \Eqsref{eq:LSFLSFasldkasdk1} and \eqref{eq:LSFLSFasldkasdkL}, but with an arbitrary $\switchgen\in[0,1]$ and $\sourcegen = (F^{(1)}, F^{(2)}, F^{(3)})$.
Pick an arbitrary smooth function $\lambda(x)>0$ and an arbitrary integer $k\ge 0$. Then
\begin{align}
  \label{eq:LSFnormenergybla82}
  &\SobolevdeltaNorm{t^{-\lambda}\vnkgen(t,\cdot)}{k}^2
  +\SobolevdeltaNorm{t^{-\lambda}\unkgen(t,\cdot)}{k}^2
  \le C\Bigl(\energy{\varphi}{k}{\lambda+\epsilon}(t_*)
  +\sup_{s\in[t_*,t]}\energy{\vnkgen}{k}{\lambda+3\epsilon}(s)
\notag\\
&\qquad+
\int_{t_*}^t\Bigl(\|s^{-\lambda} \sourcegen^{(2)}\|^2_{\delta,H^k(M)}
+\|s^{-\lambda} \sourcegen^{(3)}\|^2_{\delta,H^k(M)}      
  \Bigr) s^{-1-4\epsilon} ds
\Bigr),
\end{align}
for all $t\in[t_*,\finaltime]$.
The constant $C$ here may depend on $\finaltime$, $k$, $\lambda$, $\epsilon$ and $\Gamma$.

Given this general estimate, we pick now an arbitrary smooth
\begin{equation}
  \label{eq:LSFLSFlambda1}
  \lambda>\lambda_c,
\end{equation}
where $\lambda_c$ is defined in \Eqref{eq:deflambdac},
and set $t_*=\matchtime$ and $t_*=\tilde\matchtime$ (respectively). \Eqref{eq:LSFforwards1} applied to the two Cauchy problems \Eqsref{eq:abstracteqw2} and \eqref{eq:abstracteqw3} yields (assuming $\matchtime\le\tilde\matchtime$)
\begin{align}
\label{eq:LSFLSFsdflkjsdflkj}
    &\energyswitch{\unkd^{(1)}, \unkd^{(2)}}\switcht{k}{\lambda}(t)\\
&\le  C
      \int^t_{\matchtime}\Bigl(\|s^{-\lambda} F_3^{(1)}\|^2_{\delta,H^k(M)} 
      +\|s^{-\lambda} F_3^{(2)}\|^2_{\delta,H^k(M)} 
      +\|s^{-\lambda} F_3^{(3)}\|^2_{\delta,H^{k+1}(M)}\Bigr) s^{-1+2\epsilon} ds,\notag
\end{align}
and 
\begin{align}
  \label{eq:LSFakldfc93jf93}
    &\energyswitch{\unkd^{(1)}-\tilde\unkd^{(1)}, \unkd^{(2)}-\tilde\unkd^{(2)}}\switcht{k}{\lambda}(t)\\
&\le  C\Bigl(\int^{\tilde\matchtime}_{\matchtime}\Bigl(\|s^{-\lambda} F_3^{(1)}\|^2_{\delta,H^k(M)} 
      +\|s^{-\lambda} F_3^{(2)}\|^2_{\delta,H^k(M)} 
      +\|s^{-\lambda} F_3^{(3)}\|^2_{\delta,H^{k+1}(M)}\Bigr) s^{-1+2\epsilon} ds\notag\\
      &+\int^t_{\tilde\matchtime}\Bigl(\|s^{-\lambda} F_4^{(1)}\|^2_{\delta,H^k(M)} 
      +\|s^{-\lambda} F_4^{(2)}\|^2_{\delta,H^k(M)} 
      +\|s^{-\lambda} F_4^{(3)}\|^2_{\delta,H^{k+1}(M)}\Bigr) s^{-1+2\epsilon} ds 
      \Bigr),\notag
\end{align}
for all $t\in [\matchtime,\finaltime]$ and $t\in [\tilde\matchtime,\finaltime]$.
We combine  these now  with \Eqref{eq:LSFnormenergybla82} 
\begin{align}
  \label{eq:LSFjvnskeodfjalj}
  &\SobolevdeltaNorm{t^{-\lambda}\unkd^{(1)}(t,\cdot)}{k}^2
+\SobolevdeltaNorm{t^{-\lambda}\unkd^{(2)}(t,\cdot)}{k}^2
  \\
&\le  C
\int^t_{\matchtime}\Bigl(\|s^{-\lambda} F_3^{(1)}\|^2_{\delta,H^k(M)} 
      +\|s^{-\lambda} F_3^{(2)}\|^2_{\delta,H^k(M)} 
      +\|s^{-\lambda} F_3^{(3)}\|^2_{\delta,H^{k+1}(M)}\Bigr) s^{-1-4\epsilon} ds,\notag
\end{align}
provided \Eqref{eq:LSFLSFlambda1} holds (notice that $\lambda_c>0$; see \Eqref{eq:deflambdac}), and then find
\begin{align*}
  &\SobolevdeltaNorm{t^{-\lambda}(\unkd^{(1)}(t,\cdot)-{\tilde\unkd^{(1)}}(t,\cdot))}{k}^2
+\SobolevdeltaNorm{t^{-\lambda}(\unkd^{(2)}(t,\cdot)-{\tilde\unkd^{(2)}}(t,\cdot))}{k}^2
  \\
&\le  C\Bigl(
(\tilde\matchtime^{2\epsilon}-\matchtime^{2\epsilon})
  \int^{\tilde\matchtime}_{\matchtime}\Bigl(\|s^{-\lambda} F_3^{(1)}\|^2_{\delta,H^k(M)} 
      +\|s^{-\lambda} F_3^{(2)}\|^2_{\delta,H^k(M)} 
      +\|s^{-\lambda} F_3^{(3)}\|^2_{\delta,H^{k+1}(M)}\Bigr) s^{-1-4\epsilon} ds 
      \\
&+\int^{\tilde\matchtime}_{\matchtime}\Bigl(\|s^{-\lambda} F_3^{(1)}\|^2_{\delta,H^k(M)} 
      +\|s^{-\lambda} F_3^{(2)}\|^2_{\delta,H^k(M)} 
      +\|s^{-\lambda} F_3^{(3)}\|^2_{\delta,H^{k+1}(M)}\Bigr) s^{-1-4\epsilon} ds\\
      &+\int^t_{\tilde\matchtime}\Bigl(\|s^{-\lambda} F_4^{(1)}\|^2_{\delta,H^k(M)} 
      +\|s^{-\lambda} F_4^{(2)}\|^2_{\delta,H^k(M)} 
      +\|s^{-\lambda} F_4^{(3)}\|^2_{\delta,H^{k+1}(M)}\Bigr) s^{-1-4\epsilon} ds 
\Bigr).
\end{align*}
In both cases, the constant $C$ may depend on $\finaltime$, $k$, $\lambda$, $\epsilon$ and $\Gamma$.
We wish to combine these estimates with \Eqsref{eq:LSFestF3sum} and \eqref{eq:LSFestF4sum}. In total we therefore  find the restrictions \Eqref{eq:LSFmucond2} and
\begin{equation}
  \label{eq:LSFlambdcond1N}
  \lambda_c<\lambda<2\kappa+\min\{\lambda_s,2(1-q_{max}),2A^2\}.
\end{equation} 
The energy terms remaining in \Eqsref{eq:LSFestF3sum} and \eqref{eq:LSFestF4sum}
can be estimated with \Eqref{eq:LSFLSFLOTestimate2} under the condition that 
\[\lambda<\mu+\min\{\lambda_s,2\kappa,2A^2\}.\]
This implies that \Eqsref{eq:LSFtheominequFinal},  \eqref{eq:LSFLSFmodLem293482} and \eqref{eq:LSFLSF1cond4} must hold.
Continuing to assume that $\epsilon>0$ is sufficiently small (and replacing multiples of $\epsilon$ by smaller multiples if convenient)
and applying \Eqref{eq:LSFLSFLOTestimate2}, we generate the conclusion
\begin{align*}
  &\SobolevdeltaNorm{t^{-\lambda}(\unkd^{(1)}(t,\cdot)-{\tilde\unkd^{(1)}}(t,\cdot))}{k}^2
+\SobolevdeltaNorm{t^{-\lambda}(\unkd^{(2)}(t,\cdot)-{\tilde\unkd^{(2)}}(t,\cdot))}{k}^2
  \\
&\le  C\Bigl(
(\tilde\matchtime^{2\epsilon}-\matchtime^{2\epsilon})
  \Bigl(
\|\AD{\vnk}\|^2_{\delta,H^{k+4+k_0}(M)}+ \|\AD{\unk}\|^2_{\delta,H^{k+4+k_1}(M)}\\
    &\quad+\int_0^\finaltime\Bigl(\SobolevdeltaNorm{s^{-\lambda_s} f^{(1)}}{k+4}^2 
+\SobolevdeltaNorm{s^{-\lambda_s} f^{(2)}}{k+4}^2    
    +\SobolevdeltaNorm{s^{-\lambda_s+2\kappa} f^{(3)}}{k+5}^2\Bigr)s^{-1}ds
\Bigr)
      \\
&+\int^{\tilde\matchtime}_{\matchtime}
\Bigl(\SobolevdeltaNorm{s^{-\lambda_s} f^{(1)}}{k+4}^2 
+\SobolevdeltaNorm{s^{-\lambda_s} f^{(2)}}{k+4}^2    
    +\SobolevdeltaNorm{s^{-\lambda_s+2\kappa} f^{(3)}}{k+5}^2\Bigr) s^{-1} ds\\      
 &   +\|\AD{\vnk}-\AD{\tilde \vnk}\|^2_{\delta,H^{k+4+k_0}(M)}+ \|\AD{\tilde \unk}-\AD{\unk}\|^2_{\delta,H^{k+4+k_1}(M)}  
\Bigr).
\end{align*}
This holds for any smooth $\lambda$ satisfying \Eqref{eq:LSFLSFmodLem29348} provided \Eqref{eq:LSFLSFmodLem293482} holds.
Evaluating this at $t=\finaltime$, transferring all norms of $\CDfinaltime{\vnk}$, $\CDfinaltime{\tilde\vnk}$, $\CDfinaltime{\unk}$, $\CDfinaltime{\tilde\unk}$ from the left side to the right side and then estimating them in terms of $\AD{\vnk}$, $\AD{\tilde \vnk}$, $\AD{\unk}$, $\AD{\tilde \unk}$  exploiting the continuity of $\Phi$, this
 leads to the required   uniform continuity estimate for the finite matching map.
This map \Eqref{eq:LSFLSFfinitematch} thus has a unique continuous extension to the domain $[0,\finaltime]\times H^{k+4+k_0}(M)\times H^{k+4+k_1}(M)$ and the co-domain $(H^{k}(M))^2$. The extended map evaluated at $\matchtime=0$ is referred to as $\AsymptoticMatchingMapRegularityAbstr{\switch\rightarrow \switcht}{k}: H^{k+4+k_0}(M)\times H^{k+4+k_1}(M)\rightarrow (H^{k}(M))^2$ which we claim to be the asymptotic matching map asserted in the proposition. 
It is  continuous in the sense
\begin{equation*}
\begin{split}
&\|\AsymptoticMatchingMapRegularityAbstr{\switch\rightarrow \switcht}k(\AD{\vnk}, \AD{\unk})
-\AsymptoticMatchingMapRegularityAbstr{\switch\rightarrow \switcht}k(\AD{\tilde \vnk}, \AD{\tilde \unk})\|_{H^k(M)}
\\
&\le C \Bigl(
\|\AD{\vnk}-\AD{\tilde \vnk}\|^2_{\delta,H^{k+4+k_0}(M)}+ \|\AD{\tilde \unk}-\AD{\unk}\|^2_{\delta,H^{k+4+k_1}(M)}  
\Bigr),
\end{split}
\end{equation*}
where the constant $C>0$ may depend on $\finaltime$, $k$, $\lambda$, $\Phi$ and $\Gamma$.
Even though $\AsymptoticMatchingMapRegularityAbstr{\switch\rightarrow \switcht}k $  clearly depends on $k$, standard arguments imply that  its restriction to the dense sub-domain  $(C^\infty(M))^2$ does \emph{not} depend on the choice of $k$ and 
therefore yields the map
$\AsymptoticMatchingMapAbstr{\switch\rightarrow \switcht}: (C^\infty(M))^2 \rightarrow (C^\infty(M))^2$.
This map satisfies the continuity estimate \Eqref{eq:LSFLSFmodLem23098}.
\Eqref{eq:LSFLSFmodLem23132} follows if we apply the same chain of arguments which we have used to obtain the continuity estimate now
to 
\Eqref{eq:LSFjvnskeodfjalj} and then take the limit $\matchtime\searrow 0$.

Let us next investigate the uniqueness statement \Eqref{eq:LSFLSFmodLem1239} of \Propref{prop:LSFresult1}. Pick any smooth Cauchy data $(\AD{\vnk}, \AD{\unk})$ and let  $(\vnk,\nunk,\unk)$ be the corresponding solution, and let $(\vnkt,\nunkt,\unkt)$ be the solution determined by the Cauchy data $(\CDfinaltime{\vnkt}, \CDfinaltime{\unkt})
=\AsymptoticMatchingMapAbstr{\switch\rightarrow \switcht} (\AD{\vnk}, \AD{\unk})$ as above. Then given any other (possibly different) smooth solution $(\hat\vnkt,\hat\nunkt,\hat\unkt)$ of the same target equation we define
\[\omega=(\omega^{(1)}, \omega^{(2)}, \omega^{(3)})=(\vnkt,\unkt,\nunkt)-(\hat\vnkt,\hat\unkt,\hat\nunkt).\]
This is therefore a smooth solution of the $\switchgen=\switcht$-version of the equation with \emph{zero source term}. 
Given any $\lambda>\lambda_c$ and sufficiently small $\epsilon>0$, \Eqsref{eq:LSFnormenergybla82} together with \Eqref{eq:LSFforwards1} yields
\begin{equation*}
  \begin{split}
  &\SobolevdeltaNorm{t^{-\lambda+5\epsilon}\omega^{(1)}(t,\cdot)}{0}^2
+\SobolevdeltaNorm{t^{-\lambda+5\epsilon}\omega^{(2)}(t,\cdot)}{0}^2\\
  \le &C\Bigl(
  t_*^{8\epsilon}\left\|t_*^{-\lambda}\omega^{(2)}(t_*,\cdot)\right\|_{L^2(M)}^2
  +t_*^{2\epsilon}\Bigl(\left\|t_*^{-\lambda+\epsilon}\omega^{(1)}(t_*,\cdot)\right\|_{L^2(M)}^2
    +\left\|t_*^{-\lambda}\omega^{(2)}(t_*,\cdot)\right\|_{H^1(M)}^2
\Bigr)
\Bigr)
\end{split}
\end{equation*}
for any $t_*\in (0,\finaltime]$ and for all $t\in [t_*,\finaltime]$. Assuming that $\lambda$ is now in the range specified by \Eqref{eq:LSFLSFmodLem29348}, it follows that the right-hand side vanishes in the limit $t_*\searrow 0$. This implies  uniqueness.

Finally let us pick any two smooth pairs $(\AD{\vnk}, \AD{\unk})$ and $(\AD{\tilde \vnk}, \AD{\tilde \unk})$. Supposing that
\begin{equation}
  \label{eq:LSFLSFinjproof2}
  \AsymptoticMatchingMapAbstr{\switch\rightarrow \switcht}(\AD{\vnk}, \AD{\unk})
=\AsymptoticMatchingMapAbstr{\switch\rightarrow \switcht}(\AD{\tilde \vnk}, \AD{\tilde \unk})
=(\CDfinaltime{\vnkt},\CDfinaltime{\unkt}),
\end{equation}
we need to establish that $\LOT{\unk}=\LOT{\tilde \unk}$. 
Let $(\vnk,\nunk,\unk)$, $(\tilde\vnk,\tilde\nunk,\tilde\unk)$ and $(\vnkt,\nunkt,\unkt)$ be the solutions of the corresponding Cauchy problems. For any smooth $\lambda<\min\{\lambda_s,2(1-q_{max}),2A^2\}$ and for all $t\in (0,\finaltime]$ we have
\begin{align*}
  &\left\|t^{-\lambda}\left(
  \LOT{\unk}-\LOT{\tilde\unk}
  \right)\right\|_{H^k(M)}\\
  \le &
  \left\|t^{-\lambda}\left(
  \unk(t,\cdot)-\LOT{\unk}
  \right)\right\|_{H^k(M)}
  +
  \left\|t^{-\lambda}\left(
  \tilde\unk(t,\cdot)-\LOT{\tilde\unk}
  \right)\right\|_{H^k(M)}\\
  & +
  \left\|t^{-\lambda}\left(
  \unk(t,\cdot)-\unkt(t,\cdot)
  \right)\right\|_{H^k(M)}
  +
  \left\|t^{-\lambda}\left(
  \tilde\unk(t,\cdot)-\unkt(t,\cdot)
  \right)\right\|_{H^k(M)}.
\end{align*}
It is a consequence of \Eqref{eq:LSFLSFLOTestimate} and of \Eqref{eq:LSFLSFmodLem23132} that $\lambda$ can be chosen such that the right-hand side approaches zero in the limit $t\searrow 0$. 
This completes the proof of \Propref{prop:LSFresult1}.
\end{proof}

\section*{Acknowledgments}
EA thanks the Knut and Alice Wallenberg foundation for support. FB
would like to thank the Departments of Mathematics of the University
of Oregon and of Kungliga Tekniska högskolan (KTH) for their
hospitality during his sabbatical visits where some of the work for
this article was done. JI was supported by NSF grants NSF Division of
Physics award number 1707427.

\addcontentsline{toc}{section}{References}

\end{document}